\documentclass[onluarkali,ingilizce,yukseklisans,bez,fenbilimleri]{itutez}

\usepackage{listings}
\usepackage{courier}
\usepackage{amsmath}

\yazar{Nazmi Burak}{BUDANUR} 
\ogrencino{509101132}

\unvan{Physicist}
 
\anabilimdali{Fizik M\"uhendisli\u{g}i Anabilim Dal{\i}}{Department of Physics Engineering}
\programi{Fizik M\"uhendisli\u{g}i Program{\i}}{Physics Engineering Programme}

\tarih{HAZ\.IRAN 2012}{JUNE 2012}
\tarihKucuk{Haziran 2012}{June 2012}

\tezyoneticisi{Prof. Dr. Cenap \c{S}ahabettin \"{O}ZBEN}{Istanbul Technical University}   

\baslik{GAZLI \.IYON\.IZASYON DEDEKT\"{O}RLER\.INDEK\.I D\.IREN\c{C}L\.I YAPILARDA}{Y\"{U}K TA\c{S}INIMININ BENZET\.IM \c{C}ALI\c{S}MALARI}{}

\title{SIMULATION STUDIES OF CHARGE TRANSPORT ON RESISTIVE}{STRUCTURES IN GASEOUS IONIZATION DETECTORS}{}

\tezvermetarih{17 May{\i}s 2012}{17 May 2012} 

\tezsavunmatarih{8 Haziran 2012}{8 June 2012}

\esdanismani{Prof. Dr. Serkant Ali \c{C}ET\.IN}{Do\u{g}u\c{s} University}   

\juriBir{Prof. Dr. Nazmi POSTACIO\u{G}LU}{Istanbul Technical University}

\juriIki{Assoc. Prof. Dr. Taylan AKDO\u{G}AN}{Bo\u{g}azi\c{c}i University}

\juriUc{Assoc. Prof. Dr. Veysi Erkcan \"{O}ZCAN}{Bo\u{g}azi\c{c}i University}

\juriDort{}{}

\juriBes{}{}

\usepackage{color}
\usepackage{times}
\usepackage{amssymb,amsmath,mathptmx,amsbsy,bm}
\usepackage{caption}            
\usepackage{graphics}
\usepackage{wrapfig}
\usepackage{epsfig}
\usepackage{enumerate}
\usepackage{rotating}
\usepackage{multirow}
\usepackage{subfigure}
\usepackage{colortbl}
\usepackage{pstricks}
\usepackage{pst-plot}
\usepackage{cite}%
\usepackage{latexsym}           %
\usepackage{rotating}

\def\be{\begin{equation}} %
\def\ee{\end{equation}}%
\def\beq{\begin{eqnarray}}%
\def\eeq{\end{eqnarray}}%
\def\bse{\begin{subequations}}%
\def\ese{\end{subequations}}%
\def\[{\left[}
\def\]{\right]}
\def\({\left(}
\def\){\right)}

\ithaf{To my parents and sister,}

\kisaltmalistesi{\hspace{-3mm}
\begin{tabular}{p{2cm}l}
{\bf{ALICE}} & {\bf:} A Large ION Collider Experiment\\
{\bf ATLAS} & {\bf:} A Torroidal LHC Apparatus\\
{\bf CERN} & {\bf:} European Organization for Nuclear Research\\
{\bf CMS} & {\bf:} Compact Muon Selonoid\\
{\bf COMPASS} & {\bf:} Common Muon and Proton Apparatus for Structure and Spectroscopy\\
{\bf CSC} & {\bf:} Cathode Strip Chamber\\
{\bf DESY} & {\bf:} Deutsches Elektronen-Synchrotron\\
{\bf EM} & {\bf:} Electromagnetic\\
{\bf GEM} & {\bf:} Gas Electron Multiplier\\
{\bf HERA-B} & {\bf:} Hadron-Elektron-Ringanlage-B\\
{\bf LAr} & {\bf:} Liquid Argon\\
{\bf LEP} & {\bf:} Large Electron-Positron Collider\\
{\bf LHC} & {\bf:} Large Hadron Collider\\
{\bf LHCb} & {\bf:} Large Hadron Collider Beauty\\
{\bf LINAC} & {\bf:} Linear Accelerator\\
{\bf MAMMA} & {\bf:} Muon ATLAS MicroMegas Activity\\
{\bf MPGD} & {\bf:} Micropattern Gaseous Detectors\\
{\bf MSGC} & {\bf:} Microstrip Gas Chamber\\
{\bf MWPC} & {\bf:} Multiwire Proportional Chamber\\
{\bf MDT} & {\bf:} Monitored Drift Tube\\
{\bf PCB} & {\bf:} Printed Circuit Board\\
{\bf PS} & {\bf:} Proton Synchrotron\\
{\bf R\&D} & {\bf:} Research and Development\\
{\bf RPC} & {\bf:} Resistive Plate Chamber\\
{\bf SC} & {\bf:} Synchrocyclotron\\
{\bf SPS} & {\bf:} Super Proton Synchrotron\\
{\bf TGC} & {\bf:} Thin Gap Chamber\\
{\bf T2K} & {\bf:} Tokai to Kamioka\\
\end{tabular}

}

\onsoz{This thesis is written during an exciting time for particle physics. The Large Hadron Collider at CERN (Geneva) has been operational for longer than two years and huge amount of data have been taken. At the end of the last year, it was announced that the existence or nonexistence of the Higgs boson, the last missing piece of the Standard Model of the elementary particles, is going to be confirmed with the collision data of 2012 by the end of the year. At the end of the year 2012, a two-year shut down period will begin to prepare the collider for its full performance goals. \newline
LHC is the largest and highest energy particle accelerator and collider that has been built until now and it is going to be upgraded to achieve even higher luminosities. This upgrade project is called the “Super Large Hadron Collider” project and it is scheduled between 2013 – 2018. The studies I present in this thesis are indirectly motivated by the sLHC.\newline
MAMMA (Muon ATLAS MicroMegas Activity) group at CERN conducts R\&D on micromegas-type detectors for use in the ATLAS detector at the LHC after the sLHC luminosity upgrade. In these detectors, high resistivity materials are used as anode electrodes for spark protection. Resistivity and dimensions of these electrodes are determined through trial and error processes. Main work of this thesis is the development of a simulation tool, for understanding the charge spread and discharge dynamics on these resistive anode strips. \newline
This simulation tool is named “Chani”. On a rectangular surface, it is possible to calculate the amount of charge transport between the different areas of the surface and the time needed for the total discharge using Chani. \newline
Although this thesis is written in the context of gaseous particle detectors, Chani can be useful for any research where the understanding of charge transport dynamics is important. \newline
I would like to thank to my parents and sister for their infinite love and support. I am and will always be grateful for having such a happy and open-minded family. I would like to thank my advisor Prof. Cenap Şahabettin Özben and co-advisor Prof. Serkant Ali Çetin for their support. I would like to thank Assoc. Prof. Veysi Erkcan \"{O}zcan for his highly beneficial suggestions and corrections on the thesis and the code. I would like to thank Prof. Nazmi Postac{\i}o\u{g}lu for his important contribution about the analytically solvable cases. I would like to acknowledge that my presence at CERN is supported by the Turkish Atomic Energy Agency.}             
\ozet{Radyasyonun gaz atomlar{\i}n{\i} iyonize etmesine dayal{\i} olarak par\c{c}ac{\i}klar{\i}n alg{\i}lanmas{\i} 19. y\"{u}zy{\i}l{\i}n sonlar{\i}ndan bu yana kullan{\i}lan bir tekniktir. O zamandan bug\"{u}ne, gazl{\i} iyonizasyon dedekt\"{o}rleri ve ilgili teknolojiler yayg{\i}n olarak geli\c{s}tirilmektedir. Bu y\"{u}z y{\i}ldan uzun s\"{u}re\c{c}teki \"{o}nemli bulu\c{s}lar Geiger-M\"{u}ller sayac{\i}, orant{\i}l{\i} say{\i}c{\i} (proportional counter), \c{c}oklu-telli orant{\i}l{\i} say{\i}c{\i} (multiwire proportional counter) ve mikrodoku gazl{\i} dedekt\"{o}rler olarak say{\i}labilir. B\"{u}t\"{u}n gazl{\i} iyonizasyon dedekt\"{o}rleri, gaz dolu bir odadan ge\c{c}en par\c{c}ac{\i}\u{g}{\i}n gaz{\i} iyonla\c{s}t{\i}rmas{\i} sonucunda olu\c{s}an serbest elektronlar{\i}n elektrodlar arac{\i}l{\i}\u{g}{\i}yla toplanmas{\i} ilkesine dayan{\i}r ancak farkl{\i} gazl{\i} dedekt\"{o}rlerin, i\c{s}lev yeterlilikleri ve \c{c}\"{o}z\"{u}n\"{u}rl\"{u}kleri b\"{u}y\"{u}k de\u{g}i\c{s}iklikler g\"{o}stermektedir. \"{O}rne\u{g}in, Geiger-M\"{u}ller sayac{\i}yla, yaln{\i}zca ka\c{c} tane par\c{c}ac{\i}\u{g}{\i}n dedekt\"{o}rden ge\c{c}ti\u{g}ini bulmak m\"{u}mk\"{u}n olabilirken, modern gazl{\i} dedekt\"{o}rlerde ge\c{c}en par\c{c}ac{\i}\u{g}{\i}n enerjisini ve izledi\u{g}i yolu y\"{u}ksek hassasiyetle belirlemek m\"{u}mk\"{u}nd\"{u}r.

Mikrodokulu dedekt\"{o}rlerin \"{o}ne \c{c}{\i}kanlar{\i}ndan biri 1996’da geli\c{s}tirilen micromegas (MicroMesh Gaseous Structure, mikro\"{o}rg\"{u} gazl{\i} yap{\i}) dedekt\"{o}rleridir. Micromegas dedekt\"{o}rlerinin olduk\c{c}a y\"{u}ksek enerji ve konum \c{c}\"{o}z\"{u}n\"{u}rl\"{u}\u{g}\"{u}ne sahip oldu\u{g}u, bu dedekt\"{o}rlerin par\c{c}ac{\i}k fizi\u{g}i deneylerinde ve t{\i}bbi g\"{o}r\"{u}nt\"{u}leme uygulamalar{\i}nda kullan{\i}labilece\u{g}i g\"{o}sterilmi\c{s}tir. Bunlar{\i}n yan{\i}s{\i}ra, 2006 y{\i}l{\i}nda geli\c{s}tirilen y{\i}\u{g}{\i}n (bulk) micromegas \"{u}retim tekni\u{g}i ile, dedekt\"{o}r\"{u}n \"{u}retim s\"{u}reci tek a\c{s}amaya indirilmi\c{s}, geni\c{s} alanlara sahip micromegas tipi dedekt\"{o}rlerin \"{u}retimi m\"{u}mk\"{u}n k{\i}l{\i}nm{\i}\c{s}t{\i}r. \"{O}te yandan bu dedekt\"{o}rlerde, k{\i}v{\i}lc{\i}m olu\c{s}ma s{\i}kl{\i}klar{\i}n{\i}n da y\"{u}ksek oldu\u{g}u g\"{o}r\"{u}lm\"{u}\c{s}t\"{u}r. K{\i}v{\i}lc{\i}m, dedekt\"{o}r\"{u}n aktif b\"{o}lgesindeki iyonizasyon yo\u{g}unlu\u{g}unun artarak gaz{\i}n iletkenle\c{s}mesi, bunun sonucunda da y\"{u}ksek gerilim elektronlar{\i}n{\i}n k{\i}sa devre olmas{\i}n{\i} takiben ani bir y\"{u}k ak{\i}\c{s}{\i}yla sonu\c{c}lanan durumdur. Micromegas dedekt\"{o}r\"{u}nde k{\i}v{\i}lc{\i}m olu\c{s}tu\u{g}unda, \"{o}rg\"{u} elektrodundaki b\"{u}t\"{u}n y\"{u}k anoda akar. Bu da \"{o}rg\"{u} elektrodu tekrar y\"{u}klenene kadar (\"{o}l\"{u} zaman) dedekt\"{o}r\"{u}n i\c{s}levsiz kalmas{\i}na sebep olur. Bu durum mikromegas dedekt\"{o}r\"{u}n\"{u}n, B\"{u}y\"{u}k Hadron \c{C}arp{\i}\c{s}t{\i}r{\i}c{\i}s{\i} deneyleri gibi y\"{u}ksek par\c{c}ac{\i}k ak{\i}\c{s}{\i} olan deneylerde kullan{\i}m{\i}n{\i} olanaks{\i}z k{\i}lmaktad{\i}r.

B\"{u}y\"{u}k Hadron \c{C}arp{\i}\c{s}t{\i}r{\i}c{\i}s{\i}`ndaki {\i}\c{s}{\i}n parlakl{\i}\u{g}{\i}, 2013-2018 y{\i}llar{\i} aras{\i}nda s\"{u}recek olan sBH\c{C} (S\"{u}per B\"{u}y\"{u}k Hadron \c{C}arp{\i}\c{s}t{\i}r{\i}c{\i}s{\i}) projesi kapsam{\i}nda, \c{s}u anki hedeflenen d\"{u}zeyinden bir mertebe (10$^{\textrm{34}}$ cm$^{\textrm{-2}}$s$^{\textrm{-1}}$’den 10$^{\textrm{35}}$ cm$^{\textrm{-2}}$s$^{\textrm{-1}}$’e) y\"{u}kseltilecektir. I\c{s}{\i}n parlakl{\i}\u{g}{\i}ndaki y\"{u}kselme, proton – proton \c{c}arp{\i}\c{s}malar{\i} sonunda olu\c{s}an son \"{u}r\"{u}nlerin s{\i}kl{\i}\u{g}{\i}nda da kabaca 10 katl{\i}k bir art{\i}\c{s}a sebep olacakt{\i}r. Bu sebeple, BH\c{C} \"{u}zerindeki genel ama\c{c}l{\i} dedekt\"{o}rler olan ATLAS ve CMS’in bu d\"{u}zeydeki par\c{c}ac{\i}k ak{\i}\c{s}lar{\i}n{\i} yeterli iyilikte alg{\i}layamayacak k{\i}s{\i}mlar{\i}nda da y\"{u}kseltmeler yap{\i}lacakt{\i}r.

CERN b\"{u}nyesindeki Muon ATLAS Micromegas Activity (MAMMA) grubu, ATLAS dedekt\"{o}r\"{u}n\"{u}n y\"{u}kseltilmesinde kullan{\i}lmak \"{u}zere Micromegas tipi dedekt\"{o}rler geli\c{s}tirmektedir. K{\i}v{\i}lc{\i}m problemlerinin \"{u}stesinden gelmek i\c{c}in, MAMMA grubu yapt{\i}\u{g}{\i} dedekt\"{o}rlerde anod elektrodlar{\i} olarak y\"{u}ksek diren\c{c}li hatlar kullanm{\i}\c{s}t{\i}r. Bu dedekt\"{o}rlerde y\"{u}ksek diren\c{c}li hatlar ve okuma hatlar{\i} elektriksel olarak yal{\i}t{\i}lm{\i}\c{s} bir \c{s}ekilde \"{u}st\"{u}ste durup, ayn{\i} deseni takip etmektedir. Bu sayede, k{\i}v{\i}lc{\i}m olu\c{s}up \"{o}rg\"{u} elektrodu anod elektroduna k{\i}sa devre oldu\u{g}unda dahi, anod elektrodu boyunca bulunan y\"{u}ksek diren\c{c} sebebiyle \"{o}rg\"{u} elektrodundaki y\"{u}kler bo\c{s}almadan k{\i}v{\i}lc{\i}m ortadan kalkmakta, ayn{\i} zamanda dedekt\"{o}r okuma elektroni\u{g}inin ba\u{g}l{\i} oldu\u{g}u okuma hatlar{\i} da k{\i}v{\i}lc{\i}mla birlikte gelen y\"{u}klerden yal{\i}t{\i}lm{\i}\c{s} olmaktad{\i}r. Diren\c{c}li anod elektrodlar{\i}n{\i}n kullan{\i}m{\i} RPC (Resistive Plate Chamber) ve ATLAS dedekt\"{o}r\"{u}nde kullan{\i}lan TGC (Thin Gap Chamber) gibi farkl{\i} gazl{\i} dedekt\"{o}r teknolojilerinde de g\"{o}r\"{u}lmektedir ancak bu dedekt\"{o}rlerde diren\c{c}li anod bir katman halindedir. MAMMA grubunun geli\c{s}tirdi\u{g}i dedekt\"{o}rlerde ise, bunlardan farkl{\i} olarak y\"{u}ksek diren\c{c}li elektrodlar da \c{s}eritler halindedir. Bu \c{s}ekilde y\"{u}k yay{\i}l{\i}m{\i}na ba\u{g}l{\i} olarak olu\c{s}abilecek \c{c}apraz etkile\c{s}im (crosstalk) ve bunun sonucunda da olu\c{s}abilecek sahte sinyalleri engellemek ama\c{c}lanm{\i}\c{s}t{\i}r. MAMMA grubu, diren\c{c}li anodlar kulland{\i}klar{\i} dedekt\"{o}rlerin y\"{u}ksek kazan\c{c}larda k{\i}v{\i}lc{\i}ma dayan{\i}kl{\i} olarak \c{c}al{\i}\c{s}t{\i}\u{g}{\i}n{\i} g\"{o}stermi\c{s}tir.

Y\"{u}ksek diren\c{c}li anod elektrodu kullan{\i}m{\i}n{\i}n getirdi\u{g}i sak{\i}nca ise, bu tekni\u{g}in, y\"{u}k bo\c{s}almas{\i} i\c{c}in gereken zaman{\i} uzatmas{\i}d{\i}r. Dedekt\"{o}re gelen par\c{c}ac{\i}k s{\i}kl{\i}\u{g}{\i}, buna ba\u{g}l{\i} olarak da anod elektroduna gelen elektronlar{\i}n s{\i}kl{\i}\u{g}{\i} artt{\i}k\c{c}a, gelen elektronlar{\i}n ortamdan uzakla\c{s}t{\i}r{\i}lamadan yeni elektronlar{\i}n geldi\u{g}i bir durum ortaya \c{c}{\i}kabilir. B\"{o}yle bir durumda, dedekt\"{o}r\"{u}n gaz kazanc{\i}nda d\"{u}\c{s}meye sebep olacak bir “y\"{u}klenme” olu\c{s}acakt{\i}r. Bu sebeple, anod elektrodlar{\i}n{\i}n boyutlar{\i} ve \"{o}zdirenci bir eniyileme s\"{u}recinden ge\c{c}melidir. Ancak bu \c{s}ekilde, k{\i}v{\i}lc{\i}ma duyars{\i}z ama y\"{u}ksek par\c{c}ac{\i}k ak{\i}\c{s}{\i}n{\i} alg{\i}layabilen dedekt\"{o}rler \"{u}retmek m\"{u}mk\"{u}n olabilir. Bu tezde sunulan ana \c{c}al{\i}\c{s}ma, bu motivasyonla geli\c{s}tirilmi\c{s} olan, dikd\"{o}rtgen \c{s}eklindeki y\"{u}zeylerde y\"{u}k ak{\i}\c{s}{\i} ve y\"{u}k bo\c{s}alma s\"{u}re\c{c}lerinin zamana ba\u{g}l{\i} benzetimini yapmaya olanak sa\u{g}layan bir ara\c{c}t{\i}r.

Geli\c{s}tirilen benzetim arac{\i} “Chani” olarak isimlendirilmi\c{s}tir. Bu isimlendirmenin sebebi s\"{o}zc\"{u}\u{g}\"{u}n, \.Ingilizcede “y\"{u}k” anlam{\i}na gelen “charge” s\"{o}zc\"{u}\u{g}\"{u}n\"{u} \c{c}a\u{g}r{\i}\c{s}t{\i}rmas{\i} ve de bilim kurgu roman{\i} Dune’daki bir karakterin ismi olmas{\i}d{\i}r. Benzetici ROOT veri analizi ortam{\i}nda \c{c}al{\i}\c{s}mak \"{u}zere bir “makro” olarak kodlanm{\i}\c{s}t{\i}r. Basit olarak, Chani dikd\"{o}rtgen \c{s}eklindeki y\"{u}zeyi alt dikd\"{o}rtgenlere b\"{o}ler; y\"{u}zeylerdeki y\"{u}k da\u{g}{\i}l{\i}m{\i}n{\i}n sebep oldu\u{g}u potansiyeli ve alt h\"{u}creler aras{\i}ndaki ak{\i}mlar{\i} ayr{\i}k anlar boyunca hesaplar. Hesaplanan ak{\i}mlar{\i}n{\i}n iki an aras{\i}ndaki zaman aral{\i}\u{g}{\i}yla \c{c}arp{\i}lmas{\i} bir h\"{u}creden di\u{g}er h\"{u}creye ta\c{s}{\i}nacak olan toplam y\"{u}k\"{u} verir. Y\"{u}zeye ula\c{s}an her yeni y\"{u}k, y\"{u}zey boyuncaki potansiyel da\u{g}{\i}l{\i}m{\i}n{\i} de\u{g}i\c{s}tirir ve y\"{u}zey ak{\i}mlar{\i}n{\i}n olu\c{s}mas{\i}na sebep olur. Y\"{u}k bo\c{s}alma s\"{u}re\c{c}leri ise baz{\i} h\"{u}crelerin “g\"{u}\c{c} kayna\u{g}{\i} ba\u{g}lant{\i} noktas{\i}” olarak (belirli bir ba\u{g}lant{\i} direnciyle) tan{\i}mlanmas{\i} sayesinde olur, kayna\u{g}a giderek ortamdan uzakla\c{s}acak olan y\"{u}kler, bu noktadaki ak{\i}mlar{\i}n benzer \c{s}ekilde zaman aral{\i}\u{g}{\i}yla \c{c}arp{\i}lmas{\i} sonucunda elde edilir.

Chani, her alt dikd\"{o}rtgenin potansiyelini her zaman \"{o}rne\u{g}inde “momentler y\"{o}ntemi” olarak adland{\i}r{\i}lan bir teknikle hesaplar. Momentler y\"{o}ntemi, kendine e\c{s}lenik operat\"{o}rlerle tan{\i}mlanan homojen olmayan diferansiyel denklemlerin, matris denklemlerine d\"{o}n\"{u}\c{s}t\"{u}r\"{u}lerek, bilgisayar yard{\i}m{\i}yla yakla\c{s}{\i}k say{\i}sal \c{c}\"{o}z\"{u}mlerine olanak sa\u{g}layan tekniklerin genel ad{\i}d{\i}r. Modern bilgisayarlar{\i}n y\"{u}ksek hesaplama g\"{u}c\"{u} sayesinde, bu teknikle yap{\i}lan yakla\c{s}{\i}k hesaplardan pratik ama\c{c}lar i\c{c}in yeteri kadar do\u{g}ru sonu\c{c}lar almak m\"{u}mk\"{u}n olmaktad{\i}r. Momentler y\"{o}ntemiyle elektrik potansiyelin hesaplanmas{\i}nda, her bir alt h\"{u}crenin \"{u}zerindeki y\"{u}k da\u{g}{\i}l{\i}m{\i} sabit kabul edilir; bir alt h\"{u}credeki y\"{u}k\"{u}n di\u{g}er alt h\"{u}crede olu\c{s}turaca\u{g}{\i} potansiyel ise, alth\"{u}credeki b\"{u}t\"{u}n y\"{u}k\"{u}n merkezde topland{\i}\u{g}{\i}nda di\u{g}er alt h\"{u}crenin merkezinde olu\c{s}turaca\u{g}{\i} potansiyel olarak hesaplan{\i}r. Alt h\"{u}crenin kendi \"{u}zerinde olu\c{s}turaca\u{g}{\i} potansiyel ise, bir dikd\"{o}rtgen boyunca sabit bir y\"{u}k da\u{g}{\i}l{\i}m{\i}n{\i}n kendi merkezinde olu\c{s}turaca\u{g}{\i} potansiyeldir. Bu varsay{\i}mlar \c{c}er\c{c}evesinde, benzetimlerin ba\c{s}{\i}nda, h\"{u}crelerdeki y\"{u}klerin di\u{g}er h\"{u}crelerde sebep olu\c{s}turaca\u{g}{\i} potansiyellerin hesaplanmas{\i}n{\i} sa\u{g}layan bir tesir matrisi hesaplan{\i}r. Y\"{u}ksek \"{o}l\c{c}ekli hesaplamalar s\"{o}z konusu oldu\u{g}unda, tesir matrisinin hesaplanmas{\i} uzun s\"{u}rmektedir. Ancak, tesir matrisi, benzetim i\c{c}in bir defa hesapland{\i}ktan sonra, y\"{u}k da\u{g}{\i}l{\i}mlar{\i}n{\i}n sebep oldu\u{g}u elektrik potansiyelin hesaplanmas{\i} \c{c}arpma ve toplama i\c{s}lemlerine indirgenmektedir. Bu da \c{c}ok say{\i}daki zaman \"{o}rne\u{g}i boyunca, her defas{\i}nda elektrik potansiyelin h{\i}zl{\i} bir \c{s}ekilde hesaplanmas{\i}n{\i} m\"{u}mk\"{u}n k{\i}lmaktad{\i}r.

Bu tezin giri\c{s} k{\i}sm{\i}nda Avrupa N\"{u}kleer Ara\c{s}t{\i}rmalar Enstit\"{u}s\"{u} (CERN) k{\i}saca tan{\i}t{\i}l{\i}p, B\"{u}y\"{u}k Hadron \c{C}arp{\i}\c{s}t{\i}r{\i}c{\i}s{\i} ve b\"{u}nyesindeki deneylerden bahsedilmi\c{s}tir. Daha sonra ATLAS dedekt\"{o}r\"{u}n\"{u}n k{\i}s{\i}mlar{\i} k{\i}saca a\c{c}{\i}klanm{\i}\c{s} ve MAMMA grubunun \c{c}al{\i}\c{s}mlarana de\u{g}inilmi\c{s}tir. Giri\c{s} k{\i}sm{\i}n{\i}n ard{\i}ndan gazl{\i} dedekt\"{o}rlerin tarihsel geli\c{s}iminden bahsedilen ikinci k{\i}s{\i}m gelir. Bu k{\i}s{\i}mda, mikrodokulu dedekt\"{o}rlere gelene kadarki gazl{\i} dedekt\"{o}rler, iyonizasyon odas{\i}ndan ba\c{s}lanarak tarihsel olarak tan{\i}t{\i}lm{\i}\c{s} ve genel davran{\i}\c{s}lar{\i} a\c{c}{\i}klanm{\i}\c{s}t{\i}r. \.Ikinci k{\i}sm{\i}n son b\"{o}l\"{u}m\"{u}nde, bu tezdeki ana \c{c}al{\i}\c{s}man{\i}n da motivasyonu olan micromegas tipi dedekt\"{o}rlerin genel \"{o}zellikleri, kullan{\i}m alanlar{\i} ve \"{u}retim teknikleri gibi konulara de\u{g}inilmi\c{s}, son olarak da MAMMA grubunun geli\c{s}tirmekte oldu\u{g}u diren\c{c}li anodlu micromegas dedekt\"{o}rleri \"{u}zerinde durulmu\c{s}tur. Takip eden k{\i}s{\i}mda, geli\c{s}tirilen y\"{u}k ta\c{s}{\i}n{\i}m{\i} benzetim arac{\i} a\c{c}{\i}klan{\i}r. Momentler metodunun matematiksel tan{\i}t{\i}m{\i}n{\i}n ard{\i}ndan, bu tekni\u{g}in elektrik potansiyeli belirleyecek olan Poisson denklemine uygulan{\i}\c{s}{\i}na yer verilmi\c{s}tir. Sonraki b\"{o}l\"{u}mlerde, benzetim arac{\i}n{\i}n \c{c}al{\i}\c{s}ma prensipleri detayl{\i} olarak anlat{\i}lm{\i}\c{s} ve \"{o}rnek hesaplamalar ve \"{o}ztutarl{\i}l{\i}k testleri sunulmu\c{s}tur.

Yer verilen ilk hesaplama \"{o}rne\u{g}inde 2 cm’ye 2 cm boyutlar{\i}nda kare \c{s}eklinde, \"{o}zdirenci 10$^{\textrm{5}}$ $\Omega / \square$ olan bir y\"{u}zeyin merkezi etraf{\i}nda k\"{u}melenmi\c{s} 10$^4$ elementer y\"{u}ke e\c{s}it miktarda y\"{u}k ula\c{s}mas{\i} durumu ele al{\i}nm{\i}\c{s}t{\i}r. Modellemede y\"{u}zey, iki do\u{g}rultada 15 par\c{c}aya b\"{o}l\"{u}nmek suretiyle 225 alt y\"{u}zeye ayr{\i}lm{\i}\c{s}t{\i}r. Benzetim, 200 ns’ye kar\c{s}{\i}l{\i}k gelen 2000 zaman ad{\i}m{\i}nda ger\c{c}ekle\c{s}tirilmi\c{s}tir. Merkezdeki y\"{u}k\"{u}n kenarlara ve k\"{o}\c{s}elere yay{\i}lma s\"{u}reci ve y\"{u}k\"{u}n tamamen bo\c{s}almas{\i} i\c{c}in gereken sure g\"{o}zlemlenebilmektedir. Y\"{u}k bo\c{s}alma s\"{u}recinin RC devresine benzer \c{s}ekilde \"{u}stel bir karaktere sahip oldu\u{g}u g\"{o}r\"{u}lm\"{u}\c{s}t\"{u}r. Toplam y\"{u}k\"{u}n yar{\i}s{\i}n{\i}n bo\c{s}almas{\i} i\c{c}in gereken s\"{u}re yakla\c{s}{\i}k 20 ns olurken, y\"{u}k\"{u}n y\"{u}zde doksan{\i}n{\i}n bo\c{s}almas{\i} i\c{c}in gereken s\"{u}re yakla\c{s}{\i}k 90 ns olmu\c{s}tur. Buna ek olarak, y\"{u}zeyin s{\i}\u{g}as{\i}n{\i}n ve y\"{u}zeyin merkezindeki y\"{u}klerin bo\c{s}alma s\"{u}relerinin benzetimde kullan{\i}lan alt h\"{u}cre say{\i}s{\i}yla de\u{g}i\c{s}imi de hesaplanarak benzetimlerin yak{\i}nsakl{\i}\u{g}{\i} do\u{g}rulanm{\i}\c{s}t{\i}r.

Sonsuz boyutlu d\"{u}zlemdeki periyodik y\"{u}k da\u{g}{\i}l{\i}m{\i}n{\i}n zamanla relaksasyonu problemi analitik olarak \c{c}\"{o}z\"{u}lebilmektedir. Kare \c{s}ekilli d\"{u}zlemlerdeki periyodik y\"{u}k da\u{g}{\i}l{\i}m{\i}n{\i}n zamanla de\u{g}i\c{s}iminin benzetimi Chani arac{\i}l{\i}\u{g}{\i}yla yap{\i}lm{\i}\c{s} ve elde edilen sonu\c{c}lar analitik \c{c}\"{o}z\"{u}mden beklenen sonu\c{c}la kar\c{s}{\i}la\c{s}t{\i}r{\i}lm{\i}\c{s}t{\i}r. Benzetim parametreleri do\u{g}ru se\c{c}ildi\u{g}inde, sonsuz d\"{u}zlem i\c{c}in beklenen relaksasyon zaman{\i} ile Chani'den elde edilen zaman aras{\i}ndaki g\"{o}reli fark \%3.5'ten az olmu\c{s}tur. B\"{o}ylece Chani'nin anlaml{\i} sonu\c{c}lar verdi\u{g}i do\u{g}rulanm{\i}\c{s}t{\i}r. 

Yer verilen \"{u}\c{c}\"{u}nc\"{u} benzetimde ise MAMMA grubunun geli\c{s}tirmi\c{s} oldu\u{g}u diren\c{c}li anodlu micromegas dedekt\"{o}r prototiplerindeki diren\c{c}li \c{s}eritlerin boyutlar{\i}na ve y\"{u}zey direncine sahip bir y\"{u}zey modellenmi\c{s}tir. Yine ba\c{s}lang{\i}\c{c}taki bir y\"{u}k\"{u}n yay{\i}l{\i}m{\i} ve topra\u{g}a iletimi incelenmi\c{s}tir. Y\"{u}k bo\c{s}alma s\"{u}resi \c{c}ok daha fazla oldu\u{g}u i\c{c}in bu benzetimde bir \"{o}ncekinden 1000 kat daha fazla zaman ad{\i}m{\i} bulunmaktad{\i}r. Chani, uzun s\"{u}reli hesaplar{\i} haf{\i}za ta\c{s}{\i}m{\i} olmadan yapabilmektedir.

Bir di\u{g}er \"{o}rnekte ise, 18 mm'ye 100 mm boyutlar{\i}ndaki bir y\"{u}zey 14229 alty\"{u}zeye b\"{o}l\"{u}nerek y\"{u}k ta\c{s}{\i}n{\i}m{\i} hesab{\i} yap{\i}lm{\i}\c{s} ve Chani`nin y\"{u}ksek boyutlu matrislerle (14229 $\times$ 14229) de sorunsuz bir \c{s}ekilde \c{c}al\c{s}abild\u{g}i g\"{o}sterilmi\c{s}tir.

Bu \c{c}al{\i}\c{s}mada g\"{o}sterilmi\c{s}tir ki, geli\c{s}tirilmi\c{s} olan benzetim arac{\i} Chani ile dikd\"{o}rtgen \c{s}eklindeki bir y\"{u}zeydeki y\"{u}klerin yay{\i}l{\i}m{\i} ve y\"{u}klerin bo\c{s}almas{\i} i\c{c}in gereken toplam zaman hesaplanabilmektedir. Sonu\c{c} olarak, zamana ba\u{g}l{\i} y\"{u}k ta\c{s}{\i}n{\i}m{\i} ve bo\c{s}almas{\i}n{\i}n modelleyebilen bir benzetim arac{\i} geli\c{s}tirilmi\c{s}tir. Bu ara\c{c}, MAMMA grubunun geli\c{s}tirece\u{g}i yeni micromegas dedekt\"{o}rlerindeki diren\c{c}li yap{\i}lar{\i}n en iyilenmesinde kullan{\i}lman{\i}n yan{\i}s{\i}ra, y\"{u}k ta\c{s}{\i}n{\i}m dinamiklerinin anla\c{s}{\i}lmas{\i}n{\i}n \"{o}nem arz etti\u{g}i her ara\c{s}t{\i}rmada uygulama bulabilir.}               
\summary{Radiation detection by ionization of the gas atoms as the radiation passes through goes back to the end of the 19$^{\textrm{th}}$ century. Since then, gaseous ionization detectors and related technologies are being developed extensively. Major discoveries of this longer-then-a-century period can be counted as the discoveries of Geiger-Müller counter, proportional counter, multiwire proportional counter and micro pattern gaseous detectors. With a Geiger-Müller counter, it was only possible to count the number of particles passing through, however, now with a modern gaseous ionization detector, it is possible to measure the energy of the passing particle and reconstruct the track followed by the particle with very high precision.

One of the important technologies of the micropattern era is micromegas (micromesh gaseous structure). Micromegas detectors have been shown to have very high position and energy resolutions, however, it has also been shown that high spark rates in of these detectors makes them impossible to use in high luminosity experiments such Large Hadron Collider experiments.

The Large Hadron Collider is going to be upgraded to have one order higher luminosity (from 10$^{\textrm{34}}$ cm$^{\textrm{-2}}$s$^{\textrm{-1}}$ to 10$^{\textrm{35}}$ cm$^{\textrm{-2}}$s$^{\textrm{-1}}$) within the sLHC (Super Large Hadron Collider) project. Due to the increase in the beam luminosity, multiplicity of the particles in the collisions is also going to increase by, roughly, one order. For this reason, those components of the general-purpose detectors ATLAS and CMS  which are not capable of handling the expected rate are also going to be upgraded.

Muon ATLAS Micromegas Activity (MAMMA) group at CERN developes micromegas detectors for the upgrade of the ATLAS Muon Spectrometer. In order to overcome the spark issues, MAMMA group used high resistivity anode strips in their chambers. In these chambers, readout strips were standing below the resistive strips, both following the same pattern and isolated electrically, hence, the detector is read-out through the capacitively induced charge on the read-out strips.

Disadvantage of using high resistivity electrodes is that they requires a longer charge removal time. As the incoming particle rate, thus the rate of electrons arriving on the anode strips, gets higher, there might not be enough time for the charges to be removed. Such a case would result a charge-up that reduces the detector gain. Therefore, the resistivity and the geometry of the resistive strips must be optimized to have a spark-safe detector with high rate capability. With this motivation, a tool for the transient simulation of the charge transport on a rectangular surface is developed and presented in this thesis.

This simulation tool is named “Chani” because of the resemblance to the word “charge” and is a reference to the name of a character in the science fiction novel Dune. It is coded as a macro to run within the ROOT Data Analysis Framework. Principally, Chani divides the surface into the subrectangles and calculates their electrical potentials and the currents in between, through predefined time instances. As new charges arrive, they disturb the potential distribution of the surface thus cause currents. Discharges are also modeled by defining some of the subcells as ground connection points with user-defined connector resistances.

In Chani, electrical potential of each subrectangle is calculated in every time instance using the so-called “method of moments”. With this technique, an influence matrix is first calculated and then used to map the new charge distributions to the new potential solutions at each time. Since the matrix calculation is done only once, the technique provides fast solutions as desired.

It is shown in this thesis that Chani is capable of calculating the spread of the charge over the surface and the time needed for the total discharge. In conclusion, a functional tool for the simulation of transient charge transport and discharge is developed. Besides its application in the optimization of the resistive structures in micromegas chambers, this tool can find application in any research where the charge transport dynamics are crucial.}             

\begin{document}

\chapter{INTRODUCTION}\label{introduction}
Gaseous ionization detectors have been widely used in particle physics experiments for over hundred years. There are various types of gas-filled detectors and all of them are based on the very same principle: When an energetic enough particle goes through a chamber with gas, it ionizes the gas atoms and creates electron – ion pairs and these electrons and ions are collected on electrodes to gather the particle track information. Although the fundamental principle is the same, different type of gaseous detectors differ in many aspects such as geometry, amplification technique or production method.

One major problem that is shared by all high-gain gaseous detectors is spark. High gains are desired to achieve high sensitivities, however, when the free electrons in a chamber exceed a critical density, high voltage electrodes become effectively shorted through the electron cloud. This results in the immediate discharge of the anode through the electron cloud, which could be dangerous for the chamber itself, as well as the peripheral devices such as read-out electronics or high voltage power supplies.

High-resistivity materials have been used as anode materials of detectors like RPCs \cite{RPC} or TGCs \cite{ATLAS} in order to prevent harms caused by sparks. Recently at CERN, MAMMA group implemented the resistive anode idea to micromegas technology and built spark-resistant micromegas chambers \cite{MAMMA1}. Instead of a full resistive layer used in RPC and TGCs, these micromegas detectors have resistive strips.

MAMMA group conducts R\&D on micromegas detectors for the upcoming luminosity upgrade of LHC \cite{MAMMA2}. In their prototypes, resistive strips with different resistivities and geometries are compared to achieve a good performance. The main study in this thesis is motivated by the need for a systematic method for the optimization of the resistive structures in micromegas chambers. For this purpose, a tool for transient simulation of the charge transport on a rectangular surface with finite resistivity is developed and applied to a number of cases.

In the rest of this chapter, CERN, the Large Hadron Collider and ATLAS experiment are briefly introduced; and the MAMMA (Muon ATLAS MicroMegas Activity) group’s activities are summarized. The second chapter is a historical introduction of the gaseous detectors. In third and fourth chapters, the charge transport simulator’s working principles and applications are presented. Discussions and conclusions are given in the following chapters.

\section{CERN \& The Large Hadron Collider}\label{cernandlhc}

The European Organization for Nuclear Research (CERN) in Geneva was established in 1954 with 12 founding states. Since then, fundamental particle physics research has been conducted at CERN which yielded many important discovery and inventions for both science and technology. Significant examples of these are the invention of multiwire proportional chamber, the discovery of neutral currents, the discovery of W and Z bosons, the invention of the World Wide Web and the observation and capturing of antihydrogen atoms \cite{history}. Currently, there are 20 European member states of CERN and numerous non-member states co-operate via special agreements. Turkey holds an “observer” status \cite{members}, but has recently applied for full membership \cite{turkey}.

The first particle accelerator at CERN, the 600 MeV Synchrocyclotron (SC) started its operation in 1957, and it was followed by The Proton Synchrotron (PS) in 1959, the Super Proton Synchrotron (SPS) and the Large Electron-Positron (LEP) collider in 1989. Besides these major ones, at CERN, there are and have been different particle accelerators, which have served various physics research. Since 2008, largest ever-built particle accelerator, the Large Hadron Collider collides protons at the highest energies achieved to date \cite{history}.

\subsection{General features and current status of the LHC}

The Large Hadron Collider is a 26.7 km long circular accelerator located under the French – Swiss border near Geneva, in a depth varying from 45 m to 175 m according to the surface shape. The tunnel that LHC is built in was already there and had previously been used for the CERN LEP machine \cite{LHC}.

\begin{figure}[h]
 \centering
 \includegraphics[width=13cm,keepaspectratio=true]{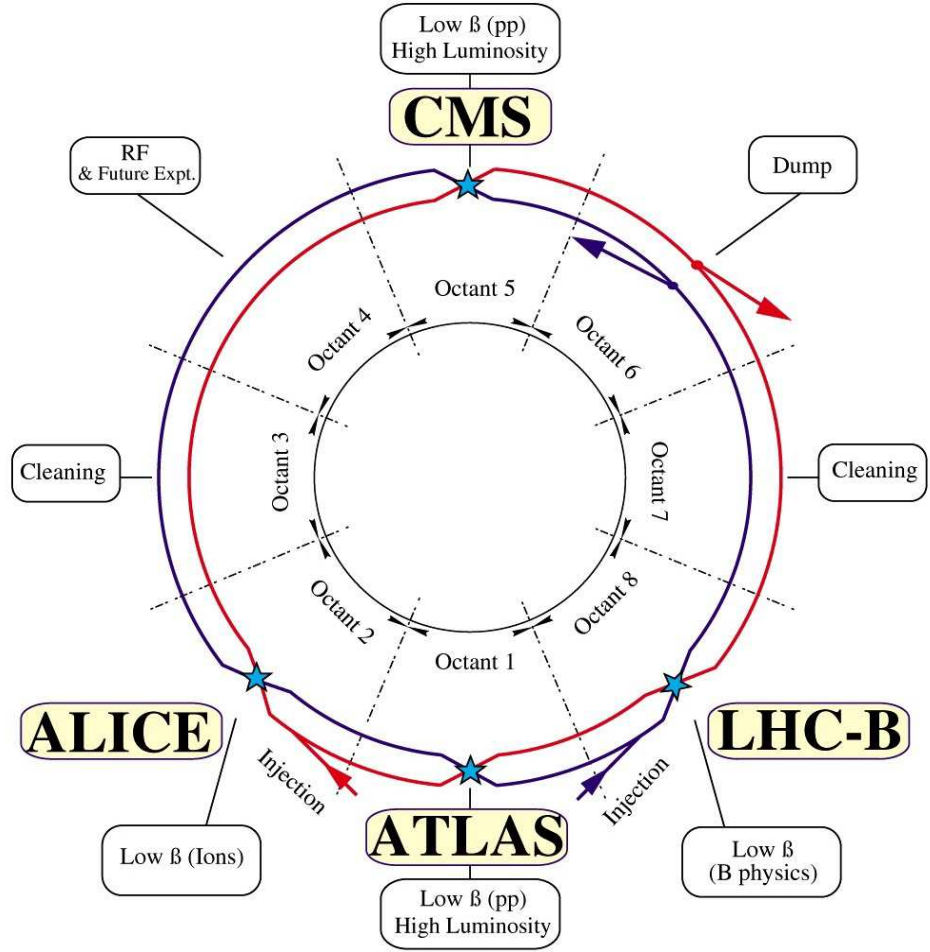}
 \caption[Schematic layout of LHC Machine and the Experiments]{Schematic layout of LHC Machine and the Experiments \cite{lhclayout}.}
\end{figure}

LHC contains two rings where protons (or ions) travel in opposite directions and eight interaction points are available from LEP construction. Four of these eight interaction points are used and equipped with the detectors and ground structures: respectively at Point 1,2, 5 and 8 ATLAS, ALICE, CMS and LHCb detectors are built \cite{LHC}. LHC layout with interaction points are illustrated in Figure 1.1.

Proton beams are injected to the LHC after four stages of acceleration in other CERN accelerators: first, they are accelerated to 50 MeV through LINACs, then to 1.4 GeV in Proton Synchrotron Booster (PSB) and to 26 GeV in Proton Synchrotron (PS) and finally to 450 GeV in Super Proton Synchrotron (SPS) \cite{LHC}. This sequence is illustrated in Figure 1.2 along with the Pb ion injection chain and former LEP injection chain. Having reached 450 GeV, proton beams are than injected to the LHC pipes through interaction points 2 and 8 in opposite directions to be accelerated to 7 TeV as shown in Figure 1.2.
\begin{figure}
 \centering
 \includegraphics[width=14cm,keepaspectratio=true]{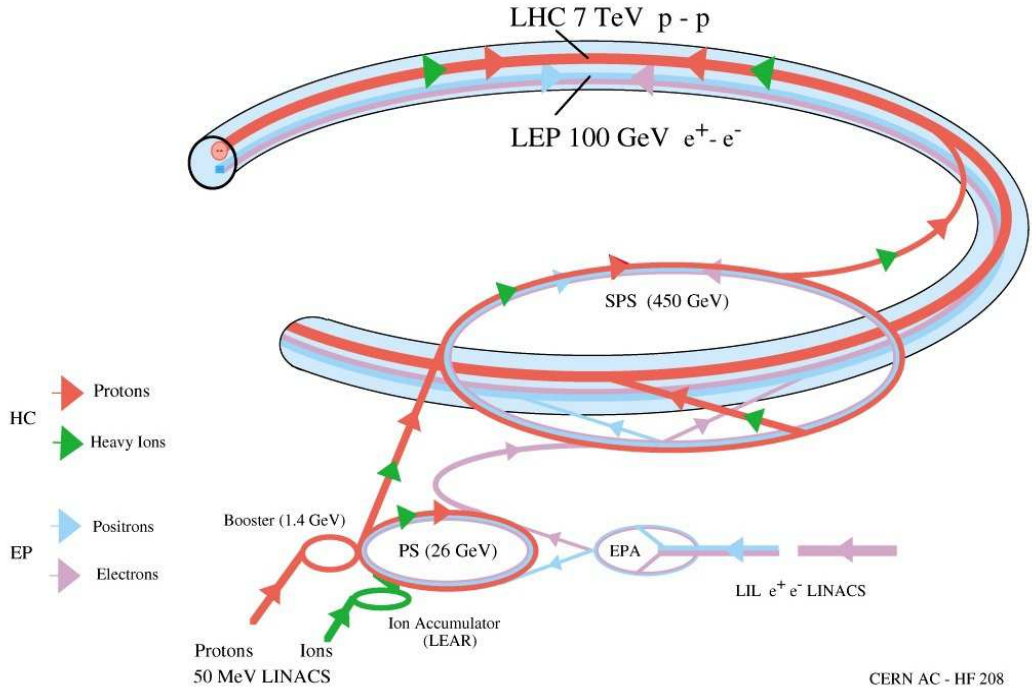}
 \caption[LHC Beam Injection Complex.]{LHC Beam Injection Complex \cite{lhcinjectioncomplex}.}
\end{figure}

Main performance goals of the Large Hadron Collider are the peak luminosity of 10$^{\textrm{34}}$ cm$^{\textrm{-2}}$s$^{\textrm{-1}}$ and the center-of-mass collision energy of 14 TeV. This corresponds to the energies of 7 TeV per proton beam, necessitating an 8.33 T dipole field which is to be achieved by superconducting magnets \cite{LHC}. 

After two years of 3.5 TeV per beam runs, recently in April 2012, LHC started its 4 TeV per proton beam runs producing 8 TeV collisions which is the current world record \cite{2012}. LHC is scheduled to begin its 2 years shutdown at the end of 2012 to get ready for 6.5 TeV per beam runs in late 2014 and finally to reach its design goal of 7 TeV per beam \cite{2012}. In 2010 runs, peak luminosity of LHC experiments was $200 \times 10^{30} cm^{-2} s^{-1}$ and as per May 2012, peak luminostity of LHC experiments is greater than $60 \times 10^{32} cm^{-2} s^{-1}$ \cite{luminosity}.

LHC and its experiments provide a powerful tool to probe the last missing piece of Standard Model of elementary particles, the Higgs boson, and the many questions beyond the Standard Model such as existence of supersymmetry or extra dimensions in nature, potential ingredients of dark matter, the cause of CP violation \cite{newphysics} and existence of new fermion families \cite{B3SMIII}. ALICE (A Large Ion Collider Experiment) is dedicated to heavy ion (Pb-Pb) collisions and LHCb (Large Hadron Collider Beauty) is focused on the physics of the bottom quark. CMS (Compact Muon Solenoid) and ATLAS (A Toroidal LHC Apparatus) are “general purpose” detectors and many questions of today’s particle physics including the ones mentioned above are studied in these experiments.

\subsection{The sLHC Project}

The Super Large Hadron Collider is the upgrade project scheduled for 2013 – 2018 with the main goal of increasing the LHC peak luminosity from 10$^{\textrm{34}}$ cm$^{\textrm{-2}}$s$^{\textrm{-1}}$ to 10$^{\textrm{35}}$ cm$^{\textrm{-2}}$s$^{\textrm{-1}}$). The upgrade scheme includes both accelerator improvements such as replacements in the injector chain and enhancement of the focusing quadrupole magnets, and detector development such as the improvement of the ATLAS and CMS \cite{sLHC}.

\section{ATLAS Experiment}\label{ATLAS}

ATLAS (A Torroidal LHC Apparatus) at Point 1 (see Figure 1.1) is one of the two “general purpose” experiments located at the LHC interaction points. Being a “general purpose” detector, ATLAS is expected to be sensitive to all physics produced by the collisions at the LHC.

\begin{figure}[h]
 \centering
 \includegraphics[width=12cm,keepaspectratio=true, trim = 0mm -5px 0mm 0mm, clip]{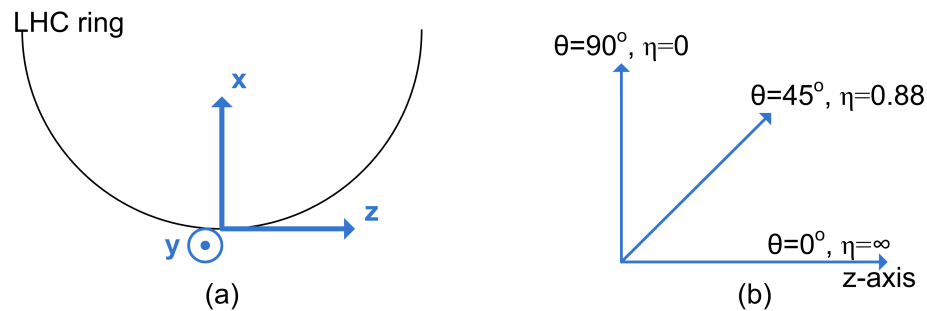}
 \caption{Detector axes. (a) Cartesian coordinates shown from interaction point. (b) Polar angle $\theta$ and pseudorapidty ($\eta$).}
\end{figure}

Before beginning to describe detector parts, the detector axes should be clearly defined. Conventionally, the beam direction is defined as the z-axis. Side-A, the side facing the Point 8 (or Geneva), is defined as the positive z and the opposite side of the detector, side-C is defined as the negative z. The x-axis points towards the center of the LHC ring from the interaction point and the y-axis points upwards. As usual, angle from the x-axis is defined as $\phi$ and the angle from z-axis is defined as $\theta$. Finally, the pseudorapidity is defined as $\eta = - \ln \tan(\theta /2)$. This coordinate system is illustrated in Figure 1.3.

A computer generated image of the ATLAS detector, where the sub detectors can be seen, is given in Figure 1.4. The detector contains three main parts: trackers, calorimeters and the muon system.

\begin{figure}
 \centering
 \includegraphics[width=14cm,keepaspectratio=true, trim = 0mm -5px 0mm 0mm, clip]{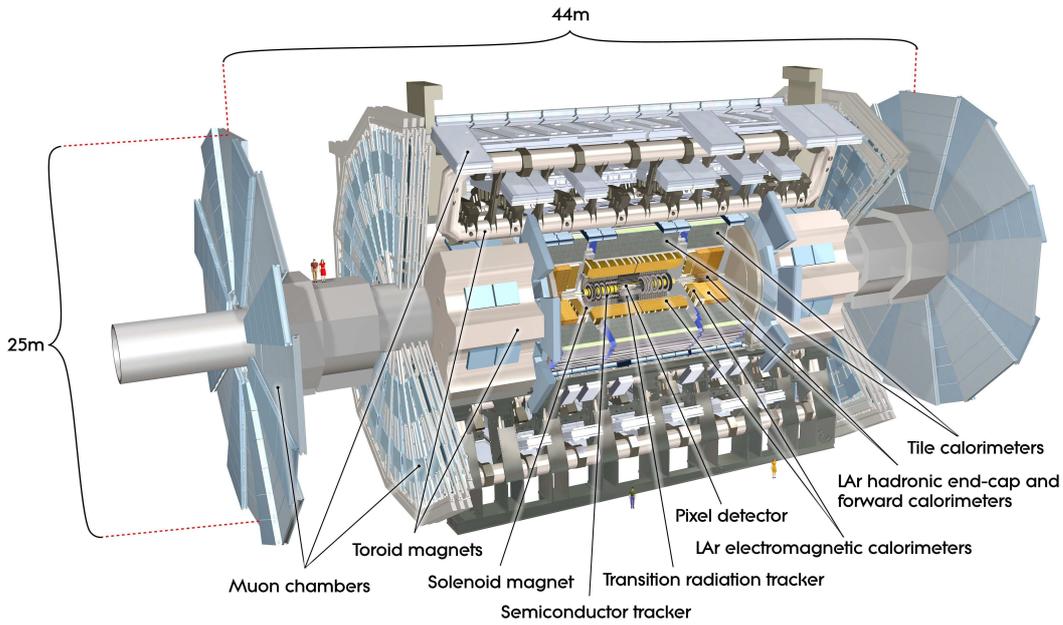}
 \caption[Computer generated image of the ATLAS detector.]{Computer generated image of the ATLAS detector \cite{atlascutaway}.}
\end{figure}

\subsection{Trackers}

When LHC run in its full performance goals, protons will collide once in every 25 ns and at each bunch crossing nearly 1000 particles will come out in $|\eta|<2.5 $ region. Momenta of these particles are detected with high resolution $(\sigma_{P_T}/P_T = 0.05\%)$ by the combination of pixel and microstrip semiconductor trackers (SCT) and drift tubes in the Transition Radiation Tracker (TRT) \cite{ATLAS}.
\newpage
\subsection{Calorimeters}

Electromagnetic (EM) and hadronic calorimeters of the ATLAS detector aim to measure the energies of electrons, photons and jets which emerge from the LHC collisions. Additionally, the calorimeter system prevents electromagnetic and hadronic showers from reaching to the muon system by absorbing their energies; hence allows the muon system to produce clean signals.

LAr (Liquid Argon) electromagnetic calorimeter covers the region $|\eta|<3.2$ and measures electron and photon energies with a resolution of $\sigma_E/E = 10\%$. Barrel and end-cap parts of the hadronic calorimeter also cover $|\eta|<3.2$ region and measure jet energies with $\sigma_E/E = 50\%$ uncertainty, and the forward calorimeters cover $3.1 < |\eta| < 4.9$ and make jet energy measurements with $\sigma_E/E = 50\%$ uncertainty \cite{ATLAS}.

\subsection{The Muon System}

The muon spectrometer of the ATLAS detector is formed by the toroid magnets (which give the ATLAS detector its name) and gaseous detectors of four different technologies. Momentum resolution of the muon spectrometer is $\sigma_{P_T}/P_T = 10\%$ \cite{ATLAS}.

The magnetic field generated by the toroid magnets bend the muon tracks to help the particle identification. Monitored Drift Tubes (MDTs) and Cathode Strip Chambers (CSCs) cover the $|\eta|<2.7$ and $2.0 < |\eta| < 2.7$ regions respectively, and perform precise measurements of the muon tracks \cite{ATLAS}. In the region where $|\eta|<1.05$ Resistive \newline Plate Chambers (RPC’s) and in the region $1.05 < |\eta| < 2.7$, Thin Gap Chambers provide trigger and second coordinate measurements.

\section{Muon ATLAS MicroMegas Activity}

The sLHC luminosity upgrade by the factor of 10 will produce roughly 10 times more particle flux on detectors. Currently installed detectors are capable of detecting the particle fluxes, at least, up to five times the expected rate from the nominal LHC conditions. An upgrade in the ATLAS Muon System is going to be necessary, especially in the forward region ($|\eta|>2$) in order to maintain a good detector performance \cite{MAMMA2}.

MAMMA (Muon ATLAS MicroMegas Activity) group, established in 2007, proposes the large area (approximately $1 m \times 2 m$) bulk-micromegas technology as a solution with the following performance goals: Counting rate capability greater then $20 kHz/cm^2$, single plane detection efficiency greater than or equal to 98\%, $100 \mu m$ spatial resolution for the incident angles less than $45^\circ$, second coordinate measurement, two-track separation at around 1-2 mm distance, approximately 5 ns time resolution, level-1 triggering capability and good aging behaviour \cite{MAMMA2}.

In 2007, MAMMA group built and tested its first medium size ($450 mm \times 350 mm$) bulk-micromegas prototype and this chamber was tested in 2008 with 120 GeV pions. In these tests, a spatial resolution of ($35 \pm 5 \mu m$) for $500 \mu m$ strip pitch and $(24 \pm 7) \mu m$ for $250 \mu m$ strip pitch was achieved \cite{MAMMA3}. Later on, in order to overcome the spark problems, MAMMA group built micromegas prototypes with resistive anodes \cite{MAMMA1}. Different from the former examples of resistive anode technique, resistive layers in these detectors were segmented like the read-out strips in order to prevent charge-spread effects.
                                       
\chapter{GASEOUS RADIATION DETECTORS}\label{CH2}

After observations of photoelectric effect on metals, first experiments in which the ionizing effects of ultraviolet light on air is performed by Lenard in late 19th century. Lenard also measured that “negative ions” of gas moved 2000 times faster than the positive ions \cite{thomson}. Besides being an early clue for the existence of electrons, this velocity asymmetry between negative and positive ions is an essential factor for the pulse shape on the output of gaseous detectors.

Schematic drawing for a simplistic gaseous radiation detector is given in Figure 2.1. When any type of radiation that is energetic enough to ionize the gas confined between the electrodes of different polarity passes through, emergent electrons and ions are drifted towards the anode and cathode respectively hence produce an electrical current.

\begin{figure}[h]
 \centering
 \includegraphics[width=9cm,keepaspectratio=true, trim = 0mm -5px 0mm 0mm, clip]{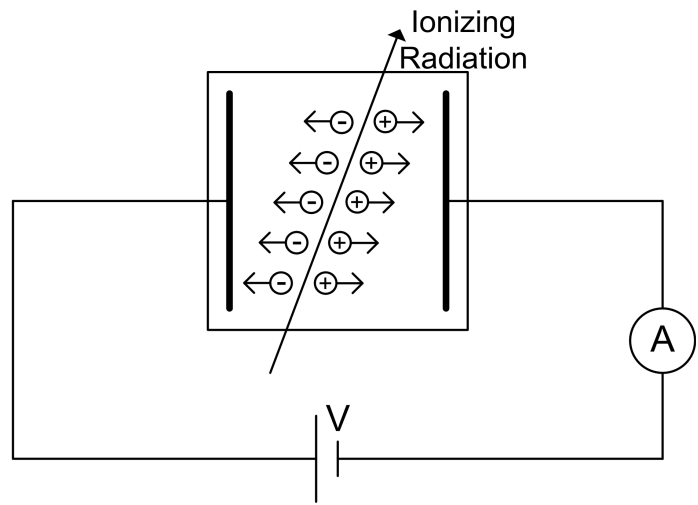}
 \caption{Schematic drawing of an ionization chamber.}
\end{figure}

When the primary electrons that are freed by the radiation are energetic enough to ionize the gas molecules again, secondary electron – ion pairs are produced. The bias voltage applied on electrodes can accelerate the electrons such that they can collide with the gas atoms and make new electron – ion pairs. As this process is repeated, an output signal proportional to the energy of the passing particle is produced via the so-called “avalanche multiplication”. Further increase in the bias voltage may result in a situation where every ionization produces sufficient number of free electrons that could make the gas effectively a conductor resulting total discharge of anode. This region of operation is called “Geiger - M\"{u}ller region” and is suitable for “counting” the number of particles that pass through the gas chamber without knowing their energies. The operating region where no secondary ionization occurs is called “ion chamber region” and the region where the bias voltage is so low that some of the primary electron – ion pairs recombine and do not reach to the electrodes is called “recombination region”. When the bias voltage is enough for the avalanche multiplication to occur, but less than the Geiger-M\"{u}ller limit, output signal strength becomes proportional to the passing particle energy; and this operating region is called “proportional region”. Dependence of the output signal strength on the passing particle energy and the bias voltage in gaseous detectors for different regions of operation is illustrated in Figure 2.2 \cite{ahmed}.

\begin{figure}
 \centering
 \includegraphics[width=14cm,keepaspectratio=true, trim = 0mm -5px 0mm 0mm, clip]{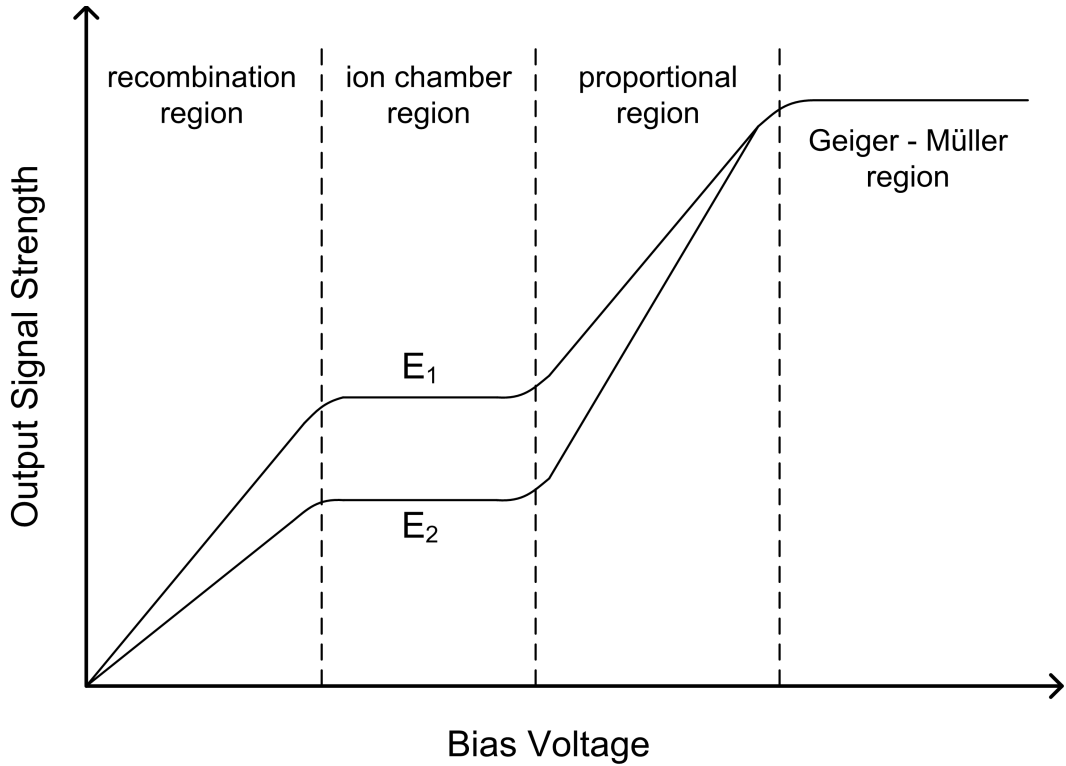}
 \caption{Dependence of the output signal shape on the bias voltage and energy of the passing particles for particles with energies $E_1$ and $E_2$ ($E_1 > E_2$).}
\end{figure}

In principle, any type of gas can be used in gas filled detectors as radiation can ionize any type of gas. In fact, as mentioned above, ionizing properties of radiation on gases were first observed on air. However, there is a certain threshold potential, depending on the type and pressure of the gas, below which the avalanche multiplication cannot occur. Since this threshold potential is lower for them, noble gases are commonly used as active gases in gaseous detectors. Most common example is argon because of its low cost. Another thing that should be taken into consideration is that, an ionized (excited) gas atom eventually recombines with an electron and emits an ultraviolet photon in this process. This emitted photon can hit the walls of the cathode electrode of the detector and free an electron via photoelectric effect and this new free electron can also get accelerated with the bias voltage and produce a “fake” signal. In order to overcome this issue, a “quenching” gas with a high absorbance for ultraviolet photons is used in the gas mixture of detector. Polyatomic gases such as $CH_4$ or $CO_2$ are common examples of quenching gases \cite{ahmed}.

\section{Early Examples}

Primitive examples of gas filled radiation detectors are ionization chambers, proportional counters and Geiger – Müller counters which work in the three different region of operation that are explained above (see Figure 2.2). Essential differences between these detectors are their region of operation and output signal characteristics.

Figure 2.1 can be thought of an example of an ionization chamber with parallel plate geometry. Since only the electrons and ions from first ionization contribute to the current; output signal in this kind of detectors is proportional to the “intensity” of the incoming radiation.

\begin{figure}
 \centering
 \includegraphics[width=12cm,keepaspectratio=true trim = 0mm -15mm 0mm 0mm, clip]{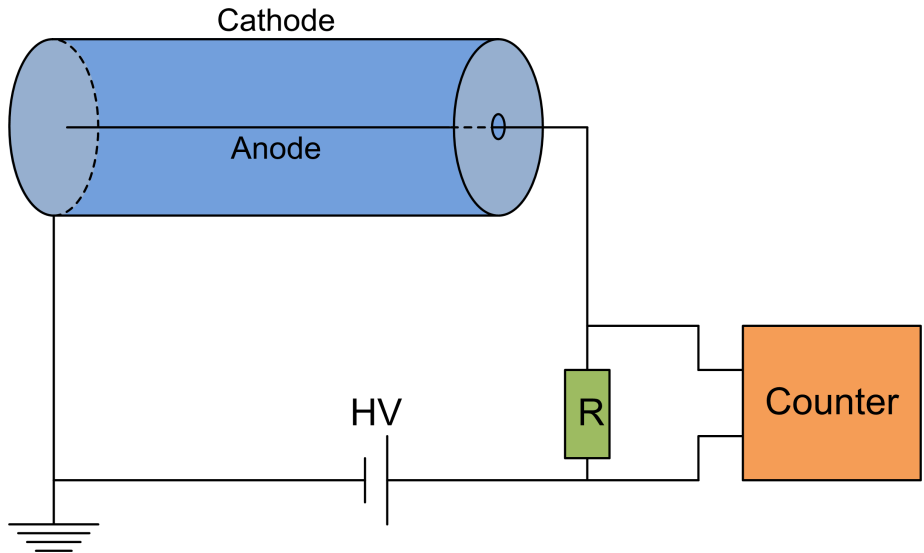}
 \newline
 \caption{Geiger-M\"{u}ller Counter.}
\end{figure}

An electrical counter for the alpha particles were first introduced by Rutherford and Geiger in 1908 \cite{RG}. Later on, in 1928, Geiger and Müller developed this device that is now called a “Geiger – M\"{u}ller counter” \cite{GM}]. A schematic view of a Geiger – Müller counter is shown in Figure 2.3. Since any ionization of the gas in a Geiger – Müller tube results in a very dense avalanche due to very high bias current, a high current flows through anode to cathode by the help of ionization electrons. Since a sharp pulse is created for each passing particle, it is possible to count the number of passing particles with a Geiger counter; however, there is no information about the energies of the incoming particles (see the “Geiger – Müller region” on Figure 2.2).

The proportional counter was invented in the late 1940s \cite{PC}. This device, in principle, looks like the Geiger – Müller counter, however it is operated with lower a bias voltage such that the output signal amplitude is proportional to the number of electrons from the initial ionization cluster. Since the number of electrons from the initial ionization is proportional to the energy of the travelling particle, it is possible to measure the energy of passing radiation with the proportional counter.

\section{Multiwire Proportional Chamber}

A revolutionary development in the area of gaseous radiation detectors was the invention of the Multiwire Proportional Chamber (MWPC) by George Charpak in 1968 \cite{MWPC}. For this invention, Georges Charpak was awarded with the Nobel Prize in Physics in 1992.

\begin{figure}
 \centering
 \includegraphics[width=14cm,keepaspectratio=true, trim = 0mm -5mm 0mm 0mm, clip]{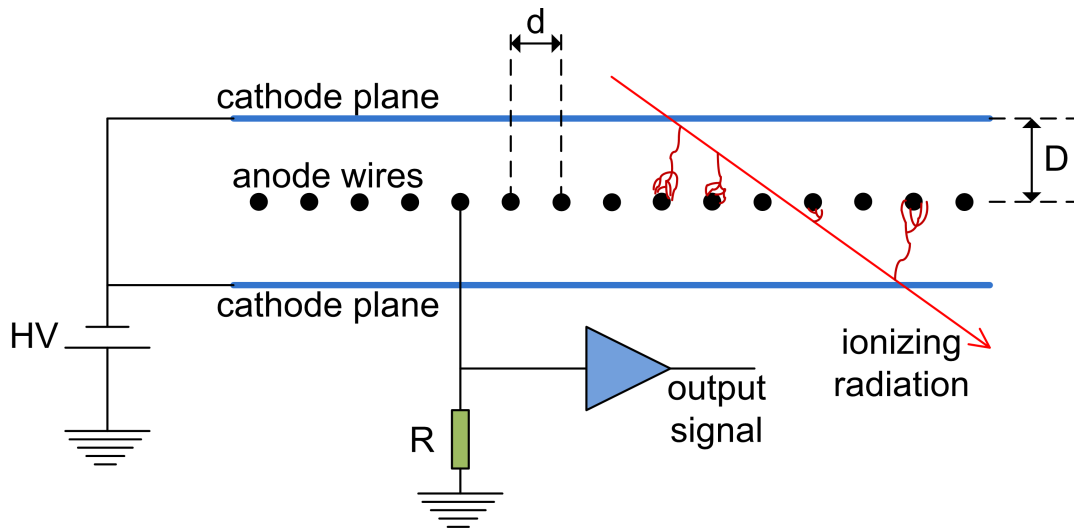}
 \caption{Multiwire Proportional Chamber. In reality, all of the anode wires are connected to amplifiers, only one is shown in this figure for the sake of clarity.}
\end{figure}

Multiwire proportional chamber, sketched on Figure 2.4, can be thought of an array of proportional counters operating in parallel. It contains a number of anode wires that are separated by a distance on the order of 1 mm and typically several centimeters above and below them, cathode planes (in the original production of Charpak, meshes are used as cathode electrodes) are present. Besides providing energy information of the passing particle, important feature of the MWPC is that it can also give the information about the track of the particle on the axis perpendicular to the anode wires. Several instances of ionization could occur during the passage of one particle which can result in avalanches on more than one anode wire (as shown in figure) and the output signal time and shape depends on the time and position (distance from the particular anode wire) of the initial ionization. These pieces of information could be combined to obtain the time and the position of the initial ionizations hence the track of the particle.

MWPCs are commonly used in particle physics experiments. One example is Thin Gap Chambers used for the triggering and the second coordinate measurement in the ATLAS muon system. In addition to the elements shown in Figure 2.4, TGCs have high resistivity (1 M$\Omega / \square$ and 0.5 M$\Omega / \square$) layers on their cathode planes for spark protection \cite{ATLAS}.

Although it is extensively developed and used, MWPCs suffered from several drawbacks; these were, mainly, difficulties in positioning the anode wires closer than a few milimeters, mechanical instabilities due to the electrostatic repulsion between the anode wires and slow removal of the positive ions from the active region of the detector hence the disturbance of the electric field due to the diffusing ions [20]. In order to overcome these limitations, new types of gaseous detectors called “Micropattern gaseous detectors” have been developed since the late 1980s.

\section{Micropattern Gaseous Detectors}

The era of the Micropattern Gaseous Detectors (MPGDs) began with the invention of the Microstrip Gas Chamber (MSGC) by Oed in 1988 \cite{MSGC}. This device had thin (10 $\mu m$) anode and thick (90 $\mu m$) cathode strips adjacently patterned on an insulating substrate and a back-plane electrode below the substrate. The structure is illustrated in Figure 2.5. 200 $\mu m$ strip pitch, one order of magnitude lower than the MWPCs is accurately achieved via photolithography production technique.

\begin{figure}[h]
 \centering
 \includegraphics[width=14cm,keepaspectratio=true, trim = 0mm -5mm 0mm 0mm, clip]{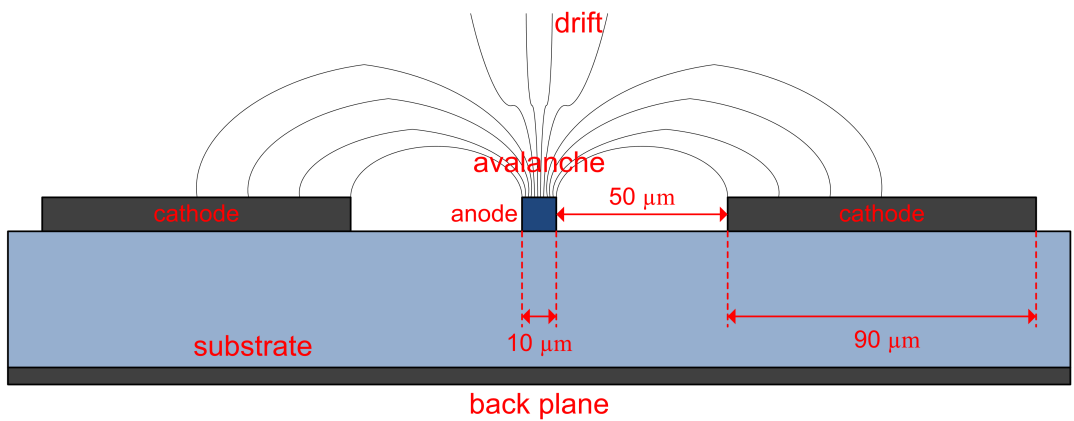}
 \caption{Schematic drawing of the Microstrip Gas Chamber. Field lines around the anode are roughly sketched.}
\end{figure}

Thin anode electrodes produce an intense electric field such that the avalanche multiplication occurs near the anode and thick cathode electrodes on both sides of the anode makes the removal of residual ions easy in MSGCs. For a proper operation, voltage on the back plane is arranged such that the field lines do not end on the insulating substrate. However, even though such an arrangement is done, some of the ions end up on the dielectric substrate and cannot be removed when the detector is operated at high gains \cite{chargeup}. This results in field-disturbing charge buildup and gain-reducing aging effects which are major issues with MSGCs. With their high read-out granularity and fast operation, MSGCs are used in several experiments such as HERA-B at DESY and CMS at CERN.

There are a few alternate micropattern structures, similar to MSGC, that will not be explained here. These are microgap chambers \cite{MG1, MG2}, smallgap chambers \cite{SG1, SG2} and microdot chambers \cite{MD1, MD2}. A detailed review on the micropattern detectors can be found in the reference \cite{MPGD}.

Since the recent developments in micromegas detectors is the motivation of the work in this thesis, the detailed explanation of micromegas and related technologies is given in a seperate section.

One last technology that should be mentioned under the micropattern gaseous detectors is the so-called Gas Electron Multiplier (GEM). Introduced by Sauli in 1996 \cite{GEM}, GEM is a three layer metal (copper, 18 $\mu m$ thick) – insulator (polymer, 25 $\mu m$ thick) - metal (copper, 18 $\mu m$ thick) structure with through-holes separated by a 100 $\mu m$ pitch. Applying a 200V potential difference between metal layers, an electric field of 40 kV/cm is achieved in the center of GEM holes. Using this technique in the parallel-plate geometries it is possible to confine the avalanche multiplication to the small region in the GEM holes hence prevent the ion diffusion as much as possible.

\section{Micromesh Gaseous Structure}

Micromegas (Micromesh Gaseous Structure) was introduced in 1996 as a new detector concept with which the problems with the MSGCs were essentially resolved \cite{MICROMEGAS}. The structure (illustrated in Figure 2.6) had 150 $\mu m$ anode strips separated by a 200 $\mu m$ pitch, deposited on a Kapton substrate via metal deposition techniques. 100 $\mu m$ above the anode strips, “the micromesh”, a grid of 3 $\mu m$ thick metals with 17 $ \times $ 17 $\mu m$ openings stands on the quartz spacers. 3 mm above the micromesh a bigger mesh is placed as the drift electrode on top.

\begin{figure}
 \centering
 \includegraphics[width=12cm,keepaspectratio=true, trim = 0mm -5px 0mm 0mm, clip]{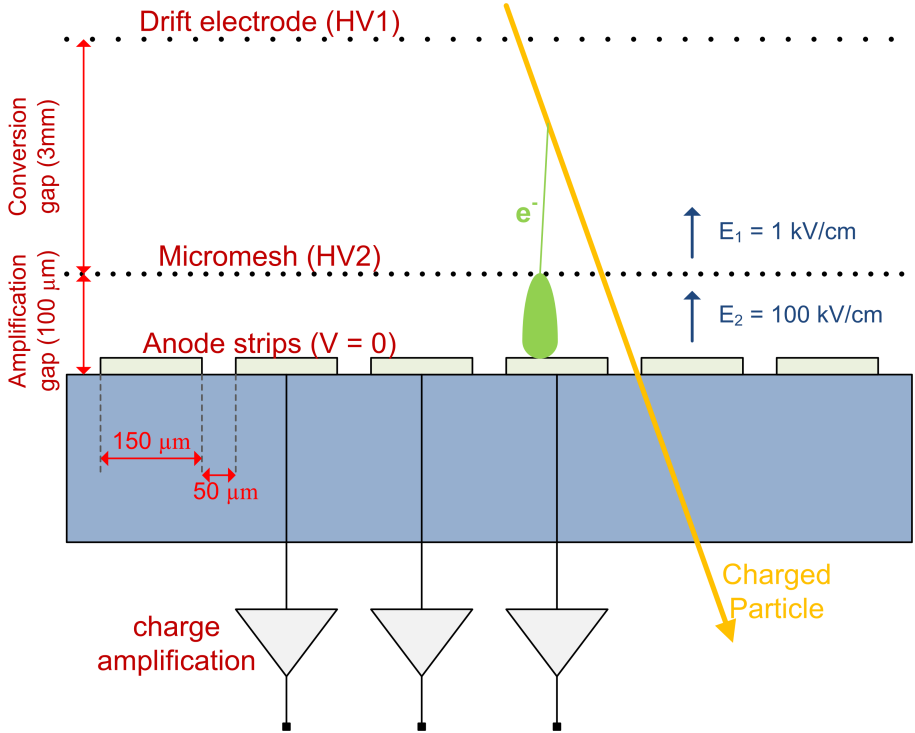}
 \caption{Schematic view of the micromegas detector.}
\end{figure}

In the micromegas detector, anode potential was set to 0 volts. Voltages of the mesh and drift electrodes were set to the negative values such that the electric field between mesh and the anode strips was 100 kV/cm and between drift and mesh were 1 kV/cm. As a result of this arrangement, avalanche multiplication of the electrons was only happening in the small amplification gap between the mesh and the anode strips. The structure provided high performance parameters: Signals faster than 1 ns and gains up to 105 are reported in \cite{MICROMEGAS}.

The challenging process in the construction of a micromegas structure was gluing the mesh on the supporting pillars which were placed on the anode plane via photolithography. Flatness of mesh, which in this process is determined by the accuracy of the pillars, is essential for the homogeneity of the gain along the detector area. In 2006, a new production method, adopted from the Printed Circuit Board (PCB) technology, was introduced with the name “bulk micromegas” \cite{muMbulk}. In this technique, anode plane with the copper readout strips is covered by a photoresistive material and then the woven wire mesh (used instead of the electroformed micromesh) is placed on top. Then the photoresistive materials were etched via photolithography to form the pillars. This technique reduced the number of production steps to one and made it possible to design large-area micromegas detectors for larger scale applications.

Several high energy physics experiments are equipped with the micromegas detectors such as COMPASS \cite{COMPASS1, COMPASS2, COMPASS3, COMPASS4, COMPASS5, COMPASS6} and CAST \cite{CAST1, CAST2, CAST3, CAST4, CAST5, CAST6, CAST7, CAST8, CAST9} experiments at CERN and T2K  \cite{T2K1, T2K2, T2K3} experiment in Japan. Besides its use in particle physics, Micromegas detectors are also proposed to be used in medical and industrial imaging \cite{IMG1, IMG2}.

It is reported for the micropattern gaseous detectors that when the number of electrons in an avalanche goes beyond a value between 10$^7$ and 10$^8$, the so-called “Raether limit” \cite{Raether}, secondary avalanches exceed the electron cloud both from the front and the back; resulting in sparks \cite{MPGDhr}. For a heavily ionizing radiation which would create approximately 10$^4$ electron-ion pairs per cm, this limit is reached with a detector gain of 10$^3$ - 10$^4$. It is shown that in the case of a spark, all the charge on the mesh of a micromegas detector is coursed down on the anode strips \cite{muMspark1}. A detailed study on the sparking in micromegas detectors showed that the spark rate in heavily ionizing environment was too big to get micromegas detectors to operate at high particle rates such as those encountered under LHC conditions without making any improvements on the basic micromegas structure \cite{muMspark2}.

In order to build a micromegas detector that would meet the conditions of the Super Large Hadron Collider, MAMMA group modified the micromegas structure by placing high-resistivity strips above the read-out strips and an insulator layer in between \cite{MAMMA1}. This structure is depicted in Figure 2.7. Both resistive and readout strips were 150 $\mu m$ wide and 100 mm long with a strip pitch of 250 $\mu m$.

\begin{figure}
 \centering
 \includegraphics[width=13cm,keepaspectratio=true, trim = 0mm -5mm 0mm 0mm, clip]{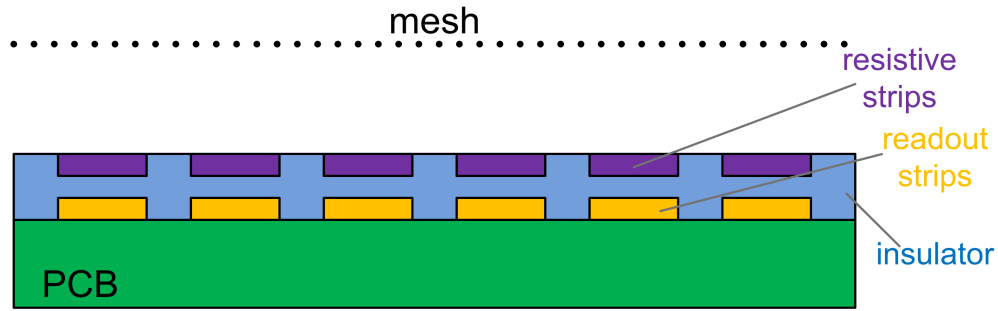}
 \caption{Micromegas structure with the resistive strips. View from the cross section along the axis perpendicular to the readout strips. Drift electrode which is not shown here stands 4 or 5 mm above the mesh.}
\end{figure}

MAMMA group presented 3 prototypes of their resistive anode bulk-micromegas detectors which were same in geometry, but had resistive strips with different resistivities. Both resistive and readout strips were at 0 potential and mesh and the drift electrodes were connected to the negative high voltages as usual. They compared their results with another prototype of same dimensions but without the resistive strips.

With the high flux experiments, MAMMA group showed that their micromegas chambers with resistive strips were stable against the sparks for gas (avalanche) gains up to 20,000. High voltage on the mesh was stable and spark currents were lesser than a few 100 nA which was more than 1 order of magnitude lower than a chamber without resistive strips under the same conditions.

MAMMA group recently started to develop large area micromegas chambers with longer resistive strips. As the length of the resistive strip increases, its capacitance and maximum resistance also increase, thus these parameters should carefully be arranged such that possible ``charge-up''s on resistive strips due to high rates could be avoided. In the next chapter, a simulation tool developed for this purpose is going to be explained.                       
\chapter{SIMULATION OF CHARGE TRANSPORT AND DISCHARGE}\label{ch:chani}

Main study of this thesis is the transient simulation of charge transport on a rectangular surface. As explained in the previous chapter, motivation of this study is to optimize the resistive structures in the micromegas chambers that is being developed by the MAMMA group at CERN.

Simulation of charge transport is carried out by the solution of the Poisson equation through the surface of interest at each time step and calculation of the currents between small subcells of the surface. Since an analytical solution of the Poisson equation does not exist for a rectangular surface of finite dimensions, an approximate method, so called “moments method” is used for the field calculations. In the next subsection, general mathematical description of method of moments and, in particular, its application to the Poisson equation is introduced. In the following section, general working principles of the simulator is explained and some calculations are presented.

\section{Electrostatics via Moment Methods}

As it is going to be explained, method of moments is the collection of methods in which the self-adjoint linear operators are written in the matrix forms. For the most cases, these matrices are infinite dimensional; however, large, but finite, dimensional estimations provides accurate enough solutions for practical purposes. General theory of linear function spaces are not going to be explained here, but can be found in many text books on applied mathematics for physics and engineering problems \cite{riley, arfken}.

Moment methods have a wide range of application in electrostatics, electrodynamics, microwaves, antennas and network problems. One can find the detailed text covering application of moment methods to all these subjects in referred monograph \cite{harrington}. Here, the general formulation of the problem and its application to the Poisson equation is presented as it is in the reference \cite{harrington}.

\subsection{Mathematical description}

Consider the following inhomogeneous equation of the self-adjoint operator \emph{L}:
\begin{equation}\label{eq:Lfg}
   Lf = g
\end{equation}
In this equation, the function \emph{g} is called the source and function \emph{f} is called the field or response. Once the matrix form of the \emph{L} is obtained and one of the functions \emph{f} or \emph{g} is known, it becomes possible to calculate the unknown function directly or after a matrix inversion.

The inner product $\langle f , g \rangle$ is a scalar with the following properties:
\begin{equation}\label{eq:prop1}
  \langle f , g \rangle = \langle g , f \rangle
\end{equation}
\begin{equation}\label{eq:prop2}
  \langle \alpha f + \beta g , h \rangle = \alpha \langle f , h \rangle + \beta \langle g , h \rangle
\end{equation}
\begin{equation}\label{eq:prop3}
    \langle f , f^* \rangle \mbox{ } \left\{ \begin{array}{ll}
		   >0 & \mbox{if } f \neq 0 \\
		   =0 & \mbox{if } f = 0
	   \end{array}
\right.
\end{equation}
here, $\alpha$ and $\beta$ are scalar constants and complex conjugation is denoted by $^*$. The adjoint operator $L^\dagger$ is defined as follows:
\begin{equation}\label{eq:adjoint}
  \langle L f , g \rangle = \langle f , L^\dagger g \rangle
\end{equation}
If the operator and its adjoint are the same, then the operator is called “self-adjoint”. Since the adjoint operator depends on the definition of the inner product and there is no unique way of defining the inner product; the inner product is mostly arranged such that the operator becomes self adjoint. A general integral expression for the inner product can be written as
\begin{equation}\label{eq:innerproduct}
  \langle f , g \rangle = \int_a^b \! w(x) f(x) g(x) \, \mathrm{d} x
\end{equation}
with the “weighting function” \emph{w(x)}.

Equations \ref{eq:prop1}-\ref{eq:innerproduct} are the fundamental objects required to formulate the method of moments. Moving to the method of moments, let the function $f$ of the Equation \ref{eq:Lfg} be expanded on the basis functions $f_1, f_2, … n$ of the space defined by \emph{L} with the expansions coefficients $\alpha_1, \alpha_1, …, \alpha_n$:
\begin{equation}
  f = \sum\limits_n \alpha_n f_n
\end{equation}
Inserting this expression into \ref{eq:Lfg}, one gets:
\begin{equation} \label{eq:alphaLfg}
  \sum\limits_n \alpha_n L f_n = g
\end{equation}
Finally, applying inner products with the weighting functions $w_1, w_2, …, w_m$ to the both sides of \ref{eq:alphaLfg} one ends up with m equations in the following form:
\begin{equation} 
  \sum\limits_n \alpha_n \langle w_m , L f_n \rangle = \langle w_m , g \rangle
\end{equation}
which can be re-written in the matrix form as:
\begin{equation} \label{eq:lmnalphag}
  [l_{mn}][\alpha_n] = [g_m]
\end{equation}
with
\begin{equation} \label{eq:lmn}
  [l_{mn}] = \begin{bmatrix}
		    \langle w_1 , L f_1 \rangle & \langle w_1 , L f_2 \rangle & \cdots \\
		    \langle w_2 , L f_1 \rangle & \langle w_2 , L f_2 \rangle & \cdots \\
		    \vdots & \vdots & \ddots
	     \end{bmatrix}
\end{equation}
\begin{equation} 
  [\alpha_n] = \begin{bmatrix}
		    \alpha_1 \\
		    \alpha_2 \\
		    \vdots
	     \end{bmatrix}
\end{equation}
\begin{equation} 
  [g_m] = \begin{bmatrix}
		    \langle w_1 , g \rangle \\
		    \langle w_2 , g \rangle \\
		    \vdots
	     \end{bmatrix}
\end{equation}
By the help of the matrix equation \ref{eq:lmnalphag}, if the expansion coefficients, $\alpha_n$, are known, it is possible to calculate the source, \emph{g}, which generates that field; and also if the l-matrix given with \ref{eq:lmn} is non-singular, hence its inverse exists, one can calculate the expansion coefficients corresponding to a source via the inverse equation:
\begin{equation} 
 [\alpha_n] =  [l_{mn}]^{-1} [g_m]
\end{equation}
So far, nothing has been stated about the boundaries and the weighting function of the inner product and the basis functions for the expansion of f and in fact, proper choice of these according to the nature of the problem of interest is essential in the application of this method.

One approximation that is going to be used for the Poisson equation is the so-called point matching technique. Within the point matching approximations, one chooses the weighting functions such that \ref{eq:Lfg} is satisfied on a discrete set of points of interest. This corresponds to choosing the weighting functions simply as Dirac delta functions:
\begin{equation} 
 w_m = \delta (x - x_m)
\end{equation}
Another approximation method that is also going to be used is the choice of subsectional bases. In this technique, each basis function $f_n$ is non-zero over a certain $n^{th}$ region and zero, outside that region; thus, each expansion coefficient $\alpha_n$ is effective over a certain region. In general, subsectional bases can be expressed as follows:
\begin{equation}
    f_n (x) = \left\{ \begin{array}{ll}
		   f(x) & \mbox{if x is in the $n^{th}$ region} \\
		   0 & \mbox{if x is outside the $n^{th}$ region}
	   \end{array}
\right.
\end{equation}

These concepts of subsectional bases and point matching are going to be used in the solution of the Poisson equation which is explained in detail in the next subsection.

\subsection{Solution of the Poisson equation using moment methods}

Electrical potential in two dimensions is determined with the Poisson equation which, in differential form, can be written as,
\begin{equation}
  \nabla^2 V = - \frac{\sigma}{\epsilon} \delta(z)
\end{equation}
here, $\nabla^2$ is the Laplace operator, $\sigma$ is the surface charge density at \emph{z = 0} and $\epsilon$ is the electric permittivity of the medium. Dirac delta function of \emph{z} trivally transforms the volume charge distribution to the surface charge distribution on \emph{z = 0}. Assuming a conducting plate lying on the \emph{x-y} plane of the coordinate system, the well known integral solution for the electrical potential on \emph{x-y} plane is:
\begin{equation}
  V(x,y) = \iint \frac{ \sigma (x', y') }{4 \pi \epsilon \sqrt{(x-x')^2 + (y-y')^2}} \, \mathrm{d} x' \mathrm{d} y'
\end{equation}
In the operator formalism, this equation can be written as,
\begin{equation}\label{eq:LsigmaV}
  L \sigma (x' , y') = V (x , y)
\end{equation}
where,
\begin{equation}
 L = \iint \frac{\mathrm{d} x' \mathrm{d} y'}{4 \pi \epsilon \sqrt{(x-x')^2 + (y-y')^2}}
\end{equation}
Equation \ref{eq:LsigmaV} is the equation that is going to be written in the discrete form via the method of moments. A rectangular surface with a width of $l_x$ and length of $l_y$, divided in $n_x$ subsections in $x$ and $n_y$ subsections in $y$ is depicted in Figure 3.1. As it is shown in the Figure, each subsection is labeled with the numbers from 1 to $n_x \times n_y$ starting from the cell on the bottom left, to the cell on the top right.

\begin{figure}
 \centering
 \includegraphics[width=12cm,keepaspectratio=true]{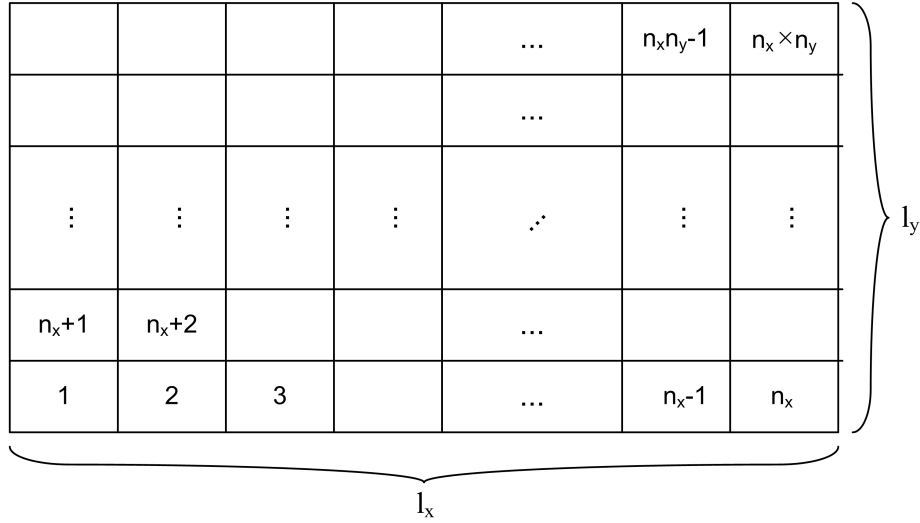}
 \caption{$l_x$ by $l_y$ rectangular surface with $n_x \times n_y$ subsections.} 
\end{figure}

A simple approximation is to expand the surface charge distribution over the constant functions which are only nonzero in one cell, namely:
\begin{equation}
  f \approx \sum\limits_n \alpha_n f_n
\end{equation}
\begin{equation}\label{eq:fnpoisson}
    f_n (x',y') = \left\{ \begin{array}{ll}
		   1 & \mbox{if (x' , y') is in the $n^{th}$ cell} \\
		   0 & \mbox{if (x' , y') is outside the $n^{th}$ cell}
	   \end{array}
\right.
\end{equation}
This, physically, corresponds to assuming that the surface charge density over the area of the $n^{th}$ cell is constant and equal to $\alpha_n$.
The second approximation to be done is the point matching approximation which is mentioned in the previous section. Denoting the center coordinates of each cell with $(x_m, y_m)$, if the weighting function is chosen as
\begin{equation} 
 w_m = \delta (x - x_m) \delta (y - y_m)
\end{equation}
one can write the approximation of Equation \ref{eq:LsigmaV} in the form of \ref{eq:lmnalphag}; the matrix elements $l_{mn}$ and the elements of the $g$ vector can be written as follows (note that the function $g$ of \ref{eq:Lfg} corresponds to $V(x,y)$ of \ref{eq:LsigmaV}):
\begin{align*}
 l_{mn} &= \langle w_m , L f_n \rangle = \iint \mathrm{d} x \mathrm{d} y \,  w_m L f_n \\
	&= \iint \mathrm{d} x \mathrm{d} y \, \delta (x - x_m) \delta (y - y_m) \\
	&\times \iint_{in \Delta S_n} \frac{\mathrm{d} x' \mathrm{d} y'}{4 \pi \epsilon \sqrt{(x-x')^2 + (y-y')^2}}
\end{align*}
\begin{equation} \label{eq:lmnintegral}
 l_{mn} = \iint_{in \Delta S_n} \frac{\mathrm{d} x' \mathrm{d} y'}{4 \pi \epsilon \sqrt{(x_m-x')^2 + (y_m-				y')^2}}
\end{equation}
\begin{equation*} 
 g_m = \langle w_m ,g \rangle = \iint \mathrm{d} x \mathrm{d} y \,  \delta (x - x_m) \delta (y - y_m) V(x,y)
\end{equation*}
\begin{equation} 
 g_m = V(x_m,y_m)
\end{equation}
Expression “in $\Delta S_n$” under the integral sign in \ref{eq:lmnintegral} indicates that the integral should be carried out through the area of the $n^{th}$ cell. This is because the basis functions $f_n$ \ref{eq:fnpoisson} are zero outside the $n^{th}$ cell.

Physical interpretation of the above equations is intuitive: $l_{mn}$ given with the \ref{eq:lmnintegral} is the influence of the unit surface charge distribution of the $n^{th}$ cell, on the electrical potential at the center of the $m^{th}$ cell.

One last, harmless, approximation can be done in the calculation of the integral $l_{mn}$ for different cells is to assume that the distance between the cells (square root in the denominator of \ref{eq:lmnintegral}) is constant and equal to the distance between center of the cells. Denoting the area of the $n^{th}$ cell with $\Delta S_n$, matrix element can be written as,
\begin{equation} \label{eq:lmn}
 l_{mn, \, m \neq n} = \frac{\Delta S_n}{4 \pi \epsilon R_{mn}}
\end{equation}
where the distance between $m^{th}$ and $n^{th}$ cells, $R_{mn}$ (see Figure 3.2) is:
\begin{equation} 
 R_{mn} = \sqrt{(x_m - x_n)^2 + (y_m - y_n)^2}
\end{equation}

\begin{figure}
 \centering
 \includegraphics[width=10cm,keepaspectratio=true]{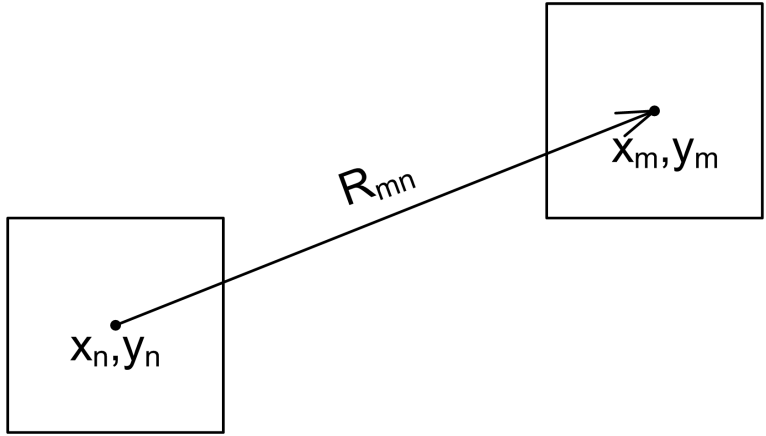}
 \newline
 \caption{Illustration of the $l_{mn}$ and $n_{th}$ cells and the $R_{mn}$ vector.}
\end{figure}

Diagonal elements of the l-matrix $l_{nn}$ given with \ref{eq:lnn}, this physically corresponds to the influence of the unit charge distribution through the $n^{th}$ surface on the electrical potential at its center. Denoting the unit lengths of the cells in $x$ and $y$ direction with $a_x=l_x/n_x$ and $a_y = l_y/n_y$, $l_{nn}$ integral is computed by the help of Wolfram Mathematica Online Integrator \cite{mathematica} as:
\begin{equation*} 
 l_{nn} = \int_{-a_x/2}^{a_x/2} \mathrm{d} x \int_{-a_y/2}^{a_y/2} \mathrm{d} y \frac{1}{4 \pi \epsilon \sqrt{x^2 + y^2}}
\end{equation*}
\begin{equation} \label{eq:lnn}
 l_{nn} = \frac{1}{4 \pi \epsilon} \left[ a_x \ln \left( \frac{\sqrt{a_x^2 + a_y^2} + a_y}{\sqrt{a_x^2 + a_y^2} - a_y} \right) +  a_y \ln \left( \frac{\sqrt{a_x^2 + a_y^2} + a_x}{\sqrt{a_x^2 + a_y^2} - a_x} \right) \right] 
\end{equation}

Since $\alpha_n$ and $g_m$ have physical meanings in the approximate solution of the Poisson equation explained above, the approximate equation can be rewritten with the new symbols according to their physical meanings as follows:
\begin{equation} \label{eq:lmnsigmaV}
  [l_{mn}][\sigma_n] = [V_m]
\end{equation}

Since the physical meaning of $\alpha_n$ is the value of the constant charge distribution on the $n_{th}$ cell it is denoted by $\sigma_n$ instead. Similarly, since the physical meaning of the $g_m$ is the magnitude of the electric potential at the center of the $m_{th}$ cell, it is denoted by $V_m$.

Combining \ref{eq:lmn} and \ref{eq:lnn}, elements of the l-matrix can be written as:
\begin{equation}
    l_{mn} = \left\{ \begin{array}{ll}
		   \frac{\Delta S_n}{4 \pi \epsilon \sqrt{(x_m - x_n)^2 + (y_m - y_n)^2}} & m \neq n \\
		   \frac{1}{4 \pi \epsilon} \left[ a_x \ln \left( \frac{\sqrt{a_x^2 + a_y^2} + a_y}{\sqrt{a_x^2 + a_y^2} - a_y} \right) +  a_y \ln \left( \frac{\sqrt{a_x^2 + a_y^2} + a_x}{\sqrt{a_x^2 + a_y^2} - a_x} \right) \right]  & m = n
	   \end{array}
\right.
\end{equation}
  
It should be noted that the elements of the l-matrix only depends on the geometrical parameters, hence, once the elements of the l-matrix are calculated, computing the electrical potential corresponding to a charge distribution is only multiplications and additions. This is desired, as through the transient simulation, this calculation has to be done many times.

It is also possible to calculate the charge distribution corresponding to a certain potential by using the inverse equation of \ref{eq:lmnsigmaV}:
\begin{equation} \label{eq:sigmalinvV}
  [\sigma_n] = [l_{mn}]^{-1}[V_m]
\end{equation}

For a constant potential, charge distribution and, hence, the total charge can also be calculated using \ref{eq:sigmalinvV}. One can find the capacitance of the single plate by calculating the total charge corresponding to the unit potential.

\section{Chani}

The developed simulation tool for the charge transport and discharge on a rectangular surface is named “Chani”. This is due to both words “charge” and “Chani” sharing the first two letters “ch” and referral to the name of a character from the science fiction novel Dune \cite{dune}.

Chani is developed to be run on the ROOT Framework \cite{ROOT} with which running C++ code as a macro (without compiling) is possible via CINT, C++ interpreter \cite{CINT}. Chani also takes the advantage of the ROOT vector, matrix and histogram classes and the libraries for visualization.

The main functions of the Chani are declared in the header file, Chani.h. These main functions and their jobs are explained as follows:

\emph{parameters(lx, nx, ly, ny, Nm, sigmas)}: Defines the surface with dimensions lx and ly with surface resistivity sigmas. Tells the simulator to divide the surface in x and y to nx and ny respectively. States that the l matrix is going to be calculated in Nm $\times$ Nm parts. \newline                                                                                    \emph{computelmn()}: Calculates the l-matrix.                 \newline
\emph{addConnector(x, y, R)}: Adds a ground connection with connector resistance R to the nearest cell to (x, y). \newline                                                                                                                                                                                                                                           \emph{addCharge(x, y, q, tStep)}: Adds q amount of charge to the nearest cell to (x, y) at the time instance tStep.\newline                                                                                                                                                                                                                                             \emph{transient1(tStepi, nofSteps, Deltat)}: Performs the transient calculation starting from the time step "tStepi" for "nofSteps" steps of "Deltat" seconds. Electrical potential and charge distribution objects are written to and read from hard drive at each step. Suitable for large-scale calculations, where, memory is not enough to hold all of the objects.\newline
\emph{transient2(tStepi, nofSteps, Deltat)}: Performs the transient calculation starting from the time step "tStepi" for "nofSteps" steps of "Deltat" seconds. Electrical potential and charge distribution objects are held in the memory. Suitable for small-scale calculations.  \newline                                                                                                                                                                   \emph{capacitance()}: Calculates and outputs the capacitance of the surface. \newline                                                                                                                                                                                                                                                                       \emph{getV(i)}: Gets the i-th element of electrical potential vector in function transient1(). \newline                                                                                                                                                                                                                                                 \emph{getQtotal(tStep)}: Gets the total amount of charge at the time step tStep. \newline   
\emph{getQxy(x, y, tStep)}: Gets the amount of charge on the nearest cell to (x, y) at the time instance tStep.

Header file, main functions and main classes are given in Appendix A.

A rectangular surface is handled by the simulator as explained in the previous section (see Figure 3.1). Through the transient simulation electric potential of each cell and the currents between neighboring cells are calculated. Surface resistivity ($\rho_s$), assumed to be constant over the surface, is an input of the simulator. For a cell with the label $i$, equivalent resistance between the $i^{th}$ cell and the cell on its right (cell number: $i + 1$) is:
\begin{equation} 
  R_x = (\rho_s \times a_x) / a_y
\end{equation}

Similarly, the resistance between $i^{th}$ cell and the upper cell (cell number: $n_x + i$) is:
\begin{equation} 
  R_y = (\rho_s \times a_y) / a_x
\end{equation}

Hence the currents between these cells are simply calculated from Ohm’s law as:
\begin{equation} 
  I_{x_i} = (V_i - V_{i+1}) / R_x
\end{equation}
\begin{equation} 
  I_{y_i} = (V_i - V_{n_x+i}) / R_y
\end{equation}
Here, $V_i$’s are electrical potential. These currents and resistances are illustrated in Figure 3.3.
\begin{figure}
 \centering
 \includegraphics[width=140px,keepaspectratio=true]{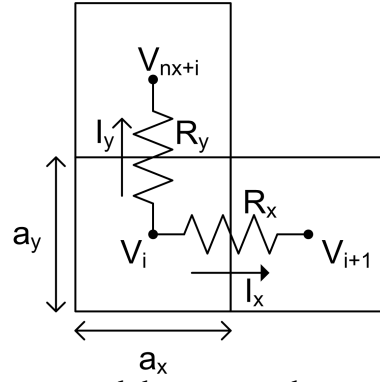}
 \caption{Resistances and the currents between neighboring cells.}
\end{figure}

These currents are assumed to be constant during one time step. Hence the amount of the migrating charge ($\Delta q$) from on cell to the other cell is simply calculated by multiplying the current with the length of the time step ($\Delta t$):
\begin{equation} 
  \Delta q = I \times \Delta t
\end{equation}

Finally, resistances at the ground connection points and the cells that the ground connection is made, are:
\begin{equation} 
  R_g = R_c + [\rho_s \times (a_{x(y)} / 2)]/a_{y(x)}
\end{equation}
If the connector is in x (y) direction. In the equation above, $R_c$ is the “connector resistance” which is the resistance between the power supply and the connection point, and it is summed with the resistance coming from the half-length of the connection cell. Discharge current and the amount of charge removed is calculated similarly. 

In the next section, the charge transport simulation on a $2 \times 2 \, cm^2$ surface carried out by Chani is explained.

\begin{figure}[h]
 \centering
 \includegraphics[width=10cm,keepaspectratio=true, trim = 0mm -2mm 0mm 0mm, clip]{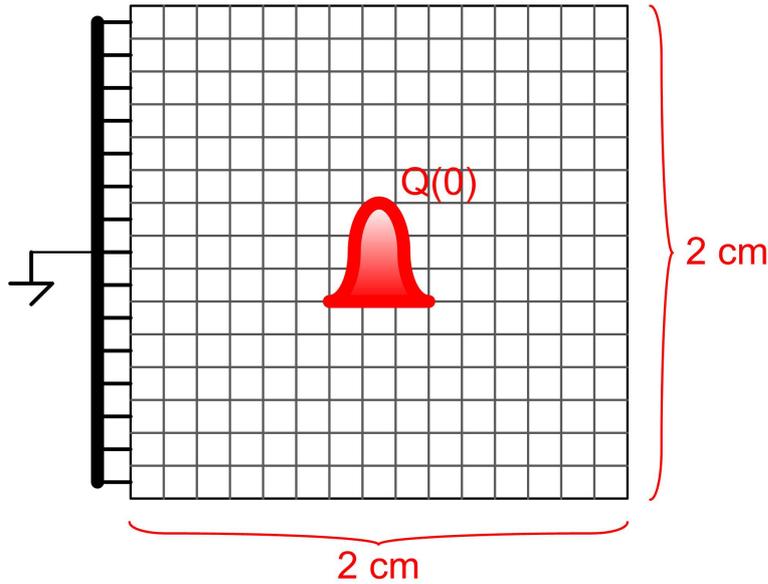}
 \caption{$2 \times 2 \, cm^2$ surface with total number of 225 cells.}
\end{figure}

\section{An Example}

Illustration of the surface on which the charge transport is simulated using Chani is in Figure 3.4. It is a $2 \times 2 \, cm^2$ square surface with the constant resistivity of $10^5 \Omega / \square$. In the simulator, it is divided to 15 in both directions hence the total number of cells for the simulation is 225. The surface is grounded from the left end of it without any connector resistance and initially $10^4$ elementary charges arrive on the center and the neighboring eight cells of the surface. 

Transient simulation with Chani is performed for 2000 time steps with one step being $10^{-10}s$ long, thus physically for $200 ns$. Surface definition and simulation files are given in Appendix B \& B.1 respectively.

Change in the total amount of charge with time is plotted in Figure 3.5 where exponential-like discharge is clearly seen. It should be noted that no capacitance parameter is input to the simulation but the effect of self-capacitance of the plate intrinsically calculated and thus RC-like discharge behavior is obtained.

\begin{figure}[h]
 \centering
 \includegraphics[width=11cm, keepaspectratio=true]{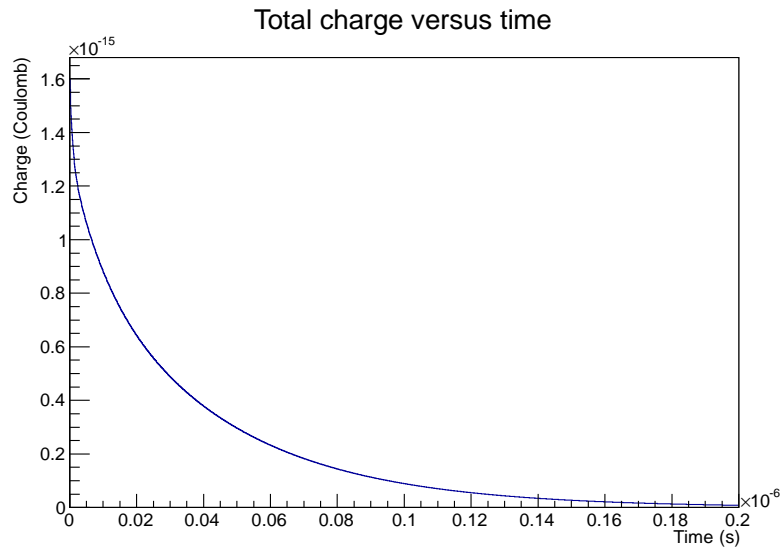}
 \caption{Total charge on the surface versus time.}
\end{figure}

Figure 3.6 shows the surface charge density at four time instances. It can be seen that the charge, initially at the center is spread very quickly and at t = 20 ns, charge distribution becomes almost uniform except the edges and the ground connection point as expected.

\section{Self-Consistency Tests}

Two self-consistency tests are performed in order to proove the convergence of the calculations and to estimate the order of errors. In this section, these tests are explained and example calculation results on the surface studied in the previous section is presented.

\begin{landscape}
\thispagestyle{empty}
 \begin{figure}
 \centering
 \includegraphics[width=700pt,keepaspectratio=true]{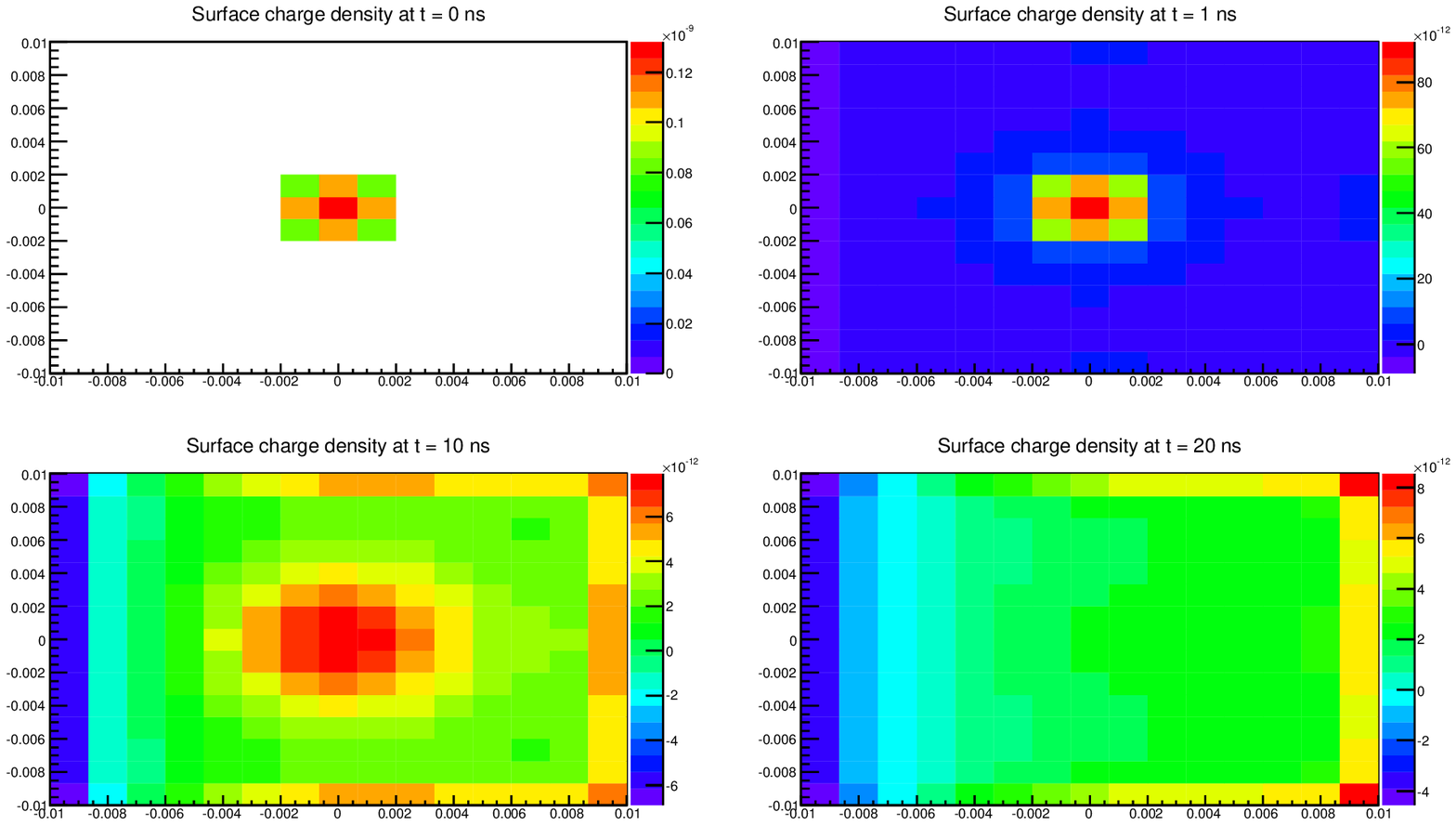}
 \caption{Color plot of the surface charge density at four different times.}    
      \vspace{0.5cm}
      \hspace{0cm}\pageref{fig:examplecolz}
      \label{fig:examplecolz}
\end{figure}
\end{landscape}

\subsection{Capacitance versus number of cells}

One way to check the accuracy of the calculations is to check the capacitance of the surface with the increasing number of cells. One should expect to find capacitance value converging to a certain real value as the number of divisions increases. Although it is not directly related with the transient calculations, convergence of capacitance is a reference point to check whether the simulation is meaningful.

As an example, capacitance of the $ 2 \times 2 \, cm^2 $ surface is iteratively calculated while increasing the number of divisions. After each iteration relative error on capacitance is calculated as follows:
\begin{equation} \label{eq:errorC}
  \text{ Error }= \frac{ | C_{\text{present}} - C_{\text{previous}} | }{C_{\text{previous}}}
\end{equation}

\begin{figure}[h]
 \centering
 \includegraphics[width=11cm, keepaspectratio=true]{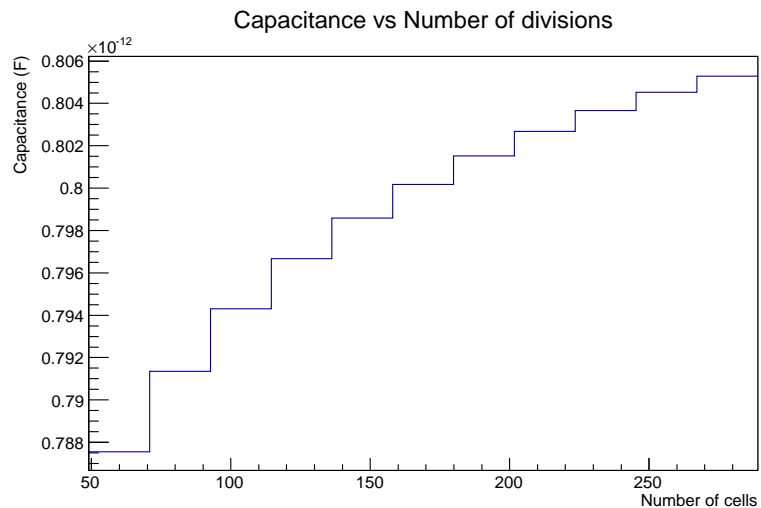}
 \caption{Capacitance versus number of cells. \label{fig:CvsN}}
\end{figure}

\begin{figure}[h]
 \centering
 \includegraphics[width=11cm, keepaspectratio=true]{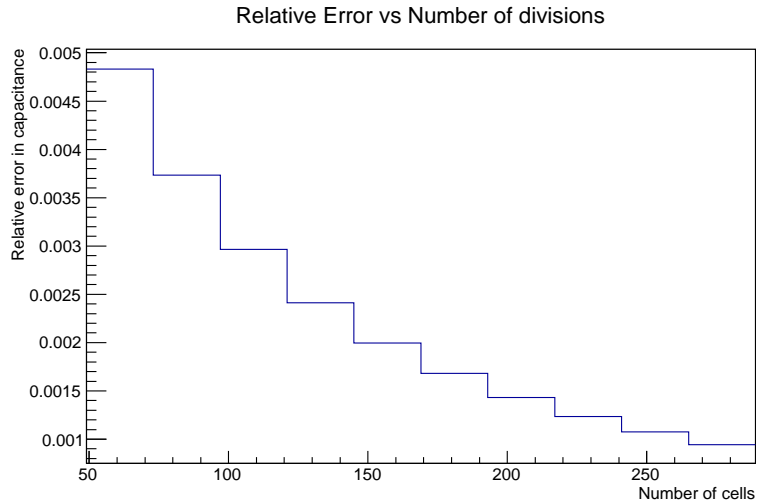}
 \caption{Relative error versus number of cells. \label{fig:ECvsN}}
\end{figure}

Figure~\ref{fig:CvsN} shows the variation of capacitance against number of divisions used in the calculation and the variation of error defined with Equation \ref{eq:errorC} is shown in Figure~\ref{fig:ECvsN}. Iterative calculation of capacitance is performed until the relative error goes below the tolerance 0.001. Source file can be found in Appendix B.2. The tolerance is satisfied with 17 divisions in each direction hence 289 total number of cells. It is clearly seen from the figures 3.7 and 3.8 that the capacitance approaches to a certain value with decreasing amount of error as the number of cells increases.

\begin{figure}[h]
 \centering
 \includegraphics[width=11cm, keepaspectratio=true]{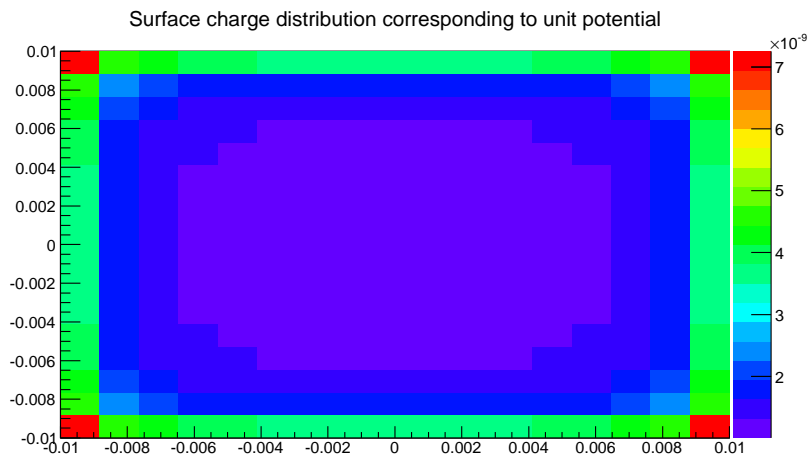}
 \caption{Surface charge density distribution corresponding to the unit constant potential. \label{fig:unitcolz}}
\end{figure}

Another “sanity check” may be to check the charge distribution corresponding to constant potential. capacitance() function of Chani also saves the charge distribution corresponding to the unit potential. Surface charge density distribution of the $ 2 \times 2 \, cm^2 $ corresponding to unit consant potential is shown in Figure~\ref{fig:unitcolz}. It is clearly seen that the more charge stands on the corners than the central points and this is indeed expected since the charge on a conductor accumulates on the sharp edges.

\subsection{Discharge time versus number of cells}

As a second self-consistency check, the time needed for 90\% of the inital charge to be removed is calculated iteratively with the increasing number of subdivisions. The number of cells is increased until the relative error between the previous discharge time is less than 0.001. Denoting the discharge time with $t_d$ Relative error is defined as follows:

\begin{equation} 
  \text{ Error }= \frac{ | t_{\text{d}} - t_{\text{d, previous iteration}} | }{t_{\text{d, previous iteration}}}
\end{equation}

$ 2 \times 2 \, cm^2 $ surface with $10^5 \Omega / \square$ surface resistivity is first divided by 7 in each direction resulting total 149 cells and initially $10^4$ elementary charge is added in the center similar to the simulation described previously (see Figure 3.2). The surface is grounded from the left end again in the same way as shown in Figure 3.2. Transient simulation is performed until the 90\% of the initial charge is removed and then the number of divisions in each direction is increased by 2 and all other steps are repeated. Related code is in Appendix B.3. This process is iterated until the relative error (as defined with Equation 3.39) between discharge times from two adjacent iterations becomes less than 0.001. Why the number of divisions are increased by 2 in every iteration is that if they were increased by 1, number of divisions were going to be even in every 2 iterations and that is not desired because for even number of divisions there is no “central cell”, thus, it is not possible to put the initial charge at the same coordinates for even and odd cases.

\begin{figure}[h]
 \centering
 \includegraphics[width=11cm, keepaspectratio=true]{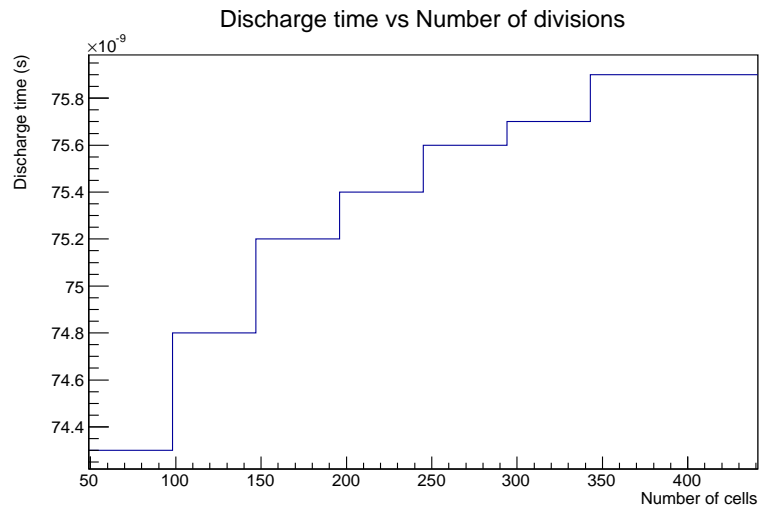}
 \caption{Discharge time versus number of cells. \label{fig:tdvsN}}
\end{figure}

\begin{figure}[h]
 \centering
 \includegraphics[width=11cm, keepaspectratio=true]{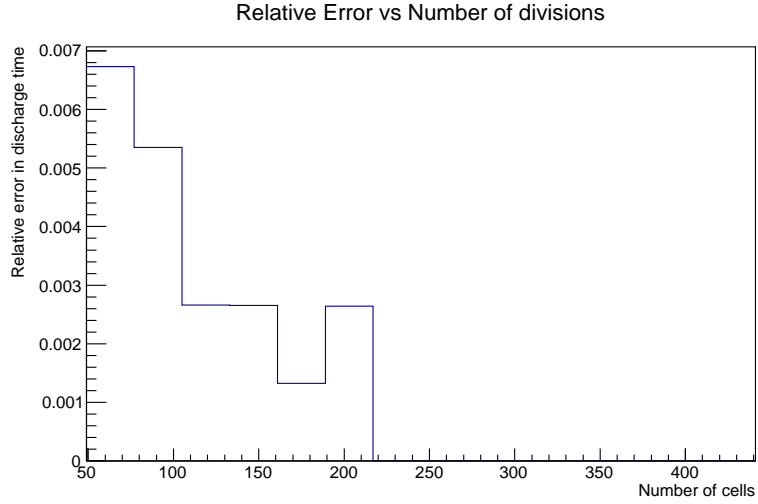}
 \caption{Relative error in discharge time versus number of cells. \label{fig:EtdvsN}}
\end{figure}

The 0.001 tolerance on the relative error is satisfied for 21 divisions in each direction (total 441 cells). Variation of discharge time and the relative error with the increasing number of cells is plotted on Figure~\ref{fig:tdvsN} and Figure~\ref{fig:EtdvsN} respectively. It can be clearly seen that discharge time asymptotically approaches to a value with decreasing amount of relative error as the number of divisions increases. 

\section{An Analytically Solvable Case}

Relaxation of a periodic charge distribution on an infinite conducting plane is described by a simple analytical solution. In the next subsection, this analytical solution is explained; and in the following subsection, results of several simulations performed by Chani, where a conducting square plate had initially a periodic charge distribution, are presented.

\subsection{Infinite conducting plane with periodic charge distribution}

The periodic surface charge distribution on an infinite plane lying on $z = 0$ is described with the following equation:
\begin{equation} \label{eq:sigmap}
  \sigma = \sigma_0 \cos(kx) \exp(-t/\tau)
\end{equation}
Here, periodicity is determined by the cosine function and the exponential is the relaxation term. Boundary condition on $z = 0$ from Gauss' Law reads:
\begin{equation} 
  E_z^+ - E_z^- = \frac{\sigma}{\epsilon_0}
\end{equation}
\begin{equation} \label{eq:BCV}
  \frac{\partial V^-}{\partial z} - \frac{\partial V^+}{\partial z} = \frac{\sigma}{\epsilon_0}
\end{equation}
In these equations, lower index $z$ denotes the $z$-component of the electrical field and upper indices $+$ and $-$ denotes ``just above'' and ``just below'' the $z =0$ surface respectively. Electrical potential satisfying Equation \ref{eq:BCV} for the charge distribution given with \ref{eq:sigmap} is:
\begin{equation} \label{eq:Vp}
  V = \frac{\sigma_0}{2 k \epsilon_0} \cos(kx) \exp(-k|z|) \exp(-t/\tau)
\end{equation}
Inserting this to \ref{eq:BCV} one can see that the boundary condition at z = 0 is indeed satisfied:
\begin{align*}
  k \frac{\sigma_0}{2 k \epsilon_0} \cos(kx) \exp(-t/\tau) - (-k) \frac{\sigma_0}{2 k \epsilon_0} \cos(kx) \exp(-t/\tau) &= \frac{\sigma_0 }{\epsilon_0} \cos(kx) \exp(-t/\tau)\\
           &= \frac{\sigma}{\epsilon_0}
\end{align*}
Continuity equation in two dimensions can be written as,
\begin{equation} \label{eq:continuity}
  \frac{\partial \sigma}{\partial t} + \mathbf{\nabla} \cdot \mathbf{K} = 0
\end{equation}
Here, \textbf{K} denotes the surface current density and $\mathbf{\nabla}$ contains derivatives in x and y. Ohm's law for surface currents is:
\begin{equation} 
  \mathbf{K} = \frac{1}{\rho_s} \mathbf{E}
\end{equation}
Here $\rho_s$ is the surface resistivity. Writing Ohm's law in terms of electrical potential and substituting it into Equation \ref{eq:continuity} one has,
\begin{equation} \label{eq:continuity2}
  \frac{\partial \sigma}{\partial t} - \frac{1}{\rho_s} \mathbf{\nabla}^2 V= 0
\end{equation}
Finally, inserting the surface charge distribution and electrical potential at $z = 0$ given with the Equations \ref{eq:sigmap} and \ref{eq:Vp} respectively into Equation \ref{eq:continuity2} one has,
\begin{equation*} 
  -\frac{1}{\tau} \sigma_0 \cos(kx) \exp(-t/\tau) + k^2 \frac{\sigma_0}{2 k \epsilon_0 \rho_s} \cos(kx) \exp(-t/\tau) = 0
\end{equation*}
which yields,
\begin{equation} \label{eq:tau}
  \tau = \frac{2 \rho_s \epsilon_0}{k}
\end{equation}
This time constant $\tau$ characterizes the relaxation time of the periodic charge distribution. In the next subsection, $\tau$ is extracted from the fits to the simulation results from Chani and compared with the values given with \ref{eq:tau}.

\subsection{Simulations of the charge relaxation}

\begin{figure}[h]
 \centering
 \includegraphics[width=11cm, keepaspectratio=true]{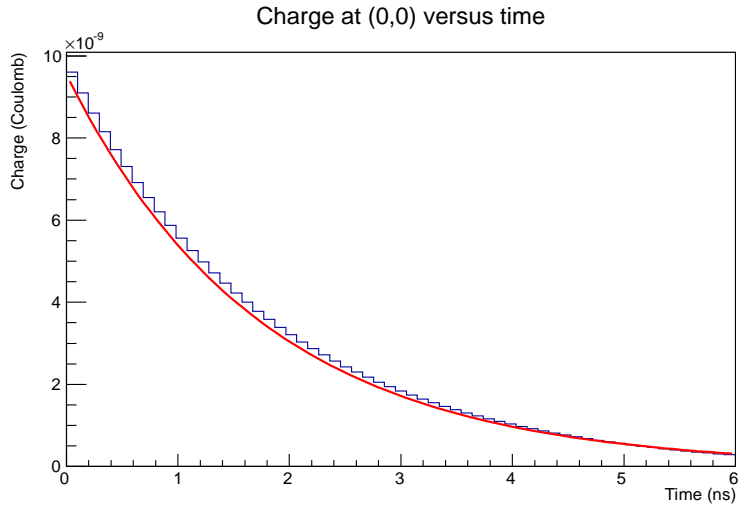}
 \caption{Decay of the total charge in the center of a square plate which initially had a periodic charge distribution. Red curve is the exponential fit. \label{fig:qcenter}}
\end{figure}

In the simulations of this section (see Appendix C for the source files), a square plane with no ground connection, initially has a periodic charge distribution given with \ref{eq:sigmap}. Transient calculation is performed and at each subcell, variation of the amount of charge with time is obtained. Decay of the charge in the middle is fitted with an exponential and the corresponding time constant is extracted. As an example, decay of the charge in the middle of the surface from the simulation of $2\,\times\,2\,\text{ cm}^2$ is plotted in Figure~\ref{fig:qcenter}. Color plots of the surface charge distribution at four time instances from the same simulation is shown in Figure~\ref{fig:periodiccolz}. 

Time constants corresponding to the relaxation of periodically distributed charge are obtained from the simulations performed by Chani for squares with different sidelengths. In these simulations, the wavenumber corresponding to the charge distribution given with \ref{eq:sigmap} is fixed to $10\,cm^{-1}$ and the surface resistivity is $10^5 \Omega / \square$. The time step is 0.01 ns. Results are presented in Table~\ref{tab:varl} where \emph{l} is the sidelength of the square plate, \emph{n} is the number of divisions in both sides (total number of subcells is $n \times n$) and $\tau_\text{Chani}$ is the time constant that is obtained from the exponential fit to the results of the simulation (as shown in Figure~\ref{fig:qcenter}). As it can be seen from Figure~\ref{fig:periodiccolz}, periodic distribution of the charge on the surface is lost with increasing time which is not the case when the plane is infinite. For this reason, exponential is fit to only the first 2 ns of the simulation assuming that during this time interval the behavior should be similar to the infinite plane solution. For $k = 10\,cm^{-1}$ and $\rho_s = 10^5 \Omega / \square$, the time constant for the infinite plane is calculated from \ref{eq:tau} is $\tau_\text{infinite plane} = 1.7708\,ns$. The \%Error is defined as follows:
\begin{equation} \label{eq:errortau}
  \text{ \%Error }= \frac{ | \tau_\text{infinite plane} - \tau_\text{Chani} | }{\tau_\text{infinite plane}} \times 100
\end{equation}

As it can be seen from the Table~\ref{tab:varl}, error between the time constants from the theory and Chani is less than 3.5\% for 5 different lengths. One might expect to see a monotonic decrease of the error as the size increases assuming that the behavior would become much similar to that of the infinite plane. However, this is not the case with these results indicating that there must be some other effect related with the simulation conditions.

\begin{table}[h]
{\setlength{\tabcolsep}{14pt}
\caption{Relaxation time constants for different sidelengths.}
\begin{center}
\vspace{-6mm}
\begin{tabular}{cccc}
\hline\hline
l (cm) & n  & $\tau_\text{Chani}$ (ns) & \%Error \\
\hline
1 & 25 & 1.7363 & 2.0 \\
2 & 49 & 1.8146 & 2.5 \\
3 & 75 & 1.8320 & 3.5 \\
4 & 99 & 1.8300 & 3.3 \\
5 & 125 & 1.8224 & 2.9 \\
\hline
\end{tabular}
\vspace{-6mm}
\end{center}
\label{tab:varl}}
\end{table}

The same simulation procedure is followed, this time, keeping the sidelength constant and equal to 5 cm while varying the number of divisions (\emph{n}). Results obtained from the fit and errors are listed in Table~\ref{tab:varn}. The wave number, k, is again fixed to 1000 and the time step is 0.01 ns.  

\begin{table}[h]
{\setlength{\tabcolsep}{14pt}
\caption{Relaxation time constants of the square plate with 5 cm sidelength with different number of divisions.}
\begin{center}
\vspace{-6mm}
\begin{tabular}{ccc}
\hline\hline
 n  & $\tau_\text{Chani}$ (ns) & \%Error \\
\hline
9 & 14.7735 & 734 \\
13 & 4.6277 & 161 \\
17 & 3.4186 & 93 \\
25 & 2.5793 & 46 \\
33 & 2.2641 & 28 \\
41 & 2.1092 & 19 \\
49 & 2.0205 & 14 \\
57 & 1.9643 & 11 \\
65 & 1.9262 & 9 \\

\hline
\end{tabular}
\vspace{-6mm}
\end{center}
\label{tab:varn}}
\end{table}

\begin{landscape}
\thispagestyle{empty}
 \begin{figure}
 \centering
 \includegraphics[width=700pt,keepaspectratio=true]{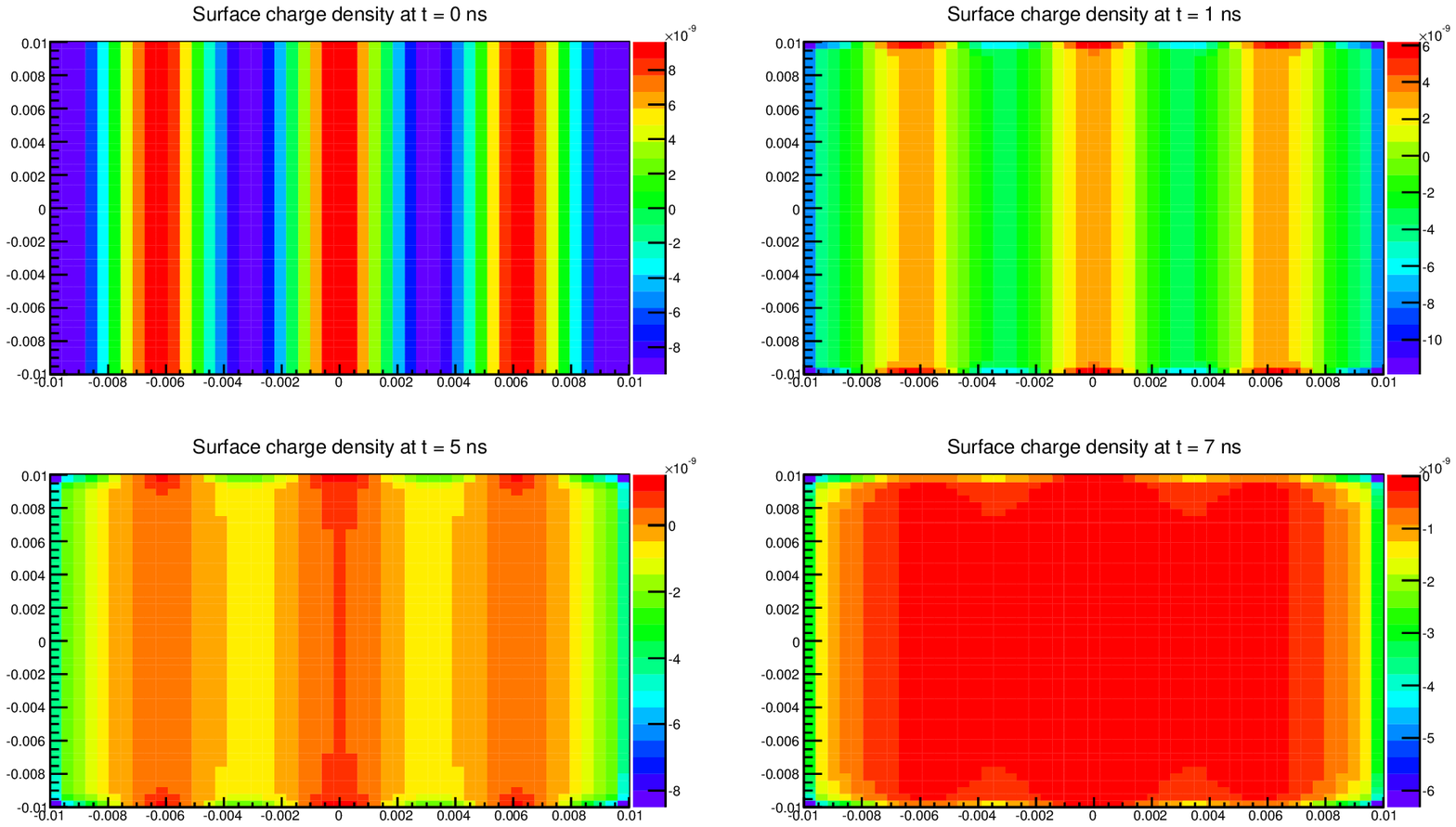}
 \caption{Color plot of the surface charge density at four different times. \label{fig:periodiccolz}}    
      \vspace{0.5cm}
      \hspace{0cm}\pageref{fig:periodiccolz}

\end{figure}
\end{landscape}

Results in Table~\ref{tab:varn} shows that the error with the theory decreases with increasing number of divisions as expected. In the first two rows of the Table~\ref{tab:varn}, the error is above 100\%. In fact, these two calculations do not make sense at all because they do not satisfy the ``Nyquist rate'' \cite{Oppenheim}. Initial charge of the surface was periodic with the wave number $k = 1000 m^{-1}$ which corresponds to the spatial frequency:
\begin{equation} \label{eq:f}
  f = k / 2\pi = 159.15 m^{-1}
\end{equation}
In essence, Nyquist - Shannon sampling theorem tells that a signal with a bandwith B, should be sampled with a sampling frequency greater than 2B in order to be able to reconstruct the original signal without aliasing. For the simulations of this section, charge distribution function \ref{eq:sigmap} and the subcells of Chani can be thought as the original signal and the sampling points respectively. Sampling frequency can be expressed as:
\begin{equation} \label{eq:fs}
  f_s = (l/n)^{-1} = n/l
\end{equation}
Where l is sidelength and n is number of divisions. Nyquist criteria reads,
\begin{align}
   f_s &> 2f \\
   n &> 2 \times 159.15\,m^{-1} \times 0.05\,m = 15.9
\end{align}
This simple calculation tells that if the number of divisions is chosen less than 16, initial charge distribution is not properly understood by the simulator hence result it gives is unreasonable. 

In the Table~\ref{tab:varn}, 93\% error is given for $n = 17$ which is just above the Nyquist limit indicating that a higher sampling frequency is required. Error with the theory falls below 10\% when $n = 65$ for which the sampling frequency is more than 4 times the Nyquist frequency.

\section{Charge Transport on a Resistive Strip}

Up to this point, presented calculations were on a square shaped surface. However, main motivation of this project was to develop a tool for the optimization of the resistive structures in the resistive-anode micromegas detectors being developed by MAMMA group and these resistive strips are rectangular. For example, prototypes presented in \cite{MAMMA1} have $10\,cm$ by $150\,\mu m$ resistive strips with surface resistivities $ 0.030, 0.075, \text{and} 0.0075 \Omega / \square $. As an example, a transient simulation of charge transport and discharge is performed with the surface parameters from \cite{MAMMA1}.

Proper choice of cell dimensions becomes essentially for calculations on rectangular surfaces with big length-to-width ratios. Several trials showed that divisions must be made such that the unit cells of the simulations have a square-like shape. Unreasonable results for charge distribution is obtained when rectangular unit cells with high length-to-width ratios are chosen.

\begin{figure}[h]
 \centering
 \includegraphics[width=14cm,keepaspectratio=true, trim = 0mm -5mm 0mm 0mm, clip]{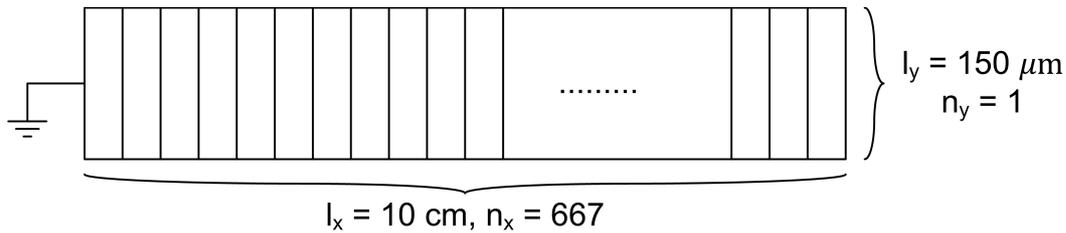}
 \caption{ Illustration of resistive strips with divisions. \label{fig:R11}}
\end{figure}

$10\,cm$ by $150\,\mu m$ rectangular surface, illustrated on Figure~\ref{fig:R11}, is divided by 667 in x and not divided in y. Initially $10^4$ elementary charge is put in the middle and the surface is grounded from the left end as it is shown in Figure \ref{fig:R11}. Surface resistivity is set to $ 0.03\, \Omega / \square $ and transient calculation is performed for 2,000,000 steps with 0.1 ns time steps. See Appendix D for source files.

\begin{figure}[h]
 \centering
 \includegraphics[width=11cm, keepaspectratio=true]{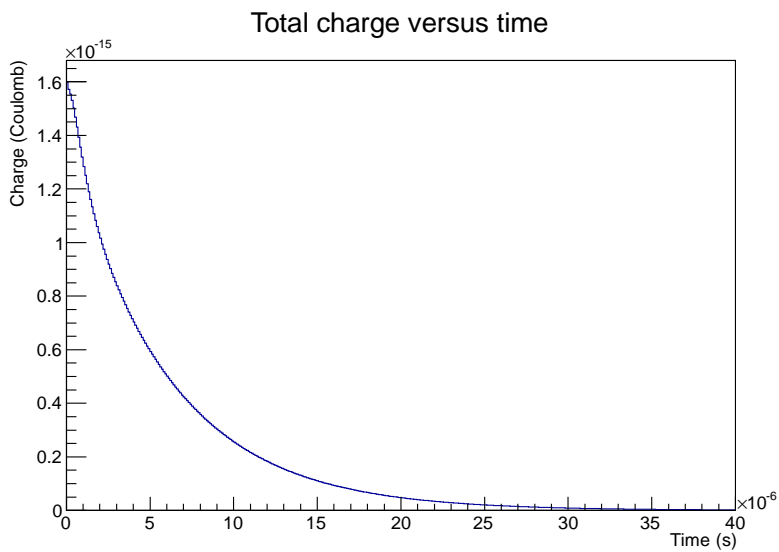}
 \caption{ Discharge plot for the rectangular surface. \label{fig:R11Qtot} }
\end{figure}

Change in the amount of total charge with time is shown on Figure~\ref{fig:R11Qtot}. As it is shown in the figure, the time needed for the removal of most of the initial charge is $30\,\mu s$. The frequency corresponding to this period is $1 / 30\,\mu s \approx 33\,kHz$. Thus one should expect a charge up on the resistive strips when the detector is put under an incoming particle rate greater than this frequency.

Color plots of the surface charge density at four simulation instances are given in Figure~\ref{fig:R11colz}. Spread of the initial charge can be seen on the plot on top right.

\section{A Large Scale Calculation}

Last simulation that is going to be presented in this section is again a resistive structure from one of the prototypes (R14) build by MAMMA group. Similar to the previous simulations, spread of the initial charge in the center is studied. This surface is a 100 $\times$ 18 mm$^2$ rectangular surface with 0.24 M$\Omega / \square$ surface resistivity. The different thing about this simulation is it has a large number of cells. It is divided by 51 in one direction and 279 in other resulting 14229 cells.

Calculation with 14229 cell corresponds to a 14229 by 14229 l-matrix and this is a huge matrix dimension to hold in the memory. Chani does the calculations with big l-matrices by dividing the matrix to submatrices. After that, calculation of the potentials are done via sub-vectors, written in the harddrive and read for the current calculations, this is done by the \emph{transient1()} function which works slower than \emph{transient2()} but has higher memory capabilities.  Number of submatrices is a user input parameter and should be carefully choosen.

Transient simulation on the 100 $\times$ 18 mm$^2$ surface is performed for 600 steps with time step 0.1 ns. Simulation files are given in Appendix E. Color plot of the surface charge density is given on Figure~\ref{fig:R14colz} where charge spread can be seen.

\begin{landscape}
\thispagestyle{empty}
 \begin{figure}
 \centering
 \includegraphics[width=700pt,keepaspectratio=true]{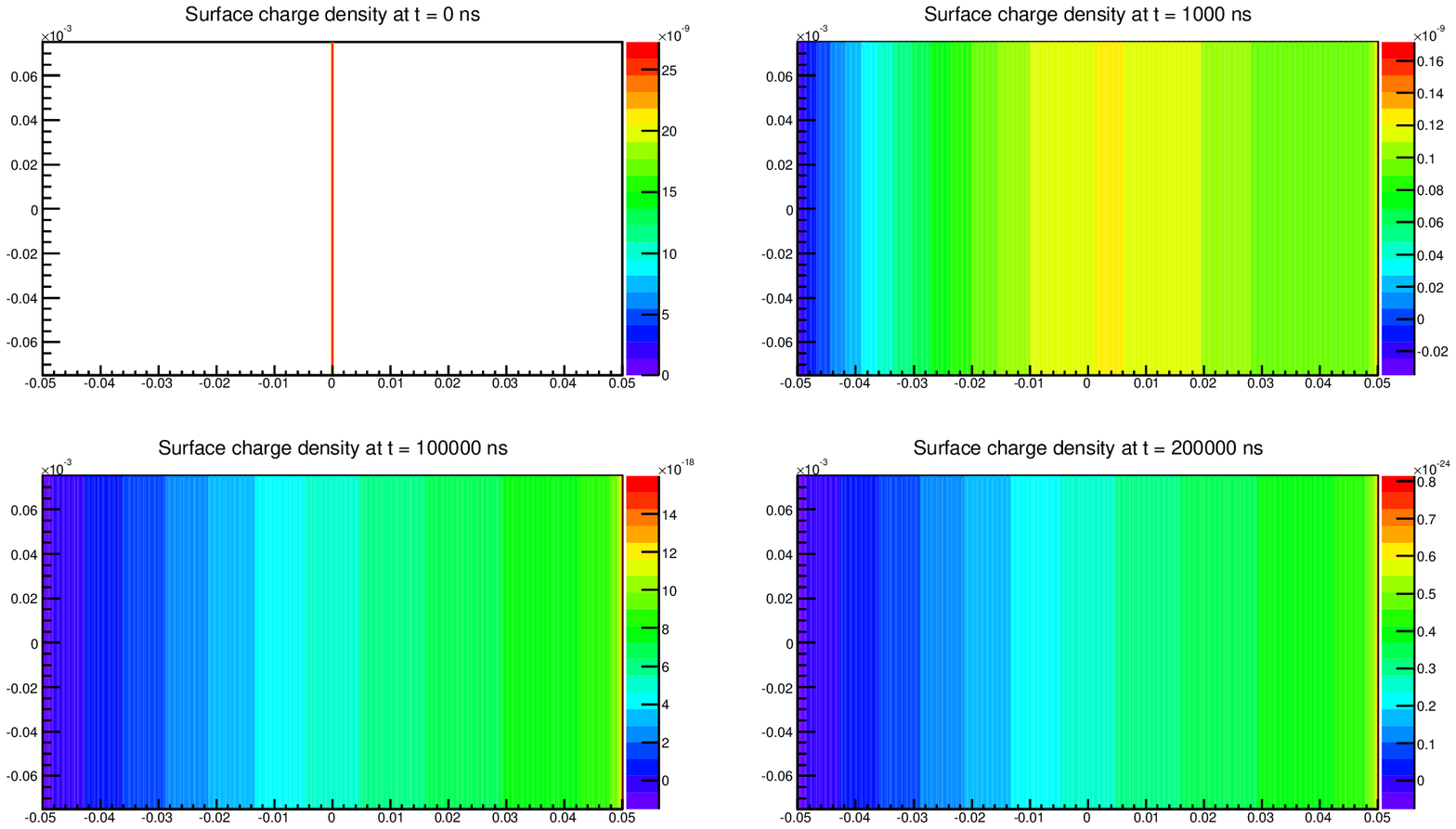}
 \caption{Color plot of the surface charge density at four different times. \label{fig:R11colz}}    
      \vspace{0.5cm}
      \hspace{0cm}\pageref{fig:R11colz}

\end{figure}
\end{landscape}

\begin{landscape}
\thispagestyle{empty}
 \begin{figure}
 \centering
 \includegraphics[width=700pt,keepaspectratio=true]{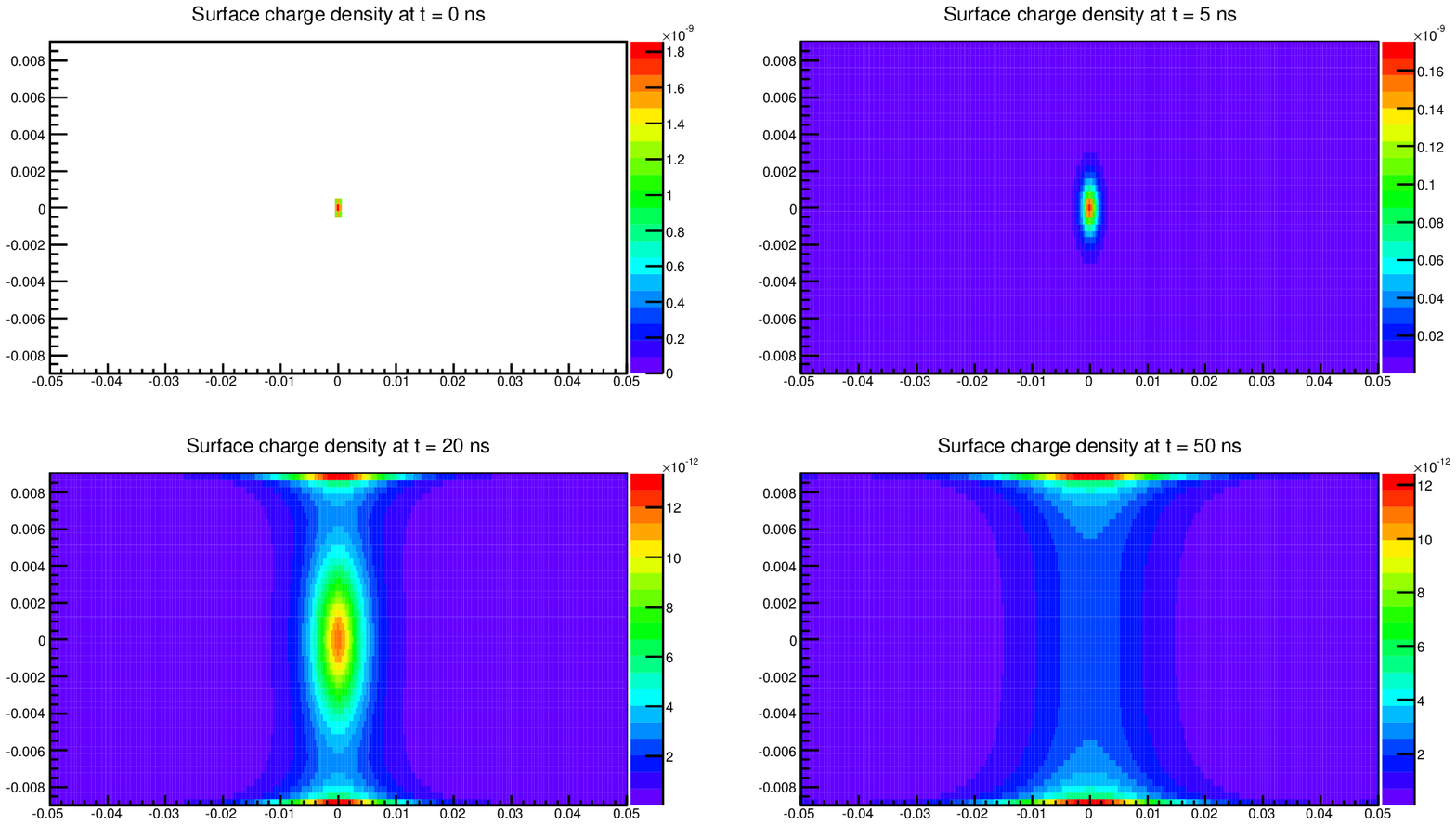}
 \caption{Color plot of the surface charge density at four different times. \label{fig:R14colz} }    
      \vspace{0.5cm}
      \hspace{0cm}\pageref{fig:R14colz}
\end{figure}
\end{landscape}

\chapter{CONCLUSIONS}\label{ch:ifnecch4}

Main goal of the study that is presented in this thesis was to understand the charge spread and discharge dynamics of the electrical charge on a rectangular surface of high resistivity in order to help the design of the micromegas detectors with resistive anodes. For this purpose, a simulator is developed and applied on several cases with different dimensions and resistivities. 

It is shown via self-consistency tests in sections 3.4.1 and 3.4.2 that the simulations are convergent if the subdivision parameters are properly chosen. 

Relaxation of periodic charge on square shaped surfaces is studied and the simulation results are compared with the theoretical expectations in section 3.5. With properly chosen parameters, error between values of the relaxation time constants from theory and Chani were below 3.5\%. 

The calculation on a resistive strip which is presented in the Section 3.6 demonstrates that the starting motivation, calculating the charge transport on a resistive strip in micromegas chambers being developed by MAMMA group is indeed achieved. This calculation also provides the fact that the simulator works for very long time steps (2,000,000) without any memory overflow or instability. 

In Section 3.7, a simulation with 14229 cells is presented, proving that the simulator can handle very large matrix operations.

In conclusion, a working simulator for the charge transport on rectangular surfaces is developed. It is possible to calculate the time needed for the total discharge and the spread of the charge over to surface using Chani. Although it is developed to help the design of the gaseous ionization detectors, Chani’s features can find applications in many fields where the understanding of charge transport properties is important. Specifically for particle detectors with resistive anodes, Chani can be used to guess the rate capabilities of the detectors from the discharge times and possible cross-talks between the readout strips due to the charge spread.


\bibliographystyle{itubib}        
\bibliography{tez}

\eklerkapak{}
\vglue20pt
\singlespacing
\textbf{APPENDIX A:} Chani.h header file, main function and classes\\
\textbf{APPENDIX B:} Definition and simulation files of Sections 3.3 \& 3.4\\
\textbf{APPENDIX B.1:} Simulation files of Section 3.3\\
\textbf{APPENDIX B.2:} Simulation files of Section 3.4.1\\
\textbf{APPENDIX B.3:} Simulation files of Section 3.4.2\\
\textbf{APPENDIX C:} Definition and simulation files of Section 3.5\\
\textbf{APPENDIX D:} Definition and simulation files of Section 3.6\\
\textbf{APPENDIX E:} Definition and simulation files of Section 3.7\\
\newpage  

\eklerbolum{5}
\chapter{APPENDIX A \textmd{Chani.h header file, main function and classes}}

\lstset{language=C++, basicstyle=\footnotesize,	breaklines=true, showstringspaces=false, tabsize=1, stepnumber=1, breakatwhitespace=false}

\begin{lstlisting}
#ifndef __CHANIH__
#define __CHANIH__

/*Header file of Chani (Charge transport simulator for rectangular s
urfaces).
 * surfaces
 * 	Created: Apr 25, 2012 by Burak Budanur (burakbudanur@gmail.c
om)
 * 	Last edited: May 21, 2012 by Burak Budanur (burakbudanur@gma
il.com)
*/

#include <iostream>
//Root libs:
#include "TMath.h"
#include "TMatrixD.h"
#include "TVectorD.h"
#include <cmath>
#include "TFile.h"
#include "TH1.h"
#include "TH2.h"
//Chani functions:
#include "functions/parameters.C"
#include "functions/computelmn.C"
#include "functions/addConnector.C"
#include "functions/addCharge.C"
#include "functions/transient1.C"
#include "functions/transient2.C"
#include "functions/getV.C"
#include "functions/getQtotal.C"
#include "functions/capacitance.C"
#include "functions/getQxy.C"
//Chani classes:
#include "classes/Readp.C"
#include "classes/Findij.C"
//Global constants
#include "globalConsts.h"

using std::cout;
using std::endl;

void parameters(double lx, int nx, double ly, int ny, int Nm, double
 sigmas);

void computelmn();

void addConnector(double x, double y, double R);

void addCharge(double x, double y, double q, int tStep);

void transient1(int tStepi, int nofSteps, double Deltat);

void transient2(int tStepi, int nofSteps, double Deltat);

double capacitance();

double getV(int i);

double getQtotal(int tStep);

double getQxy(double x, double y, int tStep);

#endif // __CHANIH__
\end{lstlisting}
\begin{lstlisting}
#ifndef __GLOBALCONSTSH__
#define __GLOBALCONSTSH__

#include "TMath.h"
const double pi = TMath::Pi();
const double epsilon0 = 8.8541878176e-12; // F / m
const char * dir = "/$ROOTSYS/burak/Chani0v3sine/output/"; /
/ DIRECTORY TO HOLD OUTPUT FILES. EDIT THIS LINE ACCORDING T
O YOUR WORKING DIRECTORY.

#endif
\end{lstlisting}
\begin{lstlisting}
#include "../Chani.h"
#include "../globalConsts.h"

void parameters(double lx, int nx, double ly, int ny, int Nm, double
 sigmas) {

 
 TString fnameg = TString::Format("%sparameters.root", dir);
 
 TString fnamev = TString::Format("%sxy.root", dir);
 
 TString fnameh = TString::Format("%sh.root", dir);
 
 double ax = lx/nx; // unit length in x
 double ay = ly/ny; // unit length in y
  
 int N = nx*ny;
 std::cout << "The total number of subsections is N = " << N << std:
:endl;

 std::cout << "The l-matrix is going to be calculated in "; 
 std::cout << Nm*Nm << " parts." << std::endl;


 // Assigning the x & y coordinates to the subsections

 TVectorD x(1,N);
 TVectorD y(1,N);

 if (nx%2 == 0 && ny%2 == 0){
   for (int j = 1; j < ny+1; j++){
     for (int i=1; i < nx+1; i++){
       
       x(i+nx*(j-1)) = -ax*(nx/2) + ax/2 + (i-1)*ax;
       y(i+nx*(j-1)) = -ay*(ny/2) + ay/2 + (j-1)*ay;
       
     }
   }
 }

 else if (nx%2 == 1 && ny%2 == 1){
   for (int j = 1; j < ny+1; j++){
     for (int i=1; i < nx+1; i++){

       x(i+nx*(j-1)) = -ax*((nx-1)/2) + (i-1)*ax;
       y(i+nx*(j-1)) = -ay*((ny-1)/2) + (j-1)*ay;

     }
   }
 }

 else {
   std::cout << "number of subsections for x and y should be both ev
en or both odd." << std::endl;
   return;
 }

 TH2 *sigmah0;
 sigmah0 = new TH2D("sigma0(x,y)", "Surface charge density at t = 0"
, nx, -lx/2, lx/2, ny, -ly/2, ly/2) ;
 
 TH2 *connh;
 connh = new TH2D("conn(x,y)", "Connectors histogram", nx, -lx/2, lx
/2, ny, -ly/2, ly/2) ;
 
 for (int j = 1; j < ny+1; j++){
  for (int i=1; i < nx+1; i++){
  
   connh->SetCellContent(i, j, 1e20); // Assign a nonzero dummy valu
e to the histogram contents
   
  }
 }


 TFile f(fnameg,"recreate"); // open the output file. 

 TH1 *params; // The histogram to hold lx, nx, ax, ly, ny, ay, Nm va
lues in
 params = new TH1D("params","lx, nx, ax, ly, ny, ay, Nm",8,1,8);
 
 params->SetBinContent(1, lx);
 params->SetBinContent(2, nx);
 params->SetBinContent(3, ax);
 params->SetBinContent(4, ly);
 params->SetBinContent(5, ny);
 params->SetBinContent(6, ay);
 params->SetBinContent(7, Nm);
 params->SetBinContent(8, sigmas);

 params->Write(); // Write params histogram in file

 delete params;

 f.Close(); // Close the output file

 TFile f2(fnamev,"recreate"); // open the output file for x & y. 

 x.Write("x-vector"); // Write x coordinates in the file
 y.Write("y-vector"); // Write y coordinates in the file
 
 f2.Close();
 
 TFile f3(fnameh,"recreate"); // open the output file for histograms

 sigmah0->Write("sigmah0"); // Blank histogram of the surface
 connh->Write("connh"); // Connector histogram of the surface
 
 f3.Close();

 delete connh;
 delete sigmah0;

}
\end{lstlisting}
\begin{lstlisting}
#include "../Chani.h"
#include "../globalConsts.h"
#include "../classes/Readp.h"

void computelmn(){

 Readp rp;
 double lx = rp.Getlx();
 double ly = rp.Getly();
 double ax = rp.Getax();
 double ay = rp.Getay();
 int nx = rp.Getnx();
 int ny = rp.Getny();
 int Nm = rp.GetNm();

 int N = nx*ny;

 TString fnamev = TString::Format("%sxy.root", dir);

 // Reading of x & y vectors: 
 TFile f(fnamev); 
  
 TVectorD *xp; // Pointer for x-vector
 f.GetObject("x-vector",xp);
 TVectorD x(1,N);
 x = *xp;
 delete xp;

 TVectorD *yp; // Pointer for y-vector
 f.GetObject("y-vector",yp);
 TVectorD y(1,N);
 y = *yp;
 delete yp;
 
 f.Close();
  
 double DeltaSn = ax*ay; //Area of a subsection

 std::cout << "Area of a subsection DeltaSn (m^2) = " << DeltaSn << 
std::endl;

 std::cout << "Calculation of the l(m,n) matrix ..." << std::endl;

 int nsub = N / Nm;
 TMatrixD lsub(1,nsub,1,nsub);

 for (int i = 1; i <= Nm; i++){
  for (int j = 1; j<= Nm; j++){
  
   TString fname = TString::Format("%slmatrix%i.root",dir, i+(j-1)*N
m);
  
   TFile f(fname,"recreate"); // open the output file. 

   for (int m = (j-1)*nsub+1; m <= j*nsub; m++){
    for (int n = (i-1)*nsub+1; n<= i*nsub; n++){
    
     int mm = m - (j-1)*nsub;
     int nm = n - (i-1)*nsub;

     if (n == m){
      lsub(mm,nm) = (1/(4*pi*epsilon0))
         *(ax*log((sqrt(pow(ay,2) + pow(ax,2)) + ay)/(sqrt(pow(ay,2)
 + pow(ax,2)) - ay)) 
         + ay*log((sqrt(pow(ax,2) + pow(ay,2)) + ax)/(sqrt(pow(ax,2)
 + pow(ay,2)) - ax)));
        }  
     else{
      lsub(mm,nm) = DeltaSn/(4*pi*epsilon0*sqrt(pow((x(m)-x(n)),2)+p
ow((y(m)-y(n)),2))) ;
      }  
     
     //std::cout << "l(" << m << ", " << n << ") = " << lsub(mm,nm) 
<< " ";
    }
   }

   char label[12];
   sprintf(label, "l-matrix%i", i+(j-1)*Nm);
   lsub.Write(label);
   f.Close(); // Close the output file
 }
}
 
}
\end{lstlisting}
\begin{lstlisting}
#include "../Chani.h"
#include "../globalConsts.h"
#include "../classes/Findij.h"

void addConnector(double x, double y, double R){
 
 //Names of the files to be read

 TString fnameh = TString::Format("%sh.root", dir);
 
 Findij fij(x,y);
 int i = fij.Geti();
 int j = fij.Getj(); 
 
 
 TFile f(fnameh, "UPDATE"); // open the output file for histograms

 TH2 *connh; // The histogram to hold lx, nx, ax, ly, ny, ay values 
in
 
 f.GetObject("connh",connh);
 
 connh->SetCellContent(i, j, R); // Write the resistance value in th
e connector cell
 connh->Write("connh"); // Connector histogram of the surface
  
 delete connh; 
  
 f.Close();

}
\end{lstlisting}
\begin{lstlisting}
#include "../Chani.h"
#include "../globalConsts.h"
#include "../classes/Findij.h" 

void addCharge(double x, double y, double q, int tStep){
 
 //File name:
 TString fnameh = TString::Format("%sh.root", dir);

 Findij fij(x,y);
 int i = fij.Geti();
 int j = fij.Getj();
 int nx = fij.Getnx();
 int ny = fij.Getny();
 double ax = fij.Getax();
 double ay = fij.Getay();
 double sigmaadd = q/(ax*ay); // additional surface charge density
  
 TFile f(fnameh, "UPDATE"); // open the output file for histograms
 //f.GetListOfKeys()->Print(); // Print to see if the objects are re
trieved properly
 
 TH2 *sigmah; // The temporary surface charge density histogram
 
 char objName[16];
 sprintf(objName, "sigmah%i", tStep); 
 
 f.GetObject(objName, sigmah);
 
 sigmah->SetCellContent(i, j, sigmah->GetCellContent(i,j)+sigmaadd);
 
 sigmah->Write(objName); // Write the histogram back
 
 f.Close();
 
}
\end{lstlisting}
\begin{lstlisting}
#include "../Chani.h"
#include "../classes/Readp.h"
#include "getV.C"

void transient1(int tStepi, int nofSteps, double Deltat){
 
 Readp rp;
 double lx = rp.Getlx();
 double ly = rp.Getly();
 double ax = rp.Getax();
 double ay = rp.Getay();
 int nx = rp.Getnx();
 int ny = rp.Getny();
 int Nm = rp.GetNm();
 double sigmas = rp.Getsigmas(); 
 
 int N = nx*ny; // number of sub-cells
 int nsub = N / Nm; // dimension of the sub matrices
 int Nl = Nm * Nm; // number of l-matrices
 
 double DeltaSn = ax*ay; // unit area
 
 //Create and write Vi & sigmai vectors:
 
 for (int i = 1; i <= Nm; i++){
  
  TVectorD V(1,nsub);
  TVectorD sigma(1,nsub);

  TString fname = TString::Format("%sVsigma%i", dir, i);
    
  TFile f(fname, "recreate"); // open the output file
  
  char labelV[5];
  char labelsigma[10];
  
  sprintf(labelV, "V%i", i);
  V.Write(labelV);
  
  sprintf(labelsigma, "sigma%i", i);  
  sigma.Write(labelsigma);
  
  f.Close();
 }
 
 //Read the initial surface charge density and connector histogram: 
 
 TH2 *sigmaht; // Surface charge density histogram
 sigmaht = new TH2D("SigmahtStep(x,y)", "Surface charge density" , n
x, -lx/2, lx/2, ny, -ly/2, ly/2) ;
 
 TH2 *connh; //The connector histogram:
 connh = new TH2D("Connectors", "Connector histogram" , nx, -lx/2, l
x/2, ny, -ly/2, ly/2) ;
 
 TString fnameh = TString::Format("%sh.root", dir);

 TFile f(fnameh); // open the output file for histograms
 
 char objName[16];
 sprintf(objName, "sigmah%i", tStepi); 
 
 TH2 *sigmahtd;
 TH2 *connhd;
   
 f.GetObject(objName, sigmahtd); // Read the initial surface charge 
density histogram
 f.GetObject("connh", connhd); // Read the connector histogram
 
 for (int j = 1; j<=ny; j++){ 
  for (int i = 1; i<=nx; i++){ 
   sigmaht->SetCellContent(i, j, sigmahtd->GetCellContent(i,j));
  }
 }

 for (int j = 1; j<=ny; j++){ 
  for (int i = 1; i<=nx; i++){ 
   connh->SetCellContent(i, j, connhd->GetCellContent(i,j));
  }
 }
 
 delete sigmahtd;
 delete connhd;
 
 f.Close();
 
 //Write the initial surface charge density histogram to the sigma v
ectors:
 
 int dummy;
 
 for (int isigma = 1; isigma <= Nm; isigma++){
 
 TString fname = TString::Format("%sVsigma%i", dir, isigma);
 
 TFile f2(fname, "UPDATE"); // open the output file
 //f.GetListOfKeys()->Print();
 
 char labelsigma[10];
 sprintf(labelsigma, "sigma%i", isigma);   
 
 TVectorD * sigmav;
 f2.GetObject(labelsigma, sigmav);
  
  for (int j = 1; j <= ny; j++){
   for (int i = 1; i <= nx; i++){
    
    dummy = ((j-1)*nx+i - 1)/nsub + 1;
    
    if (dummy == isigma){
      //std::cout << "cell (" << i << ", " << j << ") belongs to the
 vector "<< isigma << std::endl;
      (*sigmav)((j-1)*nx+i - (isigma-1)*nsub) = sigmaht->GetCellCont
ent(i, j);
     
    }
    
   } 
  }  
  
 sigmav->Write(labelsigma); 
 delete sigmav;
 f2.Close(); 
  
 }
 
 double Kt, It, Rcon; //Surface current density, Discharge current
 TVectorD Vi(1,nsub); // i-th V vector
 TVectorD sigmavj(1,nsub); // j-th sigma vector
 TMatrixD lj(1,nsub,1,nsub); // [(i-1)*Nm+j]-th l-matrix
    
 //Transient loop:
 
 for (int it = tStepi + 1; it <= tStepi + nofSteps; it++){
  
  std::cout << "Time step " << it << std::endl;
 
  //Calculation of the new potential:
 
  for (int i = 1; i<=Nm; i++){
   
   Vi.Zero();
   
   for (int j = 1; j<=Nm; j++){
 
    //std::cout << "i, j = " << i << ", " << j << std::endl;
    
    TString fname = TString::Format("%sVsigma%i", dir, j);
     
    TFile f1(fname); // open the output file of vectors
    
    TVectorD *sigmavp; // pointer for the sigma-vector   

    char labelsigma[10];
    sprintf(labelsigma, "sigma%i", j);  
    f1.GetObject(labelsigma, sigmavp);
    sigmavj = *sigmavp;
   
    delete sigmavp;
   
    f1.Close();
    
    //sigmavp->~sigmavp();    
    //delete sigmavp;

    TString fnamel = TString::Format("%slmatrix%i.root",dir, (i-1)*N
m+j);

    TMatrixD *ljp; // pointer for the (i-1)Nm+jth l-matrix     
     
    TFile f(fnamel); // open the output file of l(i-1)Nm+j

    //f.GetListOfKeys()->Print(); // 
     
     char label[10];
     sprintf(label, "l-matrix%i", (i-1)*Nm+j);     
     f.GetObject(label, ljp);
     //*/
    f.Close();
    
    lj = *ljp;
    delete ljp;
    
    Vi = Vi + lj * sigmavj;

   }

   //std::cout << " Vi(38) = " << Vi(38) << std::endl; 

   TString fname = TString::Format("%sVsigma%i", dir, i);

   TFile f(fname, "UPDATE"); // open the output file
  
   char labelV[5];
   sprintf(labelV, "V%i", i);
   
   Vi.Write(labelV); 
   
   f.Close();

  }
  
  //Update the surface charge density histogram due to the currents:
  // Handling the currents in x:

  for (int j = 1; j<=ny; j++){
   for (int i = 1; i<=nx-1; i++){ 
  
    Kt = ((-1)*(getV(i+1+nx*(j-1)) - getV(i+nx*(j-1)))*sigmas)/ax;
    sigmaht->SetCellContent(i, j, sigmaht->GetCellContent(i,j) - (Kt
*Deltat*ay)/DeltaSn);   
    sigmaht->SetCellContent(i+1, j, sigmaht->GetCellContent(i+1,j) +
 (Kt*Deltat*ay)/DeltaSn); 
  
   }
  }

  // Handling the currents in y:
  for (int j = 1; j<=ny-1; j++){
   for (int i = 1; i<=nx; i++){ 
  
    Kt = ((-1)*(getV(i+nx*j) - getV(i+nx*(j-1)))*sigmas)/ay;
    sigmaht->SetCellContent(i, j, sigmaht->GetCellContent(i,j) - (Kt
*Deltat*ax)/DeltaSn);   
    sigmaht->SetCellContent(i, j+1, sigmaht->GetCellContent(i,j+1) +
 (Kt*Deltat*ax)/DeltaSn); 
  
   }
  }
  //*/

  //Discharge currents:
  for (int j = 1; j<=ny; j++){
   for (int i = 1; i<=nx; i++){ 
   
    if(connh->GetCellContent(i,j) < 1e20) {
     
     Rcon = (1/(2*sigmas)) + connh->GetCellContent(i,j);
     It = ((-1)*(getV(1+nx*(j-1)) - 0)) / Rcon;
     
     sigmaht->SetCellContent(i, j, sigmaht->GetCellContent(i,j) + (I
t*Deltat)/DeltaSn);  
     
     //std::cout << "It = " << It << std::endl;
     
    }
  
   }
  }

  //Update sigmavt: 
  for (int isigma = 1; isigma <= Nm; isigma++){

  TString fname = TString::Format("%sVsigma%i", dir, isigma);
 
  TFile f(fname, "UPDATE"); // open the output file
 
  char labelsigma[10];
  sprintf(labelsigma, "sigma%i", isigma);   

  TVectorD * sigmav;
  f.GetObject(labelsigma, sigmav);

   for (int j = 1; j <= ny; j++){
    for (int i = 1; i <= nx; i++){
    
      dummy = ((j-1)*nx+i - 1)/nsub + 1;
    
      if (dummy == isigma){
     
        (*sigmav)((j-1)*nx+i - (isigma-1)*nsub) = sigmaht->GetCellCo
ntent(i, j);
     
      }
    } 
   }  

  sigmav->Write(labelsigma); 
  f.Close(); 
  
  }
  
 }

 TFile f3(fnameh, "UPDATE"); // open the output file for histograms
 
 char sigmahfName[16];
 sprintf(sigmahfName, "sigmah%i", tStepi+nofSteps); 
 TH2 *sigmahf; // Final surface charge density histogram

 sigmahf = new TH2D(Form("Sigmah%i(x,y)",tStepi+nofSteps), "Surface 
charge density", nx, -lx/2, lx/2, ny, -ly/2, ly/2) ;

 //Fill the final surface charge density histogram: 
 for (int j = 1; j<=ny; j++){ 
   for (int i = 1; i<=nx; i++){ 
    sigmahf->SetCellContent(i, j, sigmaht->GetCellContent(i,j));
   }
  }

 sigmahf->Write(sigmahfName); // Write the histogram after transient
 calculation

 delete sigmahf;
 
 f3.Close(); 
 
 delete sigmaht;
 delete connh;
  
}
\end{lstlisting}
\begin{lstlisting}
#include "../Chani.h"
#include "../classes/Readp.h"

void transient2(int tStepi, int nofSteps, double Deltat){
 
 Readp rp;
 double lx = rp.Getlx();
 double ly = rp.Getly();
 double ax = rp.Getax();
 double ay = rp.Getay();
 int nx = rp.Getnx();
 int ny = rp.Getny();
 int Nm = rp.GetNm();
 double sigmas = rp.Getsigmas();

 int N = nx*ny;
 double DeltaSn = ax*ay;
 
 //std::cout << "Reading of the l-matrix" << std::endl;

 // Read the l-matrix:
 TMatrixD l(1,N,1,N);
 int nsub = N / Nm;

 for (int i = 1; i <= Nm; i++){
  for (int j = 1; j<= Nm; j++){
   
   TString fnamel = TString::Format("%slmatrix%i.root",dir, i+(j-1)*
Nm);

   TFile f(fnamel); // open the input file. 

   char label[10];
   sprintf(label, "l-matrix%i", i+(j-1)*Nm);
   
   TMatrixD *lp; // pointer of the sub matrix
   f.GetObject(label,lp);
   
   for (int m = (j-1)*nsub+1; m <= j*nsub; m++){
    for (int n = (i-1)*nsub+1; n<= i*nsub; n++){
    
     int mm = m - (j-1)*nsub;
     int nm = n - (i-1)*nsub;

     l(m,n) = (*lp)(mm,nm);
     //std::cout << "l(" << m << "," << n << ") = " << l(m,n) << " "
;

    }
   }
   
   delete lp;
   f.Close();
  }
 }
 
 //std::cout << "Reading of the l-matrix is completed" << std::endl;
 
 
 //Read the initial surface charge density histogram:
 
 TH2 *sigmaht; // Surface charge density histogram
 sigmaht = new TH2D("SigmahtStep(x,y)", "Surface charge density" , n
x, -lx/2, lx/2, ny, -ly/2, ly/2) ;
 
 TH2 *connh; //The connector histogram:
 connh = new TH2D("Connectors", "Connector histogram" , nx, -lx/2, l
x/2, ny, -ly/2, ly/2) ;
  
 char fnameh[100];
 sprintf(fnameh, "%sh.root", dir);  
 TFile f(fnameh, "UPDATE"); // open the output file for histograms
 
 TH2 *sigmahtd;
 TH2 *connhd;
 
 char objName[16];
 sprintf(objName, "sigmah%i", tStepi); 
 //f.GetListOfKeys()->Print();
 
 f.GetObject(objName, sigmahtd); // Read the initial surface charge 
density histogram
 f.GetObject("connh", connhd); // Read the connector histogram
 
  
 TVectorD sigmavt(1,N); // sigma vector
 TVectorD Vvt(1,N); // potential vector  
 
 //Fill the sigma vector:
 


 for (int j = 1; j<=ny; j++){ 
  for (int i = 1; i<=nx; i++){ 
   sigmaht->SetCellContent(i, j, sigmahtd->GetCellContent(i,j));
  }
 }

 for (int j = 1; j<=ny; j++){ 
  for (int i = 1; i<=nx; i++){ 
   connh->SetCellContent(i, j, connhd->GetCellContent(i,j));
  }
 }
 
 delete sigmahtd;
 delete connhd; 
 
 for (int j = 1; j<=ny; j++){
  for (int i = 1; i<=nx; i++){
   sigmavt(i+nx*(j-1)) = sigmaht->GetCellContent(i,j);
  }
 }
  
 double Kt; // Surface current density variable
 double It; // Discharge current variable
 double Rcon; // Connector resistance variable
 double sumt; // Charge sum variable


 for (int it = tStepi + 1; it <= tStepi + nofSteps; it++){

  std::cout << "time step " << it << std::endl;
  
  Vvt = l * sigmavt;  
  
  //Update the surface charge density histogram due to the currents:

  // Handling the currents in x:
  for (int j = 1; j<=ny; j++){
   for (int i = 1; i<=nx-1; i++){ 
  
    Kt = ((-1)*(Vvt(i+1+nx*(j-1)) - Vvt(i+nx*(j-1)))*sigmas)/ax;
    sigmaht->SetCellContent(i, j, sigmaht->GetCellContent(i,j) - (Kt
*Deltat*ay)/DeltaSn);   
    sigmaht->SetCellContent(i+1, j, sigmaht->GetCellContent(i+1,j) +
 (Kt*Deltat*ay)/DeltaSn); 
  
   }
  }
  
  // Handling the currents in y:
  for (int j = 1; j<=ny-1; j++){
   for (int i = 1; i<=nx; i++){ 
  
    Kt = ((-1)*(Vvt(i+nx*j) - Vvt(i+nx*(j-1)))*sigmas)/ay;
    sigmaht->SetCellContent(i, j, sigmaht->GetCellContent(i,j) - (Kt
*Deltat*ax)/DeltaSn);   
    sigmaht->SetCellContent(i, j+1, sigmaht->GetCellContent(i,j+1) +
 (Kt*Deltat*ax)/DeltaSn); 
  
   }
  }
  
  //Discharge currents:
  for (int j = 1; j<=ny; j++){
   for (int i = 1; i<=nx; i++){ 
     if(connh->GetCellContent(i,j) < 1e20) {
   
     Rcon = (1/(2*sigmas)) + connh->GetCellContent(i,j);
     
     It = ((-1)*(Vvt(1+nx*(j-1)) - 0)) / Rcon;
     //cout << "Rcon = " << Rcon << endl;
     //cout << "It = " << It << endl;
     sigmaht->SetCellContent(i, j, sigmaht->GetCellContent(i,j) + (I
t*Deltat)/DeltaSn);  
     
    }
  
   }
  }
  
  //Update the sigmavt 
  for (int j = 1; j<=ny; j++){
   for (int i = 1; i<=nx; i++){
    sigmavt(i+nx*(j-1)) = sigmaht->GetCellContent(i,j);
   }
  }
  
  //sumt = sigmavt.Sum()*DeltaSn;
  //cout << "sum(" << it << ") = " << sumt << endl; // Sanity check 
(Charge Conservation).
 
 }
 
 char sigmahfName[16];
 sprintf(sigmahfName, "sigmah%i", tStepi+nofSteps); 
 TH2 *sigmahf; // Final surface charge density histogram
 
  
 sigmahf = new TH2D(Form("Sigmah%i(x,y)",tStepi+nofSteps), "Surface 
charge density", nx, -lx/2, lx/2, ny, -ly/2, ly/2) ;
 
 //Fill the final surface charge density histogram: 
 for (int j = 1; j<=ny; j++){ 
   for (int i = 1; i<=nx; i++){ 
    sigmahf->SetCellContent(i, j, sigmaht->GetCellContent(i,j));
    //std::cout << sigmahtd->GetCellContent(i,j) << std::endl;
   }
  }

 
 sigmahf->Write(sigmahfName); // Write the histogram after transient
 calculation
 
 delete sigmahf;
 
 f.Close();
 
 delete sigmaht;
 delete connh;

}
\end{lstlisting}
\begin{lstlisting}
#ifndef __GETVC__
#define __GETVC__
//#include "../Chani.h"
#include "../classes/Readp.h"

// This function reads and returns the potential of the i-th cell

double getV(int i){
 
 Readp rp;
 int nx = rp.Getnx();
 int ny = rp.Getny();
 int Nm = rp.GetNm();
 double sigmas = rp.Getsigmas(); 
 
 int N = nx*ny;
 int nsub = N/Nm;
 
 int iV = (i - 1)/nsub + 1; //indice of the v sub-vector
 
 double V; // Potential value to return

 TString fname = TString::Format("%sVsigma%i", dir, iV);

 TFile f(fname); // open the output file of vectors
 
 TVectorD *Vp; // pointer for the V-vector

 char labelV[5];
 sprintf(labelV, "V%i", iV);

 f.GetObject(labelV, Vp);

 V = (*Vp)(i - (iV-1) * nsub);
 
 delete Vp; 
 f.Close();
 
 return V;
}

#endif
\end{lstlisting}
\begin{lstlisting}
#include "../Chani.h"
#include "../classes/Readp.h"

double getQtotal(int tStep){
 
 Readp rp;
 double ax = rp.Getax();
 double ay = rp.Getay();
 double nx = rp.Getnx();
 double ny = rp.Getny();
 
 double sum = 0;
 
 TString fnameh = TString::Format("%sh.root", dir);

 TFile f(fnameh); // open the output file for histograms
 
 TH2 *sigmah; // Surface charge density histogram

 char objName[16];
 sprintf(objName, "sigmah%i", tStep); 
 f.GetObject(objName, sigmah); // read the surface charge distributi
on at time = Deltat * tStep
 
 for (int j = 1; j<=ny; j++){ 
  for (int i = 1; i<=nx; i++){ 
   sum = sum + sigmah->GetCellContent(i,j);
  }
 }
 
 delete sigmah;
 
 sum = sum * ax * ay; //from charge density to total charge
 
 f.Close();
 
 return sum;
 
}
\end{lstlisting}
\begin{lstlisting}
#include "../Chani.h"
#include "../globalConsts.h"
#include "../classes/Readp.h"

double capacitance(){

 double C = 0;

 Readp rp;
 double lx = rp.Getlx();
 double ly = rp.Getly();
 double ax = rp.Getax();
 double ay = rp.Getay();
 int nx = rp.Getnx();
 int ny = rp.Getny();
 int Nm = rp.GetNm();
 
 int N = nx*ny;
 double DeltaSn = ax*ay;
 
 std::cout << "Reading the l-matrix" << std::endl;

 TMatrixD l(1,N,1,N);
 int nsub = N / Nm;

 for (int i = 1; i <= Nm; i++){
  for (int j = 1; j<= Nm; j++){
   
   TString fnamel = TString::Format("%slmatrix%i.root",dir, i+(j-1)*
Nm);
      
   TFile f(fnamel); // open the input file.

   char label[12];
   sprintf(label, "l-matrix%i", i+(j-1)*Nm);

   TMatrixD *lp; // pointer of the sub matrix
   f.GetObject(label,lp);

   for (int m = (j-1)*nsub+1; m <= j*nsub; m++){
    for (int n = (i-1)*nsub+1; n<= i*nsub; n++){

     int mm = m - (j-1)*nsub;
     int nm = n - (i-1)*nsub;

     l(m,n) = (*lp)(mm,nm);

    }
   }

   delete lp;
   f.Close();
  }
 }

 std::cout << "Reading of the l-matrix is completed" << std::endl;

 TMatrixD linv(1,N,1,N);

 linv = l;

 std::cout << "Calculating the inverse matrix" << std::endl;

 linv.Invert();

 std::cout << "Matrix inversions is completed" << std::endl;

 TVectorD V(1,N);
 TVectorD sigma(1,N);
 
 V += 1; // surface with unit potential
 
 sigma = linv * V; // surface charge density distribution
 
 C = sigma.Sum() * DeltaSn; // capacitance

 TH2D * sigmah;
 sigmah = new TH2D("sigma(x,y)", "Charge distribution corresponding 
to unit potential", nx, -lx/2, lx/2, ny, -ly/2, ly/2);
 
 for (int j = 1; j<=ny; j++){
  for (int i = 1; i<=nx; i++){
   sigmah->SetCellContent(i, j, sigma(i+nx*(j-1)));
  }
 }
 
 TString fnameh = TString::Format("%sh.root", dir);

 TFile f(fnameh, "UPDATE"); // open the output file for histograms

 sigmah->Write("sigmahVconst"); // Write the histogram after transie
nt calculation

 f.Close();
 
 delete sigmah;
 
 return C;
 
}
\end{lstlisting}
\begin{lstlisting}
#include "../Chani.h"
#include "../globalConsts.h"
#include "../classes/Findij.h" 

// This function returns the total charge on a space point x,y at th
e
// time step tStep

double getQxy(double x, double y, int tStep){

double Qxy;

Findij fijQxy(x,y);
int i = fijQxy.Geti();
int j = fijQxy.Getj();
double ax = fijQxy.Getax();
double ay = fijQxy.Getay();

TString fnameh = TString::Format("%sh.root", dir);
 
TFile f(fnameh); // open the output file for histograms

TH2 *sigmah; // Surface charge density histogram
 
char objName[16];

sprintf(objName, "sigmah%i", tStep);
 
f.GetObject(objName, sigmah); // read the surface charge distributio
n at time = Deltat * tStep

Qxy = sigmah->GetCellContent(i,j);

delete sigmah;
  
f.Close(); 

return Qxy;

}
\end{lstlisting}
\begin{lstlisting}
#ifndef __READPH__
#define __READPH__

#include "../Chani.h"

class Readp{

 public:

 double lx, ly, ax, ay, sigmas;
 int nx, ny, Nm;
 Readp( );

 double Getlx();
 double Getly();
 double Getax();
 double Getay();
 int Getnx();
 int Getny();
 int GetNm();
 double Getsigmas();

};

#endif // __READPH__
\end{lstlisting}
\begin{lstlisting}
#ifndef __READPC__
#define __READPC__

#include "../Chani.h"
#include "Readp.h"

Readp::Readp(){
  char fname[100];
  sprintf(fname, "%sparameters.root", dir);

  TFile f(fname); // open the paramsxy.root file. EDIT THIS LINE ACC
ORDING TO YOUR WORKING DIRECTORY.
  //f.GetListOfKeys()->Print(); // Print to see if the objects are r
etrieved properly

  TH1 *params; // The histogram to hold lx, nx, ax, ly, ny, ay value
s in

  f.GetObject("params",params);
  lx = params->GetBinContent(1);
  nx = params->GetBinContent(2);
  ax = params->GetBinContent(3);
  ly = params->GetBinContent(4);
  ny = params->GetBinContent(5);
  ay = params->GetBinContent(6);
  Nm = params->GetBinContent(7);
  sigmas = params->GetBinContent(8);

  f.Close();
 }

double Readp::Getlx() { return lx; }
double Readp::Getly() { return ly; }
double Readp::Getax() { return ax; }
double Readp::Getay() { return ay; }
int Readp::Getnx() { return nx; }
int Readp::Getny() { return ny; }
int Readp::GetNm() { return Nm; }
double Readp::Getsigmas() { return sigmas; }

#endif // __READPC__
\end{lstlisting}
\begin{lstlisting}
#ifndef __FINDIJH__
#define __FINDIJH__

#include "Readp.h"
#include "../Chani.h"

class Findij: public Readp{
 public:
 double x, y;
 int imin, jmin;

 Findij( );
 Findij(double, double);

 int Geti();
 int Getj();

};

#endif // __FINDIJH__
\end{lstlisting}
\begin{lstlisting}
#ifndef __FINDIJC__
#define __FINDIJC__

#include "../Chani.h"
#include "Findij.h" 

Findij::Findij(double x, double y){
 //Reading x & y vectors:  
 char fnamev[100];
 sprintf(fnamev, "%sxy.root", dir);
 
 TFile f(fnamev); // open the paramsxy.root file. EDIT THIS LINE ACC
ORDING TO YOUR WORKING DIRECTORY.
  
 TVectorD *xp; // Pointer for x-vector
 f.GetObject("x-vector",xp);
  
 TVectorD *yp; // Pointer for y-vector
 f.GetObject("y-vector",yp);
  
 int N = nx*ny;
  
 TVectorD xv(1,N);
 xv = *xp;
 delete xp;
 TVectorD yv(1,N);
 yv = *yp;
 delete yp;
  
 f.Close();
 //Dummy variables:
 double dist, distmin;
  
 for (int j = 1; j < ny+1; j++){
  for (int i=1; i < nx+1; i++){

  dist = pow((xv(i+nx*(j-1))-x),2) + pow((yv(i+nx*(j-1))-y),2);
  //cout << "dist = " << dist << " i, j = " << i << ", " << j << " x
v, yv = " << xv(i+nx*(j-1)) << " , " << yv(i+nx*(j-1)) << endl;
   
  if (i == 1 && j == 1){distmin = dist; imin = i; jmin = j;}
  if (dist <= distmin){distmin = dist; imin = i; jmin = j;} 
  // please note that, 
  // if the connector coordinates are in the middle of two cells, th
e one comes late is automatically choosen

  }
 }
 
//cout << "imin, jmin = " << imin << ", " << jmin << endl;
}

int Findij::Geti() { return imin; }
int Findij::Getj() { return jmin; }

#endif // __FINDIJC__
\end{lstlisting}

\newpage
\chapter{APPENDIX B \textmd{Definition and simulation files of Sections 3.3 \& 3.4}}

\begin{lstlisting}
#include <iostream>
#include "Chani.h"

/*
Dimensions and number of divisions are defined in this file
All lengths are in meters
Once the macro is run, l-matrix is calculated
//*/

void definitions(){

double lx = 2e-2; // length in x direction
int nx = 15; // number of divisions in x
double ly = 2e-2; // length in y direction
int ny = 15; // number of divisions in y


double sigmas = 1e-5; // surface conductivity

int Nm = 1; // l-matrix is calculated in Nm x Nm part

parameters(lx, nx, ly, ny, Nm, sigmas); // call the function for par
ameter definitions

computelmn(); // call the function that computes the l-matrix

}
\end{lstlisting}
\begin{lstlisting}
#include <iostream>
#include "Chani.h"
#include "TH1"
#include "TH2"
#include "classes/Readp.h"

void connectors(){

Readp rp;
double lx = rp.Getlx();
double ly = rp.Getly();
double ax = rp.Getax();
double ay = rp.Getay();
double nx = rp.Getnx();
double ny = rp.Getny();

double y = -ly/2+ay/2;

while  (y < 1e-2){

 addConnector(-lx/2, y, 0);
 y = y + ay;
 
}

}
\end{lstlisting}
\begin{lstlisting}
#include <iostream>
#include "Chani.h"
#include "TH1"
#include "TH2"

void initialcharge(){

Readp rp;
double lx = rp.Getlx();
double ly = rp.Getly();
double ax = rp.Getax();
double ay = rp.Getay();
int nx = rp.Getnx();
int ny = rp.Getny();
int Nm = rp.GetNm();

int N = nx*ny;
double DeltaSn = ax*ay;

double qadd = (1e4*1.6e-19); //add 1e4 elementary charge
//5 areas due to the 2d gaussian probabilities:
double a0 = (1 - ROOT::Math::normal_cdf(-ax/2, ax, 0)*2)*(1 - ROOT::
Math::normal_cdf(-ay/2, ay, 0)*2);
double a1 = ROOT::Math::normal_cdf_c(ax/2,ax,0)*(1 - ROOT::Math::nor
mal_cdf(-ay/2, ay, 0)*2);
double a2 = ROOT::Math::normal_cdf_c(ax/2,ax,0)*ROOT::Math::normal_c
df_c(ay/2,ay,0);
double a3 = ROOT::Math::normal_cdf_c(ay/2,ay,0)*(1 - ROOT::Math::nor
mal_cdf(-ax/2, ax, 0)*2);

//Add the initial charge with the gaussian probabilities:
addCharge(0, 0, qadd*a0, 0); // C
addCharge(ax, 0, qadd*a1, 0); // R
addCharge(ax, ay, qadd*a2, 0); // UR
addCharge(0, ay, qadd*a3, 0); // U
addCharge(-ax, ay, qadd*a2, 0); // UL
addCharge(-ax, 0, qadd*a1, 0); // L
addCharge(-ax, -ay, qadd*a2, 0); // DL
addCharge(0, -ay, qadd*a3, 0); // D
addCharge(ax, -ay, qadd*a2, 0); // UR

}
\end{lstlisting}

\chapter{APPENDIX B.1 \textmd{Simulation files of Section 3.3}}

\begin{lstlisting}
#include <iostream>
#include "Chani.h"
#include "TH1"
#include "TH2"
#include "connectors.C"

void main(){

 for (int i = 0; i < 2000; i++){

  transient2(i, 1, 1e-10);
 }
}
\end{lstlisting}
\begin{lstlisting}
#include <iostream>
#include "Chani.h"
#include "TH1"
#include "TH2"
#include "TCanvas"
#include "TMath"
#include "TMatrixD"
#include "TVectorD"
#include "math.h"
#include "TFile"

void getresults(){

Readp rp;
double lx = rp.Getlx();
double ly = rp.Getly();
double ax = rp.Getax();
double ay = rp.Getay();
int nx = rp.Getnx();
int ny = rp.Getny();

cout << " ax = " << ax << endl;

double Deltat = 1e-11;

int tStep1 = 0; 
int t1 = tStep1 * Deltat * 1e9;
int tStep2 = 200; 
int t2 = tStep2 * Deltat * 1e9;
int tStep3 = 500; 
int t3 = tStep3 * Deltat * 1e9;
int tStep4 = 800;
int t4 = tStep4 * Deltat * 1e9;



TH2 *sigmahd1; // Surface charge density histogram for drawing  
sigmahd1 = new TH2D("Sigmah0(x,y)", Form("Surface charge density at 
t = %i ns", t1) , nx, -lx/2, lx/2, ny, -ly/2, ly/2) ;

TH2 *sigmahd2; // Surface charge density histogram for drawing  
sigmahd2 = new TH2D("Sigmah10(x,y)", Form("Surface charge density at
 t = %i ns", t2) , nx, -lx/2, lx/2, ny, -ly/2, ly/2) ;

TH2 *sigmahd3; // Surface charge density histogram for drawing  
sigmahd3 = new TH2D("Sigmah100(x,y)", Form("Surface charge density a
t t = %i ns", t3) , nx, -lx/2, lx/2, ny, -ly/2, ly/2) ;

TH2 *sigmahd4; // Surface charge density histogram for drawing  
sigmahd4 = new TH2D("Sigmah200(x,y)", Form("Surface charge density a
t t = %i ns", t4) , nx, -lx/2, lx/2, ny, -ly/2, ly/2) ;

char fnameh[100];
sprintf(fnameh, "%sh.root", dir);  
TFile f(fnameh); // open the output file for histograms
f.GetListOfKeys()->Print(); // Print to see if the objects are retri
eved properly

TH2 *sigmah1; // Surface charge density histogram

TH2 *sigmah2; // Surface charge density histogram

TH2 *sigmah3; // Surface charge density histogram

TH2 *sigmah4; // Surface charge density histogram


 
char objName1[16];
sprintf(objName1, "sigmah%i", tStep1); 

char objName2[16];
sprintf(objName2, "sigmah%i", tStep2); 

char objName3[16];
sprintf(objName3, "sigmah%i", tStep3); 

char objName4[16];
sprintf(objName4, "sigmah%i", tStep4); 

f.GetObject(objName1, sigmah1); // read the surface charge distribut
ion at time = Deltat * tStep

f.GetObject(objName2, sigmah2); // read the surface charge distribut
ion at time = Deltat * tStep

f.GetObject(objName3, sigmah3); // read the surface charge distribut
ion at time = Deltat * tStep

f.GetObject(objName4, sigmah4); // read the surface charge distribut
ion at time = Deltat * tStep
 
//Fill the surface charge density histograms: 
for (int j = 1; j<=ny; j++){ 
 for (int i = 1; i<=nx; i++){ 
  sigmahd1->SetCellContent(i, j, sigmah1->GetCellContent(i,j));
 }
}
for (int j = 1; j<=ny; j++){ 
 for (int i = 1; i<=nx; i++){ 
  sigmahd2->SetCellContent(i, j, sigmah2->GetCellContent(i,j));
 }
}
for (int j = 1; j<=ny; j++){ 
 for (int i = 1; i<=nx; i++){ 
  sigmahd3->SetCellContent(i, j, sigmah3->GetCellContent(i,j));
 }
}
for (int j = 1; j<=ny; j++){ 
 for (int i = 1; i<=nx; i++){ 
  sigmahd4->SetCellContent(i, j, sigmah4->GetCellContent(i,j));
 }
}

f.Close(); 

sigmahd1->SetStats(0);
sigmahd2->SetStats(0);
sigmahd3->SetStats(0);
sigmahd4->SetStats(0);

TCanvas *c1 = new TCanvas("c1","c1",1000,600);
c1->Divide(2,2);
c1->cd(1);
sigmahd1->Draw("colz");

c1->cd(2);
sigmahd2->Draw("colz");

c1->cd(3);
sigmahd3->Draw("colz");

c1->cd(4);
sigmahd4->Draw("colz");

c1->cd();

cout << t4 << endl;

}
\end{lstlisting}
\begin{lstlisting}
#include <iostream>
#include "Chani.h"
#include "TH1"
#include "TH2"
#include "TCanvas"
#include "TMath"
#include "TMatrixD"
#include "TVectorD"
#include "math.h"
#include "TFile"

void totalcharge(){

Readp rp;
double lx = rp.Getlx();
double ly = rp.Getly();
double ax = rp.Getax();
double ay = rp.Getay();
int nx = rp.Getnx();
int ny = rp.Getny();
int Nm = rp.GetNm();

//Total charge histogram:


int tStepi = 0; // first time step
int tStepf = 2000; // final time step
double Deltat = 1e-10; // duration of time steps
int DeltatStep = 1; // histograms corresponding to 0, DeltatStep, 2*
DeltatStep, 3*DeltatStep... instances are recorded
int nBinTotal = (tStepf-tStepi)/DeltatStep + 1; //total number of bi
ns

double ti = tStepi*Deltat;
double tf = tStepf*Deltat;

TH1 *totalcharge;
totalcharge = new TH1D("Q vs t","Total charge versus time",nBinTotal
, ti, tf);

char fnameh[100];

sprintf(fnameh, "%sh.root", dir);  

TFile f(fnameh); // open the output file for histograms

f.GetListOfKeys()->Print(); // Print to see if the objects are retri
eved properly

for (int it = 0; it <= tStepf; it += DeltatStep){

 TH2 *sigmah; // Surface charge density histogram
 
 char objName[16];
 

 sprintf(objName, "sigmah%i", it); 
 f.GetObject(objName, sigmah); // read the surface charge distributi
on at time = Deltat * tStep

 double sum = 0;

 for (int j = 1; j<=ny; j++){ 
  for (int i = 1; i<=nx; i++){ 
   sum = sum + sigmah->GetCellContent(i,j);
  }
 }
 
 sum = sum * ax * ay; //from surface charge density to total charge
 
 totalcharge->SetBinContent(it/DeltatStep+1, sum);
 
 delete sigmah;
  
}
f.Close(); 

totalcharge->SetStats(0);
totalcharge->SetXTitle("Time (s)");
totalcharge->SetYTitle("Charge (Coulomb)");
totalcharge->Draw();
//*/
}
\end{lstlisting}

\chapter{APPENDIX B.2 \textmd{Simulation file of Section 3.4.1}}

\begin{lstlisting}
#include <iostream>
#include "Chani.h"
#include "TH1"
#include "TH2"

void CvsN(){

// Dimentions and initial number of divisons in x and y axes:

double lx = 2e-2;
double nxi = 7;
double ly = 2e-2;
double nyi = 7;

//Relative Tolerance for Capacitance 
double tol = 0.001;

//Absolut maximum number for increasing of nx & ny:
int imax = 30;

// nx/ny ratio should be held constant
double ratio = nxi/nyi; // ly < lx should be Choosen

TH1 * Chd;
Chd = new TH1D("C DUMMY", "DUMMY Capacitance with increasing number 
of divisions", imax+1, 0, imax);

TH1 * nxhd;
nxhd = new TH1I("nx DUMMY", "DUMMY Increase in nx", imax+1, 0, imax)
;

TH1 * nyhd;
nyhd = new TH1I("ny DUMMY", "DUMMY Increase in ny", imax+1, 0, imax)
;

TH1 * Nhd;
Nhd = new TH1I("N DUMMY", "DUMMY Increase in total number of divisio
ns", imax+1, 0, imax);

TH1 * Ehd;
Ehd = new TH1D("Error DUMMY", "DUMMY Error vs increasing number of d
ivisions", imax, 1, imax);

int nx, ny;
double sigmas = 1e-5; // surface conductivity (irrelevant for this c
alculation.)
double Nm = 1; // 1-Part matrix calculation
double C, error;

for (int i = 0; i <= imax+1; i++) {
  
 cout << "i = " << i << endl; 
  
 ny = nyi + i; 
 nx = nxi + ratio*i;  
   
 parameters(lx, nx, ly, ny, Nm, sigmas); // call the function for pa
rameter definitions

 computelmn(); // call the function that computes the l-matrix
 
 C = capacitance();
 
 Chd->SetBinContent(i, C);
 nxhd->SetBinContent(i, nx);
 nyhd->SetBinContent(i, ny);
 Nhd->SetBinContent(i, nx*ny);
 
 if (i >= 1){
  
  error = fabs((C - Chd->GetBinContent(i-1)))/Chd->GetBinContent(i-1
);
  Ehd->SetBinContent(i, error);
  
  
  if (error < tol) { 
   
   cout << "Tolerance for relative error is achieved for nx = " << n
x << " ny = " << ny << endl; 
   
   break; 
   
   }
   
 } 
 
}

TH1 * Ch;
Ch = new TH1D("C vs N", "Capacitance vs Number of divisions", i+1, n
xi*nyi, nx*ny); 

for (int j = 0; j <= i; j++){
 Ch->SetBinContent(j+1, Chd->GetBinContent(j));
 cout << "Ch("<< j << ")= " << Ch->GetBinContent(j) << endl;
}

delete Chd;

TH1 * nxh;
nxh = new TH1D("nx", "Increase in nx", i+1, 0, i);

for (int j = 0; j <= i; j++){
 nxh->SetBinContent(j+1, nxhd->GetBinContent(j));
}

delete nxhd;

TH1 * nyh;
nyh = new TH1I("ny", "Increase in ny", i+1, 0, i);

for (int j = 0; j <= i; j++){
 nyh->SetBinContent(j+1, nyhd->GetBinContent(j));
}

delete nyhd;

TH1 * Nh;
Nh = new TH1I("N", "Increase in total number of divisions", i+1, 0, 
i);

for (int j = 0; j <= i; j++){
 Nh->SetBinContent(j+1, Nhd->GetBinContent(j));
}

delete Nhd;

TH1 * Ehd;
Eh = new TH1D("Error", "Relative Error vs Number of divisions", i,  
nxi*nyi, nx*ny);

for (int j = 1; j <= i; j++){
 Eh->SetBinContent(j, Ehd->GetBinContent(j));
}

delete Ehd;

Ch->SetStats(0);
Ch->SetYTitle("Capacitance (F)");
Ch->SetXTitle("Number of cells");
Ch->SetTitleOffset(1.4, "Y");
nxh->SetStats(0);
nyh->SetStats(0);
Nh->SetStats(0);
Eh->SetStats(0);
Eh->SetXTitle("Number of cells");
Eh->SetYTitle("Relative error in capacitance");
Eh->SetTitleOffset(1.4, "Y");

TCanvas *c1 = new TCanvas("c1","c1",600,400);

c1->cd(1);

  Ch->Draw();

c1->cd();

TCanvas *c2 = new TCanvas("c2","c2",600,400);

c2->cd(1);

  Eh->Draw();

c2->cd();

TH2 *sigmah; // Surface charge density histogram for drawing  
sigmah = new TH2D("Sigma(x,y)", "Surface charge distribution corresp
onding to unit potential" , nx, -lx/2, lx/2, ny, -ly/2, ly/2) ;

TH2 *sigmahd; // Surface charge density histogram for drawing  

char fnameh[100];
sprintf(fnameh, "%sh.root", dir);  

TFile f(fnameh); // open the output file for histograms
 
 f.GetObject("sigmahVconst", sigmahd);
 
 for (int j = 1; j<=ny; j++){ 
  for (int i = 1; i<=nx; i++){ 
   sigmah->SetCellContent(i, j, sigmahd->GetCellContent(i,j));
  }
 }
 
f.Close();

sigmah->SetStats(0);

TCanvas *c3 = new TCanvas("c3","c3",600,600);
c3->cd(1);

 sigmah->Draw("colz");

c3->cd();

}
\end{lstlisting}

\chapter{APPENDIX B.3 \textmd{Simulation file of Section 3.4.2}}

\begin{lstlisting}
#include <iostream>
#include "Chani.h"
#include "TH1"
#include "TH2"
#include "connectors.C"
#include "initialcharge.C"

void tvsN(){

// Dimentions and initial number of divisons in x and y axes:

double lx = 2e-2;
double nxi = 7;
double ly = 2e-2;
double nyi = 7;

//Relative Tolerance for Discharge Time 
double tol = -0.001;

//Absolut maximum number for increasing of nx & ny:
int imax = 20;

// nx/ny ratio should be held constant
double ratio = nxi/nyi; // ly < lx should be Choosen

TH1 * tdhd;
tdhd = new TH1D("t DUMMY", "DUMMY Discharge time with increasing num
ber of divisions", imax+1, 0, imax);

TH1 * nxhd;
nxhd = new TH1I("nx DUMMY", "DUMMY Increase in nx", imax+1, 0, imax)
;

TH1 * nyhd;
nyhd = new TH1I("ny DUMMY", "DUMMY Increase in ny", imax+1, 0, imax)
;

TH1 * Nhd;
Nhd = new TH1I("N DUMMY", "DUMMY Increase in total number of divisio
ns", imax+1, 0, imax);

TH1 * Ehd;
Ehd = new TH1D("Error DUMMY", "DUMMY Error vs increasing number of d
ivisions", imax, 1, imax);

int nx, ny, N;
double qadd = 1e4*1.6e-19;
double sigmas = 1e-5; // surface conductivity
double Nm = 1; // 1-Part matrix calculation
double C, error, tdischarge;

double Deltat = 1e-10;

for (int i = 0; i <= imax+1; i += 2) {
  
 cout << "i = " << i << endl; 
  
 ny = nyi + i; 
 nx = nxi + ratio*i;  
   
 parameters(lx, nx, ly, ny, Nm, sigmas); // call the function for pa
rameter definitions

 computelmn(); // call the function that computes the l-matrix
 
 connectors();
 
 addCharge(0, 0, qadd, 0);
 
 cout << "qadd = " << qadd << endl;
 
 for (int it = 0; it < 2000; it++){

  transient2(it, 1, Deltat);
  double Q = getQtotal(it+1);
  
  if (Q/qadd < 0.1){ break;}
   
 }
 
 tdischarge = Deltat*it;
 
 tdhd->SetBinContent(i/2+1, tdischarge);
 nxhd->SetBinContent(i/2+1, nx);
 nyhd->SetBinContent(i/2+1, ny);
 Nhd->SetBinContent(i/2+1, nx*ny);
 
 if (i/2 >= 1){
  
  cout << "tdischarge = " << tdischarge << endl;
  cout << "tdhd( " << i/2 << ") = " << tdhd->GetBinContent(i/2) << e
ndl;
   
  
  error = (tdischarge - tdhd->GetBinContent(i/2)) / tdhd->GetBinCont
ent(i/2);
  
  error = fabs(error);
  
  cout << "error = " << error << endl;
  
  Ehd->SetBinContent(i/2, error);
  
  
  if (error < tol) { 
   
   cout << "Tolerance for relative error is achieved for nx = " << n
x << " ny = " << ny << endl; 
   
   break; 
   
   }
   
 } 

 
}

TH1 * tdh;
tdh = new TH1D("t_discharge vs N", "Discharge time vs Number of divi
sions", i/2+1, nxi*nyi, nx*ny); 

for (int j = 0; j <= i/2; j++){
 tdh->SetBinContent(j+1, tdhd->GetBinContent(j+1));
 cout << "tdh("<< j << ")= " << tdh->GetBinContent(j+1) << endl;
}

delete tdhd;

TH1 * nxh;
nxh = new TH1D("nx", "Increase in nx", i+1, 0, i);

for (int j = 0; j <= i/2; j++){
 nxh->SetBinContent(j+1, nxhd->GetBinContent(j+1));
}

delete nxhd;

TH1 * nyh;
nyh = new TH1I("ny", "Increase in ny", i+1, 0, i);

for (int j = 0; j <= i/2; j++){
 nyh->SetBinContent(j+1, nyhd->GetBinContent(j+1));
}

delete nyhd;

TH1 * Nh;
Nh = new TH1I("N", "Increase in total number of divisions", i+1, 0, 
i);

for (int j = 0; j <= i/2; j++){
 Nh->SetBinContent(j+1, Nhd->GetBinContent(j+1));
}

delete Nhd;

TH1 * Ehd;
Eh = new TH1D("Error", "Relative Error vs Number of divisions", i,  
nxi*nyi, nx*ny);

for (int j = 1; j <= i/2; j++){
 Eh->SetBinContent(j, Ehd->GetBinContent(j));
}

delete Ehd;

tdh->SetStats(0);
tdh->SetYTitle("Discharge time (s)");
tdh->SetXTitle("Number of cells");
tdh->SetTitleOffset(1.4, "Y");
nxh->SetStats(0);
nyh->SetStats(0);
Nh->SetStats(0);
Eh->SetStats(0);
Eh->SetXTitle("Number of cells");
Eh->SetYTitle("Relative error in discharge time");
Eh->SetTitleOffset(1.4, "Y");

TCanvas *c1 = new TCanvas("c1","c1",600,400);

c1->cd(1);

  tdh->Draw();

c1->cd();

TCanvas *c2 = new TCanvas("c2","c2",600,400);

c2->cd(1);

  Eh->Draw();

c2->cd();
//*/
}
\end{lstlisting}

\chapter{APPENDIX C \textmd{Definition and simulation files of Section 3.5}}

\begin{lstlisting}
#include <iostream>
#include "Chani.h"

/*
Dimensions and number of divisions are defined in this file
All lengths are in meters
Once the macro is run, l-matrix is calculated
//*/

void definitions(){

double lx = 4e-2; // length in x direction
int nx = 99; // number of divisions in x
double ly = 4e-2; // length in y direction
int ny = 99; // number of divisions in y


double sigmas = 1e-5; // surface conductivity

int Nm = 1; // l-matrix is calculated in Nm x Nm part

parameters(lx, nx, ly, ny, Nm, sigmas); // call the function for par
ameter definitions

computelmn(); // call the function that computes the l-matrix

}
\end{lstlisting}
\begin{lstlisting}
// this file adds a periodic charge distribution on the surface 
// sigma = sigma0 * sin(kx) e^(-t/tau)
// Chani will simulate the relaxation of the charge.

#include <iostream>
#include "Chani.h"
#include "TH2.h"
#include "TMath.h"

// compile and run with:
// g++ `root-config --glibs` -I `root-config --incdir` initialcharge
sin.C ; ./a.out

//void initialchargesin(){
int initialchargesin(){

Readp rp;
double lx = rp.Getlx();
double ly = rp.Getly();
double ax = rp.Getax();
double ay = rp.Getay();
int nx = rp.Getnx();
int ny = rp.Getny();
int Nm = rp.GetNm();

int N = nx*ny;
double DeltaSn = ax*ay;

TString fnamev = TString::Format("%sxy.root", dir);

// Reading of x & y vectors: 
TFile f(fnamev); 
  
TVectorD *xp; // Pointer for x-vector
f.GetObject("x-vector",xp);
TVectorD x(1,N);
x = *xp;
delete xp;

TVectorD *yp; // Pointer for y-vector
f.GetObject("y-vector",yp);
TVectorD y(1,N);
y = *yp;
delete yp;

f.Close();

double k = 1000;

cout << "k = " << k << endl;

double sigmaadd0 = (1e4*1.6e-19)/DeltaSn; //1e4 elementary charge

// add the initial periodic charge:

double sigmaadd;

TH2 *sigmah0;
sigmah0 = new TH2D("sigma0(x,y)", "Surface charge density at t = 0",
 nx, -lx/2, lx/2, ny, -ly/2, ly/2) ;
 
for (int j = 1; j <= ny; j++){
 for (int i =  1; i <= nx; i++){
 
 //std::cout << "i = " << i << std::endl;
 
 sigmaadd = sigmaadd0*cos(k*x(i+(j-1)*nx));
 sigmah0->SetCellContent(i,j,sigmaadd);
 
 }
}

TString fnameh = TString::Format("%sh.root", dir);

TFile f2(fnameh,"UPDATE"); // open the output file for histograms

sigmah0->Write("sigmah0"); // Write the surface with periodic initia
l charge
 
f2.Close();

}

#ifndef __CINT__
int main(){
 initialchargesin2();
 return 0;
}
#endif
\end{lstlisting}
\begin{lstlisting}
#include <iostream>
#include "Chani.h"
#include "TH1.h"
#include "TH2.h"
#include "connectors.C"

int main(){
 
 int tStepi = 0;
 int tStepf = 1000;
 int DeltatStep = 1;
 double Deltat = 1e-11; // 0.01 ns

 //double C = capacitance();
 
 //std::cout << "C = " << C << std::endl;

 for (int i = tStepi; i < tStepf; i = i + DeltatStep){

  transient2(i, DeltatStep, Deltat);
  //cout << "Q total = " << getQtotal(i) << endl;
 }

 return 0;
}
\end{lstlisting}
\begin{lstlisting}
#include <iostream>
#include "Chani.h"
#include "TH1"
#include "TH2"
#include "TCanvas"
#include "TMath"
#include "TMatrixD"
#include "TVectorD"
#include "math.h"
#include "TFile.h"
#include "classes/Readp.h"

void qxy(){

 int tStepi = 0; // first time step
 int tStepf = 600; // final time step
 double Deltat = 1e-11; // duration of time steps
 int DeltatStep = 10; // histograms corresponding to 0, DeltatStep, 
2*DeltatStep, 3*DeltatStep... instances are recorded
 int nBinTotal = (tStepf-tStepi)/DeltatStep + 1; //total number of b
ins

 double ti = tStepi*Deltat;
 double tf = tStepf*Deltat;

 double x = 0;
 double y = 0;

 TH1 *qxy;
 qxy = new TH1D("Q vs t","Charge at (0,0) versus time",nBinTotal, ti
*1e9, tf*1e9);

 double q;

 for (int it = tStepi; it <= tStepf; it += DeltatStep){
 
  q = getQxy(x,y,it-tStepi);
  //std::cout << "q at time step " << it << " is " << q << std::endl
;
  qxy->SetBinContent(it/DeltatStep+1, q);  
   
 }

 qxy->SetStats(0);
 qxy->SetXTitle("Time (ns)");
 qxy->SetYTitle("Charge (Coulomb)");
 
 TCanvas *c1 = new TCanvas("c1","c1",600,400);
 
 qxy->Draw();
 qxy->Fit("expo");

 TF1 *fit = qxy->GetFunction("expo");

 //fit->SetParameter(0, log(qxy->GetBinContent(tStepi)));
  
 //c1->Update();
  
 double tau ;

 tau = (-1)/(fit->GetParameter(1));

 std::cout << "tau = " << tau << std::endl;

}
\end{lstlisting}

\chapter{APPENDIX D \textmd{Definition and simulation files of Section 3.6}}

\begin{lstlisting}
#include <iostream>
#include "Chani.h"

/*
Dimensions and number of divisions are defined in this file
All lengths are in meters
Once the macro is run, l-matrix is calculated
//*/

void definitions(){

double lx = 10e-2; // length in x direction
int nx = 667; // number of divisions in x
double ly = 150e-6; // length in y direction
int ny = 1; // number of divisions in y


double sigmas = 33.3e-6; // surface conductivity

int Nm = 1; // l-matrix is calculated in Nm x Nm part

parameters(lx, nx, ly, ny, Nm, sigmas); // call the function for par
ameter definitions

computelmn(); // call the function that computes the l-matrix

}
\end{lstlisting}
\begin{lstlisting}
#include <iostream>
#include "Chani.h"
#include "TH1"
#include "TH2"

void connectors(){

addConnector(-5e-2, 0, 0);

}
\end{lstlisting}
\begin{lstlisting}
#include <iostream>
#include "Chani.h"
#include "TH1"
#include "TH2"

void initialcharge(){

Readp rp;
double lx = rp.Getlx();
double ly = rp.Getly();
double ax = rp.Getax();
double ay = rp.Getay();
int nx = rp.Getnx();
int ny = rp.Getny();
int Nm = rp.GetNm();

int N = nx*ny;
double DeltaSn = ax*ay;

double qadd = (1e4*1.6e-19); //add 1e4 elementary charge
//5 areas due to the 2d gaussian probabilities:
double a0 = (1 - ROOT::Math::normal_cdf(-ax/2, ax, 0)*2);
double a1 = ROOT::Math::normal_cdf(-ax/2, ax, 0);

//Add the initial charge with the gaussian probabilities:
addCharge(0, 0, qadd*a0, 0); // C
addCharge(ax, 0, qadd*a1, 0); // R
addCharge(-ax, 0, qadd*a1, 0); // L

}
\end{lstlisting}
\begin{lstlisting}
#include <iostream>
#include "Chani.h"
#include "TH1"
#include "TH2"

void main(){

double C;

for (int i = 0; i < 2000000; i += 1000){

  transient2(i, 1000, 1e-10);

}

}
\end{lstlisting}
\lstinputlisting{appendices/R11/getresults.C}
\begin{lstlisting}
#include <iostream>
#include "Chani.h"
#include "TH1"
#include "TH2"
#include "TCanvas"
#include "TMath"
#include "TMatrixD"
#include "TVectorD"
#include "math.h"
#include "TFile"

void totalcharge(){

Readp rp;
double lx = rp.Getlx();
double ly = rp.Getly();
double ax = rp.Getax();
double ay = rp.Getay();
int nx = rp.Getnx();
int ny = rp.Getny();
int Nm = rp.GetNm();

//Total charge histogram:


int tStepi = 0; // first time step
int tStepf = 400000; // final time step
double Deltat = 1e-10; // duration of time steps
int DeltatStep = 1000; // histograms corresponding to 0, DeltatStep,
 2*DeltatStep, 3*DeltatStep... instances are recorded
int nBinTotal = (tStepf-tStepi)/DeltatStep + 1; //total number of bi
ns

double ti = tStepi*Deltat;
double tf = tStepf*Deltat;

TH1 *totalcharge;
totalcharge = new TH1D("Q vs t","Total charge versus time",nBinTotal
, ti, tf);

char fnameh[100];

sprintf(fnameh, "%sh.root", dir);  

TFile f(fnameh); // open the output file for histograms

f.GetListOfKeys()->Print(); // Print to see if the objects are retri
eved properly

for (int it = 0; it <= tStepf; it += DeltatStep){

 TH2 *sigmah; // Surface charge density histogram
 
 char objName[16];
 

 sprintf(objName, "sigmah%i", it); 
 f.GetObject(objName, sigmah); // read the surface charge distributi
on at time = Deltat * tStep

 double sum = 0;

 for (int j = 1; j<=ny; j++){ 
  for (int i = 1; i<=nx; i++){ 
   sum = sum + sigmah->GetCellContent(i,j);
  }
 }
 
 sum = sum * ax * ay; //from surface charge density to total charge
 
 totalcharge->SetBinContent(it/DeltatStep+1, sum);
 
 delete sigmah;
  
}
f.Close(); 

totalcharge->SetStats(0);
totalcharge->SetXTitle("Time (s)");
totalcharge->SetYTitle("Charge (Coulomb)");
totalcharge->Draw();
//*/
}
\end{lstlisting}

\chapter{APPENDIX E \textmd{Definition and simulation files of Section 3.7}}

\lstinputlisting{appendices/R14/definitions.C}
\begin{lstlisting}
#include <iostream>
#include "Chani.h"
#include "TH1"
#include "TH2"

void connectors(){

double y = -9e-3;

while  (y < 10e-3){

 addConnector(-500e-3, y, 0);
 y = y + 9e-3;
 
}

}
\end{lstlisting}
\lstinputlisting{appendices/R14/initialcharge.C}
\begin{lstlisting}
#include <iostream>
#include "Chani.h"
#include "TH1"
#include "TH2"

void main(){

for (int i = 0; i < 200; i++){

  transient(i, 1, 1e-10);

 }


}
\end{lstlisting}

\ozgecmis{\vspace{10mm}

\setlength{\TPHorizModule}{10pt}
\setlength{\TPVertModule}{10pt}
\begin{textblock}{1}(40,10)
 \begin{figure}[p]
 \includegraphics[scale=0.2,keepaspectratio=true]{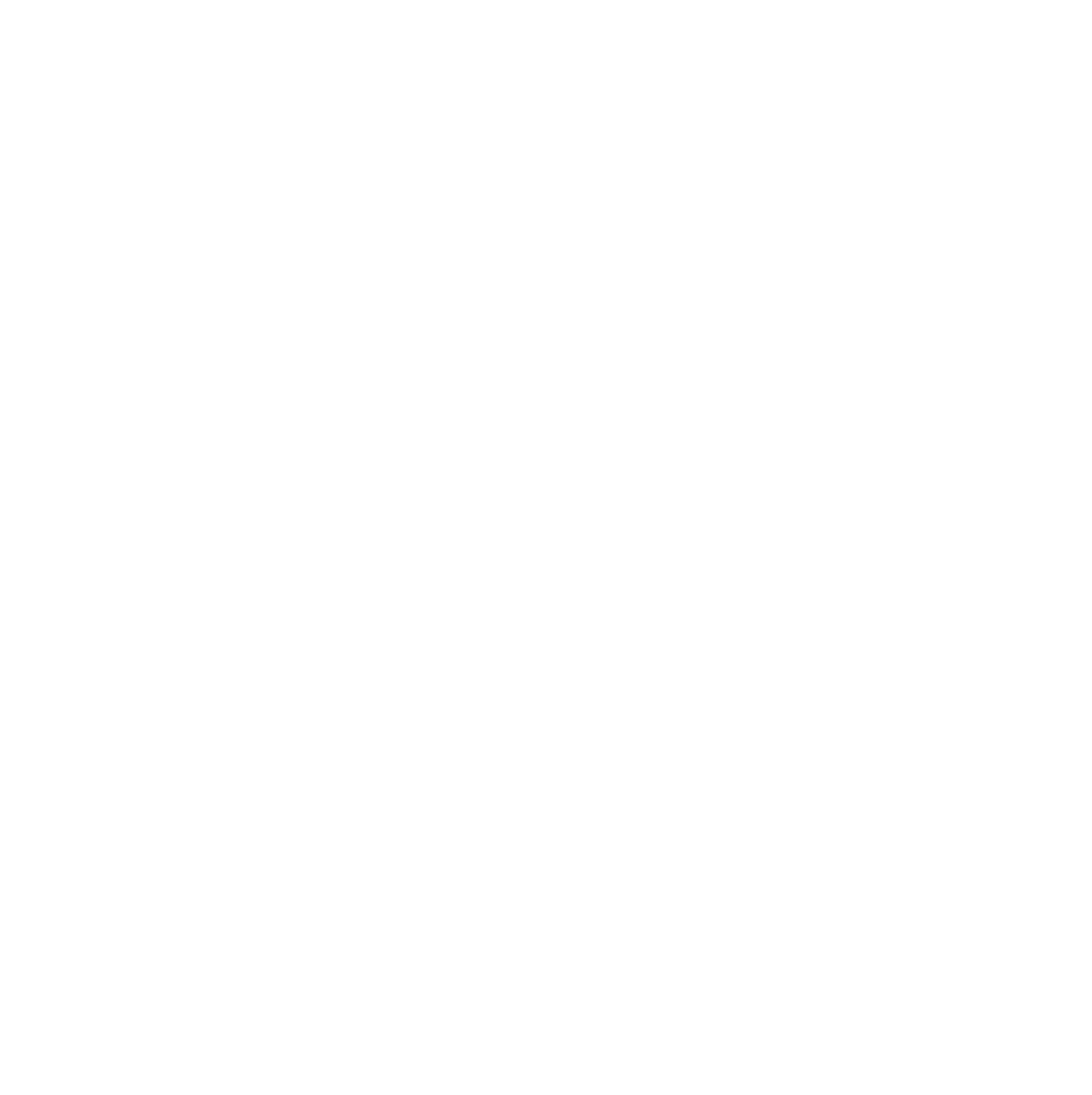}
\end{figure}

\end{textblock}
\textbf{Name Surname:} Nazmi Burak Budanur \\

\vspace{-3mm}
\textbf{Place and Date of Birth:} \.{I}zmir, 23.11.1988 \\


\vspace{-3mm}
\textbf{E-Mail:} burakbudanur@gmail.com \\

\vspace{-3mm}
\textbf{B.Sc.:} Istanbul Technical University, Electronics Engineering \& Physics Engineering \\ 

\vspace{-3mm}
\textbf{M.Sc.:} Istanbul Technical University, Physics Engineering \\

\vspace{-3mm}
\textbf{Professional Experience:} 01/2010 – Present Research and Teaching Assistant at Dogus University (Istanbul), Physics Department \\  




}
\end{document}